\documentclass[aps,prd,twocolumn,superscriptaddress,nofootinbib]{revtex4-1}

\usepackage{aas_macros}
\usepackage[utf8]{inputenc}
\usepackage[T1]{fontenc}
\usepackage[english,french]{babel}
\usepackage{mathrsfs}
\usepackage{amsmath}
\usepackage{amssymb}
\usepackage{tipa}
\usepackage{ntheorem}
\usepackage{braket}
\usepackage{graphicx}
\usepackage{tabularx}
\usepackage{bm}
\usepackage{wasysym}
\usepackage{enumitem}
\usepackage{stmaryrd}
\usepackage[breaklinks=true,colorlinks,citecolor=blue,linkcolor=blue,urlcolor=blue]{hyperref}
\usepackage{tikz}
\usepackage{footmisc}
\usetikzlibrary[patterns]
\usetikzlibrary{shapes}

\theoremseparator{.}

\makeatletter
\newskip\@bigflushglue \@bigflushglue = -100pt plus 1fil

\def\bigcentering{\let\\\@centercr\rightskip\@bigflushglue
\leftskip\@bigflushglue
\parindent\z@\parfillskip\z@skip}

\makeatother

\allowdisplaybreaks[1]

\newcommand{\dd}{\mathrm{d}}

\begin{document}

\title{Impact of dipolar magnetic fields on gravitational wave strain by galactic binaries }

\author{A. Bourgoin}
\email{adrien.bourgoin@obspm.fr}
\affiliation{SYRTE, Observatoire de Paris, PSL Research University, CNRS, Sorbonne Universit\'es, UPMC Univ. Paris 06, LNE, 61 avenue de l'Observatoire, 75014 Paris, France}
\affiliation{Département d'Astrophysique-AIM, CEA/IRFU/DAp, CNRS/INSU, Université Paris-Saclay, Université de Paris, Gif-sur-Yvette, France}

\author{C.~Le~Poncin-Lafitte}
\affiliation{SYRTE, Observatoire de Paris, PSL Research University, CNRS, Sorbonne Universit\'es, UPMC Univ. Paris 06, LNE, 61 avenue de l'Observatoire, 75014 Paris, France}

\author{S. Mathis}
\affiliation{Département d'Astrophysique-AIM, CEA/IRFU/DAp, CNRS/INSU, Université Paris-Saclay, Université de Paris, Gif-sur-Yvette, France}

\author{M.-C. Angonin}
\affiliation{SYRTE, Observatoire de Paris, PSL Research University, CNRS, Sorbonne Universit\'es, UPMC Univ. Paris 06, LNE, 61 avenue de l'Observatoire, 75014 Paris, France}


\begin{abstract}
  White dwarfs (WDs) and neutron stars (NSs) are among the most magnetized astrophysical objects in the universe, with magnetic fields that can reach up to $10^9\,\mathrm{G}$ for WDs and up to $10^{15}\,\mathrm{G}$ for NSs. The galaxy is expected to be populated with approximately one hundred million of double WD and millions of NS-WD binaries. Throughout the duration of the mission, the Laser Interferometer Space Antenna (LISA) will observe gravitational waves (GWs) emitted simultaneously by more than ten thousand of such galactic binaries. In this paper, we investigate the effect of the magnetic dipole-dipole interaction on the GW signal emitted by magnetic galactic binaries. We derive the secular equations governing the orbital and rotational motion of these objects. Then, we integrate these equations both numerically and analytically. We conclude that the overall visible effect is an additional secular drift of the mean longitude. This drift is proportional to the product of the magnetic moments and is inversely proportional to the $7/2$ power of the semi-major axis. Finally, we show that, at zeroth-order in eccentricity, the magnetic dipole-dipole interaction shifts the main frequency of the gravitational strain measured by LISA.
\end{abstract}

\maketitle

\section{Introduction}
\label{sec:int}

The Laser Interferometer Space Antenna (LISA) is the ESA L3 mission that aims at observing gravitational waves (GWs) from space \cite{LISAcoll,2017arXiv170200786A}. The observatory consists of six active laser links between three identical spacecraft in a triangular formation separated by $2.5$ million $\mathrm{km}$. This configuration will allow LISA to observe GWs in the frequency band from below $10^{-4}\,\mathrm{Hz}$ to above $10^{-1}\,\mathrm{Hz}$. Within this range, the main source of GWs are the galactic binaries (GBs). Around ten thousand of these systems should be resolvable by LISA \cite{1990ApJ...360...75H,PhysRevD.73.122001}.

Galactic binaries are comprised primarily of white dwarfs (WDs) but also neutron stars (NSs) and stellar-origin black holes. In LISA's bandwidth, the typical orbital period for GBs of WDs and NSs ranges from minutes to several hours.  This corresponds to a semi-major axis between $10^4\,\mathrm{km}$ to $10^{6}\,\mathrm{km}$ (for a total mass around $1.5\,\mathrm{M}_{\odot}$). Therefore, LISA will observe GWs emitted by GBs during the inspiral phase, which is before the merger which can be detected by ground-based GW detectors such as LIGO \cite{LIGOcoll}, Virgo \cite{Virgocoll}, KAGRA \cite{KRAGAcoll}, and the future Einstein Telescope~\cite{ETcoll}.

The first (extra-galactic) merger of a binary NS (GW170817) was observed in 2017. It was detected simultaneously using GWs by the LIGO and Virgo detectors and across the electromagnetic (EM) spectrum using its $\gamma$-ray, ultraviolet, optical, infrared, and radio band emissions \cite{PhysRevLett.119.161101,2017ApJ...848L..12A}.  The detection of this event in both the GW and EM sectors is the first direct confirmation of the existence of double compact stars mergers. It allowed the determination of the physical properties of the two stars such as their masses, radii, spins, and tidal deformability parameter \cite{2018PhRvL.121p1101A,2019PhRvX...9a1001A}, and placed strong constrains on the fundamental physics of gravity \cite{Abbott_2017}. By probing the earlier inspiral phase of the future galactic GW170817-type systems, LISA will enable scientists to anticipate merger events and perform efficient combined GWs and EM observations. This will bring much information on the long term evolution of GBs, their internal structure, and equation of state \cite{2017arXiv170200786A}.

The galaxy is expected to be populated with approximately one hundred millions of WD-WD systems and millions of NS-WD binaries \cite{2009CQGra..26i4030N}. These compact objects can have intense magnetic fields that may reach up to $10^9\,\mathrm{G}$ for WDs and up to $10^{15}\,\mathrm{G}$ for NSs \cite{2005MNRAS.356..615F}. White dwarfs with magnetic fields ranging from $10^6\,\mathrm{G}$ to $10^9\,\mathrm{G}$ should represent around $20\%$ of the total WD population \cite{2007ApJ...654..499K,2015SSRv..191..111F,2020AdSpR..66.1025F} while NSs with magnetic fields between $10^{14}\,\mathrm{G}$ to $10^{15}\,\mathrm{G}$ (i.e., the magnetars) should represent around $10\%$ of the total NS population \cite{2008MNRAS.387..897T}. The origin of these strong magnetic fields in WDs and NSs is an active area of research in astrophysics (see e.g., \citet{2021MNRAS.507.5902B}) with several scenarios having been proposed.

The first mechanism that would permit WDs and NSs to develop intense magnetic fields is the ``merging scenario''. According to \citet{2008MNRAS.387..897T}, highly magnetic WDs are formed from the merger of cataclysmic variables (i.e., binary systems consisting of a WD and a mass transferring companion). The main observational motivation justifying the merging scenario is the fact that highly magnetic WDs are generally isolated or in cataclysmic variable stars but not in binary systems with a detached low-mass main sequence companion. If highly magnetic WDs are formed from the isolated evolution of a single star, then there should be the same fraction of them observed individually and in binary systems with a detached low-mass main sequence companion, which is not in agreement with most observations. The merging scenario also explains the formation of magnetars. To do so, it relies on the merger of a binary system made of WDs \cite{1997MNRAS.292..205F,2005MNRAS.356..615F}. The main observational justification rests on the fact that magnetars are observed individually and not in binary systems. Unfortunately, only 30 magnetars have been identified so far\footnote{See e.g., McGill Online catalog at \url{http://www.physics.mcgill.ca/~pulsar/magnetar/main.html}.\label{citationurl}}, and the statistics are thus too poor to draw strong conclusions on the reliability of the merging scenario. In addition, it was recently pointed out by \citet{2020A&A...634L..10L}, that magnetic WDs in binary system with a detached main sequence companion may be rare but do exist.

The second mechanism commonly invoked to explain the occurrence of strong magnetic fields is the ``dynamo'' hypothesis \cite{1992ApJ...392L...9D,2001ApJ...559.1094C,2020SciA....6.2732R,2021arXiv211102148R}. This scenario predicts that the strong magnetic field would result from a turbulent dynamo amplification occurring primarily in the convection zone of the progenitor, as well as in differentially rotating nascent NSs. The dynamo hypothesis requires an extremely rapidly rotating nascent NS. Unfortunately, the current population of magnetars seems to favor slow rotators\footref{citationurl}.

The third possibility is the ``fossil-field'' scenario \cite{1945MNRAS.105..166C,1987MNRAS.226..297M}. It has been argued by \citet{2005MNRAS.356..615F} that the origin of strong magnetic fields could also come from progenitors main-sequence stars. The mechanism at work would imply conservation of the magnetic flux during stellar evolution off the main sequence to the degenerate phase (i.e., WD or NS). The candidates for WD progenitors would be the Ap and Bp main-sequence stars with large scale stable dipolar magnetic fields \cite{2005MNRAS.356.1576W} while the progenitors for magnetars would be the stars of spectral type O with strong effective dipolar magnetic field \cite{2005MNRAS.356..615F,2009MNRAS.396..878H}. This scenario is an attractive possibility. It must however be noted that highly magnetic WDs are mostly observed individually and not paired in a detached system with a non-degenerate star (cf. the ``merging scenario''). This is a serious challenge to the fossil-field hypothesis \cite{2015SSRv..191..111F}. This being said, numerical simulations by \citet{2004Natur.431..819B} favor the fossil-field scenario as a natural explanation for the magnetism of non-convective stars. Indeed, the authors show that stable dipolar magnetic field can develop from an arbitrary initial configuration and persist over the lifetime of the stars through magnetohydrodynamic relaxation mechanism (see also \citet{2010A&A...517A..58D}). The equilibrium configuration consists of a combination of an internal twisted toroidal field stabilizing a poloidal field that emerges from the surface of the star as an offset dipolar shape. Furthermore, \citet{2008MNRAS.386.1947B} showed, with magnetohydrodynamic simulations, that the fossil-field scenario is also compatible with the emergence of stable non-axisymmetric field configurations, in agreement with spectroscopic and spectropolarimetric observations \cite{2006MNRAS.370..629D,2007A&A...463..647B,2015SSRv..191..111F}.

In spite of the fact that their observational implications are quite different, neither one of these three scenarios can be favored or dismissed, due to a lack of observations. By increasing the number of observations made in the EM sector, and by observing simultaneously more than ten thousand galactic binary systems, LISA will most likely bring new insights into the nature of the magnetic fields within WDs and NSs. The impact of the magnetic effects on the GW signal must therefore be investigated. Indeed, the future data processing of the LISA mission will require that all observable physical effects be modeled with sufficient accuracy in order to better understand the physics of these compact objects. In addition, because the GBs will be the dominant source of GWs within the galaxy, they can potentially hide signals produced by extra-galactic sources. To avoid the contamination of the latter, removal of the galactic foreground noise from GBs during data processing must be as accurate as possible.

Preliminary studies, that aimed at modeling the GW signal emitted by GBs, have focused on the monochromatic approximation only \cite{PhysRevD.76.083006,LDCGroup}. This corresponds to the well-known circular motion in the Newtonian picture of two point-masses in gravitational interaction. However, it has been shown that a number of physical effects, such as the backreaction induced by gravitational radiation \cite{2009CQGra..26i4030N} or the dynamical tides \cite{1997ApJ...490..847L,2010ApJ...713..239W,2011MNRAS.412.1331F,*2012MNRAS.421..426F,*2013MNRAS.430..274F,*2014MNRAS.444.3488F,2020MNRAS.491.3000M,2017PhRvD..96h3005X,2019PhRvD.100f3001V,2020PhRvD.102h3005W}, can make GBs exhibit a continuous frequency shift, which can potentially be detected over the time-span of the LISA mission. The monochromatic approximation for GBs is motivated by gravitational radiation which is an efficient mechanism for orbit circularization. However, of the ten thousand sources that LISA will observe, it is expected that a non negligible amount of them might be in an eccentric orbit (see e.g., \citet{2021PhRvD.104j4023T} and references therein), and hence, might exhibit discrete frequency domain. In this context, the influence of a wide variety of physical effects on GB's eccentric orbits must be investigated, and their impact on data processing must be quantified in order to best prepare the future data analysis pipeline of the LISA mission.

In this paper, we focus on the impact of the magnetic dipole-dipole interaction on the GWs emitted by GBs, in circular and in quasi-circular orbits. We neglect the effect of dynamical tides for the sake of simplicity. We approximate the magnetic field of the stars by their dipole moments. This approximation is motivated by spectropolarimetric observations \cite{2015SSRv..191..111F} and is also coherent with the fossil-fields hypothesis as discussed by \citet{2004Natur.431..819B}. We consider both the orbital and the rotational motion of the binary system. We assume a general configuration where the magnetic moments can have arbitrary orientations and we let the system evolve under the action of the magnetic torques. In other words, we suppose that the system has not reached an exact equilibrium yet but can oscillate around its equilibrium positions. The gravitational interaction is modeled in the framework of General Relativity (GR) up to the 2.5 Post-Newtonian (PN) order (i.e., up to terms of the order of $c^{-5}$ with $c$ being the speed of light in vacuum). The spin-orbit and the spin-spin interactions are neglected for the sake of simplicity.

The paper is organized as follows. In Sect. \ref{sec:onebody}, the equations of motion including both the magnetic interaction and the GR contribution up to the 2.5PN order are computed. The secular parts of the equations of motion are derived in Sect.~\ref{sec:seceqs}. The secular equations are then solved analytically and numerically in Sect.~\ref{sec:sol}. We show that the magnetic dipole-dipole interaction generates an additional linear time variation on the evolution of the mean longitude and the longitude of the pericenter. The effect of the dipole-dipole interaction on the GW mode polarizations is derived in Sect. \ref{sec:GWsignal}. At zeroth-order in eccentricity, we show that magnetism generate a secular variation of the mean longitude. At first-order in eccentricity, magnetism can be observed through the secular variations of both the mean longitude and the longitude of the pericenter. Finally, we give our conclusions in Sect.~\ref{sec:ccl}.

\section{Dynamics of compact binaries}
\label{sec:onebody}

\subsection{Notations and reference frames}

In this paper, we consider two compact and well-separated bodies that form a binary system. The system consists of a first body (the primary) of mass $m_1$, magnetic moment $\bm{\mu}_1$, position $\mathbf x_1$, and velocity $\mathbf v_1$, and a second body (the secondary) of mass $m_2$, magnetic moment $\bm{\mu}_2$, position $\mathbf x_2$, and velocity $\mathbf v_2$.

The motion is conveniently cast in the form of an effective one-body problem by introducing the relative position $\mathbf x\equiv\mathbf x_2-\mathbf x_1$ and velocity $\mathbf v\equiv\mathbf v_2-\mathbf v_1=\dd\mathbf x/\dd t$. We introduce, the orbital separation $r=|\mathbf x|$, the direction of the secondary with respect to the primary $\hat{\mathbf n}\equiv\mathbf x/r$, and the magnitude of the relative velocity $v=|\mathbf{v}|$. We also introduce the following useful mass parameters
\begin{equation}
  m\equiv m_1+m_2\text{,} \qquad \Delta\equiv\frac{m_1-m_2}{m}\text{,} \qquad \eta\equiv\frac{m_1m_2}{m^2}\text{,}
\end{equation}
with $m$ being the total mass, $\Delta$ the relative mass difference, and $\eta$ the symmetric mass ratio. The notations used throughout this paper are summarized in Tab. \ref{tab:notations}.

We now define the different reference frames that are used hereafter. First, let $(\hat{\mathbf e}_X,\hat{\mathbf e}_Y,\hat{\mathbf e}_Z)$ be a right-handed vectorial basis which we refer to as the ``source frame'' and is used to describe the motion of the source of the GW signal. The origin of this frame is attached to the barycenter of the binary system. The $z$-axis points in the direction of the observer, assumed to be in the ``far-away wave zone'' (see \citet{2014grav.book.....P}). Accordingly, the axes of the source frame can be chosen non-rotating with respect to distant stars such that the source frame is considered inertial.

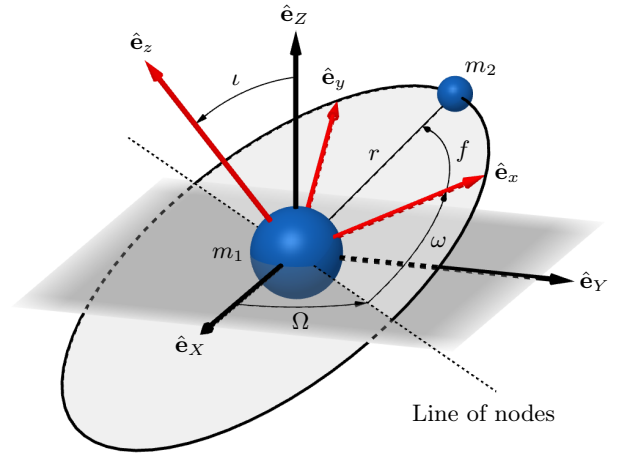
\begin{figure}
  \begin{center}
    \includegraphics[scale=0.23]{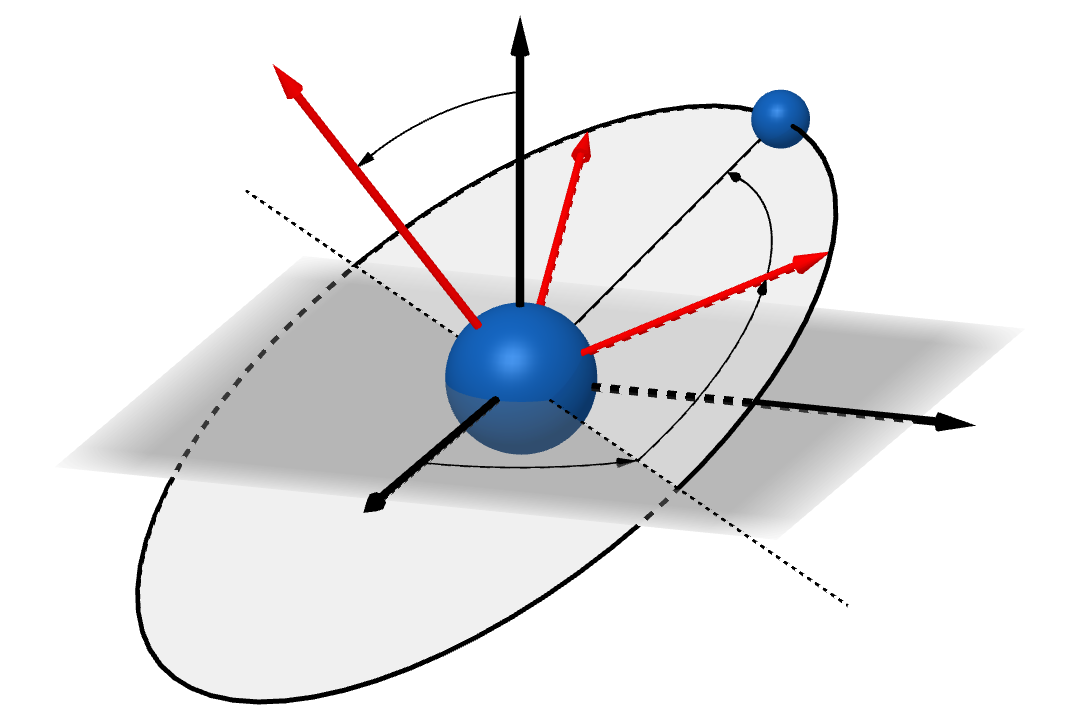}
  \end{center}
  \setlength{\unitlength}{1.0cm}
  \begin{picture}(0,0)
    \put(-1.65,2.2){\rotatebox{0}{$\hat{\mathbf e}_X$}}
    \put(3.73,3.05){\rotatebox{0}{$\hat{\mathbf e}_Y$}}
    \put(-0.25,6.6){\rotatebox{0}{$\hat{\mathbf e}_Z$}}
    \put(2.6,4.5){\rotatebox{0}{$\hat{\mathbf e}_x$}}
    \put(0.3,5.76){\rotatebox{0}{$\hat{\mathbf e}_y$}}
    \put(-2.25,6.23){\rotatebox{0}{$\hat{\mathbf e}_z$}}
    \put(-1.16,3.45){\rotatebox{0}{$m_1$}}
    \put(2.2,5.92){\rotatebox{0}{$m_2$}}
    \put(0.92,4.75){\rotatebox{0}{$r$}}
    \put(-0.9,5.75){\rotatebox{0}{$\iota$}}
    \put(-0.1,2.5){\rotatebox{0}{$\Omega$}}
    \put(1.75,3.6){\rotatebox{0}{$\omega$}}
    \put(2.08,4.8){\rotatebox{0}{$f$}}
    \put(1.5,1.28){\rotatebox{0}{$\text{Line of nodes}$}}
  \end{picture}
  \vspace{-0.2cm}
  \caption{Orientation of $(\hat{\mathbf e}_x,\hat{\mathbf e}_y,\hat{\mathbf e}_z)$, the orbit frame, in the source frame, namely $(\hat{\mathbf e}_X,\hat{\mathbf e}_Y,\hat{\mathbf e}_Z)$. The primary is shown at the center-of-mass of the binary system in order to simplify the drawing. This corresponds to the case where the mass of the secondary is negligible with respect to primary's.}
  \label{fig:orbit}
\end{figure}

Let $(\hat{\mathbf e}_x,\hat{\mathbf e}_y,\hat{\mathbf e}_z)$ be a right-handed vectorial basis defining the ``orbit frame''. The $z$-axis is orthogonal to the orbital plane, the $x$-axis is pointing toward the closest approach of the effective one-body orbit, and the $y$-axis completes the basis. For Keplerian motion the orbit frame is non-rotating with respect to distant stars since the direction of the closest approach $\hat{\mathbf e}_x$ is a first integral of motion as dictated by the conservation of the Runge-Lenz vector. For a non-Keplerian motion, the orbit frame is not inertial anymore. The orbit frame and its orientation within the source frame are depicted in Fig.~\ref{fig:orbit}.

Finally, let us introduce $(\hat{\mathbf n},\hat{\mathbf u},\hat{\mathbf e}_z)$, a right-handed vectorial basis defining the ``corotating frame'', that is to say the frame that is corotating with the effective body. The unit-vector $\hat{\mathbf u}$ is introduced such that it completes the basis. The transformation from the source frame to the corotating frame involves the orbital angles that are depicted in Fig.~\ref{fig:orbit}, namely $\iota$, the inclination of the orbit on the $(\hat{\mathbf{e}}_X,\hat{\mathbf{e}}_Y)$-plane, $\Omega$, the longitude of the ascending node measured from $\hat{\mathbf{e}}_X$, $\omega$, the argument of the pericenter measured from the ascending node, and $f$, the true anomaly measured from the closest approach.

\subsection{Orbital motion with magnetism}

We consider a binary system that is radiating GWs to infinity according to GR (cf. Sect.~\ref{sec:GWsignal}). In GR, the gravitational radiation has its own degrees of freedom so it can carry energy and angular momentum away from the source. In the framework of the PN approximation, the backreaction on the orbit due to the radiation is described by terms at the 2.5PN order. Therefore, in order to satisfy the energy and angular momentum balance equations, we consider the orbital dynamics up to 2.5PN order. The effective motion can be summarized by the following equation:
\begin{equation}
  \frac{\dd^2\mathbf x}{\dd t^2}+a_{\mathrm N}\hat{\mathbf n}=\mathbf a_{\mathrm{GR}}+\mathbf{a}_{\mathrm{M}}+\mathcal O(c^{-6})\text{,}
  \label{eq:motion}
\end{equation}
where $a_{\mathrm N}=Gm/r^2$ is the magnitude of the Newtonian acceleration with $G$ the gravitational constant, and where $\mathbf a_{\mathrm{GR}}$ contains the GR corrections up to terms proportional to $c^{-5}$ \cite{1990PhRvD..42.1123L,2014LRR....17....2B}, namely
\begin{equation}
  \mathbf a_{\mathrm{GR}}=-a_{\mathrm N}(\mathcal A\hat{\mathbf n}+\mathcal B\hat{\mathbf u})\text{.}
  \label{eq:compaccpertGR}
\end{equation}
The dimensionless coefficients $\mathcal A$ and $\mathcal B$ are given explicitly in appendix \ref{sec:PNcorr}.

In Eq. \eqref{eq:motion}, the acceleration $\mathbf a_{\mathrm{M}}$ represents the magnetic dipole-dipole interaction. To determine this term, we work in the magnetostatic approximation. In addition, we assume that the magnetic fields are frozen into the stars as dictated by the fossil-field hypothesis \cite{2004Natur.431..819B,2010ApJ...724L..34D}. Accordingly, the internal currents that generate the magnetic field of the primary are not distorted significantly by the external field of the secondary and vice-versa (see also \citet{1990MNRAS.244..731K} for a similar hypothesis). This assumption is justified by the fact that we focus on the inspiral phase where the bodies are always well-separated. As a first step and in agreement with results from \citet{2004Natur.431..819B}, we consider that the magnetic fields of both stars are dominated by their dipole moments $\bm{\mu}_1$ and $\bm{\mu}_2$, although other configurations might be stable as well \cite{2008MNRAS.386.1947B}. Since internal currents are assumed to be stationary, the magnitude of the magnetic moments is taken to be constant during the motion and we introduce the two following parameters $\mu_1=|\bm\mu_1|$ and $\mu_2=|\bm\mu_2|$. According to \citet{2019MNRAS.488...64P}, the magnitude of a magnetic moment $\mu$ is given by
\begin{equation}
  \mu=\frac{2\pi}{\mu_0}BR^3\text{,}
  \label{eq:magmom}
\end{equation}
where $\mu_0$ is the permeability of vacuum, $R$ is the equatorial radius of the star, and $B=|\mathbf{B}|$ is the magnitude of the magnetic field at the surface of the star. It is convenient to define the unit-vectors $\hat{\mathbf{s}}_1$ and $\hat{\mathbf{s}}_2$ such that $\hat{\mathbf{s}}_1\equiv \bm\mu_1/\mu_1$ and $\hat{\mathbf{s}}_2\equiv \bm\mu_2/\mu_2$.

Given these assumptions, the secondary feels a dipolar magnetic field $\mathbf B_1$ and experiences a magnetic force \cite{1990MNRAS.244..731K} that is given by $\mathbf F_{12}=\bm\nabla(\bm\mu_2\cdot\mathbf B_1)$ or more explicitly,
\begin{align}
  \mathbf F_{12}&=-\frac{\mu_0}{4\pi}\frac{\mu_1\mu_2}{r^4}\big[15(\hat{\mathbf n}\cdot\hat{\mathbf s}_1)(\hat{\mathbf n}\cdot\hat{\mathbf s}_2)\hat{\mathbf n}\nonumber\\*
  &-3(\hat{\mathbf n}\cdot\hat{\mathbf s}_1)\hat{\mathbf s}_2-3(\hat{\mathbf n}\cdot\hat{\mathbf s}_2)\hat{\mathbf s}_1-3(\hat{\mathbf s}_1\cdot\hat{\mathbf s}_2)\hat{\mathbf n}\big]\text{.}
\end{align}
The expression of the force acting on the primary is found by interchanging the subscripts ``1'' and ``2'' and changing the sign of $\hat{\mathbf n}$ in the expression of $\mathbf F_{12}$. Then the magnetic relative acceleration $\mathbf{a}_{\mathrm{M}}$ takes the form
\begin{equation}
  \mathbf{a}_{\mathrm{M}}=-\left(\frac{1}{\eta m}\right)\mathbf F_{12}\text{.}
  \label{eq:compaccpertM}
\end{equation}

Once the effective one-body motion in Eq. \eqref{eq:motion} is solved, the individual positions can be retrieved from the PN definition of the barycenter of the binary system (see e.g. \citet{2014LRR....17....2B} for a complete definition up to 3PN):
\begin{subequations}\label{eq:posabs}
  \begin{align}
    \mathbf x_1&=-\frac{m_2}{m}\mathbf x-\eta\Delta r(\mathcal P\hat{\mathbf n}+\mathcal Q\hat{\mathbf u})+\mathcal O(c^{-6})\text{,}\\
    \mathbf x_2&=\frac{m_1}{m}\mathbf x-\eta\Delta r(\mathcal P\hat{\mathbf n}+\mathcal Q\hat{\mathbf u})+\mathcal O(c^{-6})\text{,}
  \end{align}
\end{subequations}
where the dimensionless coefficients $\mathcal P$ and $\mathcal Q$ are given explicitly at the 2.5PN order in appendix \ref{sec:PNcorr}. Analogous transformations can be derived for the individual velocities by taking a time derivative of Eqs.~\eqref{eq:posabs} while keeping the appropriate PN orders in the equation of motion.

\subsection{Rotational motion with magnetism}

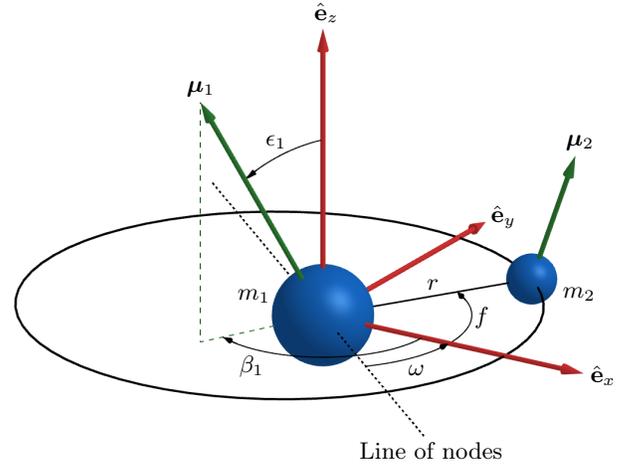
\begin{figure}
  \begin{center}
    \includegraphics[scale=0.222]{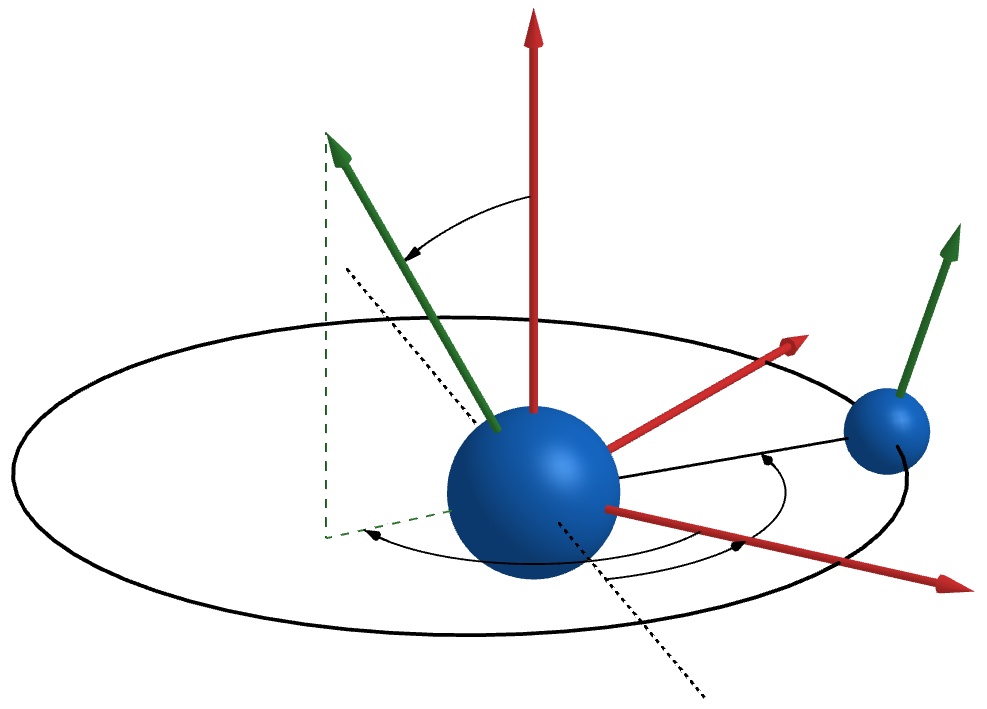}
  \end{center}
  \setlength{\unitlength}{1.0cm}
  \begin{picture}(0,0)
    \put(3.87,1.52){\rotatebox{0}{$\hat{\mathbf e}_x$}}
    \put(2.55,3.65){\rotatebox{0}{$\hat{\mathbf e}_y$}}
    \put(0.2,6.34){\rotatebox{0}{$\hat{\mathbf e}_z$}}
    \put(-1.5,5.38){\rotatebox{0}{$\bm\mu_1$}}
    \put(3.54,4.67){\rotatebox{0}{$\bm\mu_2$}}
    \put(0.8,0.5){\rotatebox{0}{$\text{Line of nodes}$}}
    \put(1.45,1.62){\rotatebox{0}{$\omega$}}
    \put(2.33,2.3){\rotatebox{0}{$f$}}
    \put(1.7,2.75){\rotatebox{0}{$r$}}
    \put(-0.8,1.65){\rotatebox{0}{$\beta_1$}}
    \put(-0.45,4.67){\rotatebox{0}{$\epsilon_1$}}
    \put(-0.82,2.62){\rotatebox{0}{$m_1$}}
    \put(3.5,2.64){\rotatebox{0}{$m_2$}}
  \end{picture}
  \caption{Orientation of the magnetic moments in the orbit frame $(\hat{\mathbf e}_x,\hat{\mathbf e}_y,\hat{\mathbf e}_z)$. The obliquity $\epsilon_1$ and the precession angle $\beta_1$ are represented for the primary only. The obliquity is a tilt between $\hat{\mathbf e}_z$, the normal to the orbital plane, and $\hat{\mathbf s}_1$, the direction of the magnetic moments. The precession angle is the angle between $\hat{\mathbf e}_x$, the direction of closest approach, and the projection of the magnetic moments on the orbital plane.}
  \label{fig:magnetic_moments}
\end{figure}

The magnetic interaction impacts not only the orbital evolution of the binary system but also the direction of the magnetic moments $\bm\mu_1$ and $\bm\mu_2$. In order to follow their evolution in space, we introduce spherical coordinates with one polar angle and one azimuth angle per compact star. The polar angles, also called obliquities, are labeled $\epsilon_1$ and $\epsilon_2$, and the azimuth angles, also called precession angles, are labeled $\beta_1$ and $\beta_2$. The angles are depicted in Fig.~\ref{fig:magnetic_moments} for the primary only. These definitions allow for a drastic simplification of the equations of motion. However, the precession angles are not defined for null obliquities, and hence, the case where the magnetic moments are exactly orthogonal to the orbital plane (i.e., $\epsilon_i=0$ or $\pi$, with $i=1$ and $2$) cannot be studied completely with these definitions. Instead, for null obliquities, the regular Cartesian coordinates must be preferred.

\citet{2019MNRAS.488...64P} have shown that the lowest stable energy is the horizontal aligned magnetic moment configuration, namely $\epsilon_1=\epsilon_2=\pi/2$, or $\epsilon_1=\epsilon_2=-\pi/2$, with $\beta_1=\beta_2=f$. This configuration can be observed when the obliquity of the field with respect to the rotation axis is large. However, as pointed out by \citet{2015MNRAS.454L...1S} the field and the rotation axis in each star are more likely to be aligned, and the directions of the magnetic fields should be parallel. In this configuration, the lowest-energy stable magnetic configuration due to the magnetic dipole-dipole interaction force is vertical
anti-aligned magnetic fields, namely $\epsilon_1=0$ and $\epsilon_2=\pi$, or $\epsilon_1=\pi$ and $\epsilon_2=0$. Hereafter, we suppose that the orientations of the magnetic moments have not reached equilibrium yet, such that the directions $\hat{\mathbf{s}}_1$ and $\hat{\mathbf{s}}_2$ can evolve under the action of dipole-dipole magnetic torques.


The direction $\hat{\mathbf s}_1$ of the magnetic moment $\bm\mu_1$ can be decomposed in the orbit frame as
\begin{equation}
  \hat{\mathbf s}_1=\hat{\mathbf e}_x\sin\epsilon_1\cos\beta_1+\hat{\mathbf e}_y\sin\epsilon_1\sin\beta_1+\hat{\mathbf e}_z\cos\epsilon_1\mathrm{.}
  \label{eq:e1comp}
\end{equation}
Differentiation with respect to time returns 
\begin{equation}
  \hat{\mathbf s}_1\cdot\frac{\dd\hat{\mathbf s}_1}{\dd t}=0\text{,}
  \label{eq:e1de1}
\end{equation}
together with
\begin{subequations}\label{eq:epsbetkin}
\begin{align}
  \sin\epsilon_1\frac{\dd\epsilon_1}{\dd t}&=-\frac{\dd\hat{\mathbf s}_1}{\dd t}\cdot\hat{\mathbf e}_z\text{,}\\
  \sin\epsilon_1\frac{\dd\beta_1}{\dd t}&=\cos\beta_1\frac{\dd \hat{\mathbf s}_1}{\dd t}\cdot\hat{\mathbf e}_y-\sin\beta_1\frac{\dd \hat{\mathbf s}_1}{\dd t}\cdot\hat{\mathbf e}_x\text{.}
\end{align}
\end{subequations}
Similar kinematic relationships can be derived for the secondary. They are directly inferred after interchanging subscripts ``1'' and ``2'' in the above equations.

Hereafter, in agreement with \citet{2015MNRAS.454L...1S}, we assume that the directions of the spins $\mathbf{S}_1$ and $\mathbf{S}_2$ are aligned with the magnetic moments, that is to say $\mathbf{S}_1/S_1=\hat{\mathbf s}_1$ and $\mathbf{S}_2/S_2=\hat{\mathbf s}_2$ with $S_1\equiv|\mathbf{S}_1|$ and $S_2\equiv|\mathbf{S}_2|$. Then, the rotational equation of motion read as
\begin{equation}
  \frac{\dd\mathbf S_1}{\dd t}=\bm\Gamma_{21}\text{,} \qquad \frac{\dd\mathbf S_2}{\dd t}=\bm\Gamma_{12}\text{,}
\end{equation}
where $\bm\Gamma_{21}$ (resp., $\bm\Gamma_{12}$) is the magnetic torque due to the secondary (resp., primary) acting on the magnetic moment of the primary (resp., secondary). The expression for the former is given by $\bm\Gamma_{21}=\bm\mu_1\times\mathbf B_2$, namely
\begin{equation}
  \bm\Gamma_{21}=\frac{\mu_0}{4\pi}\frac{\mu_1\mu_2}{r^3}\Big[3(\hat{\mathbf s}_1\times\hat{\mathbf n})(\hat{\mathbf n}\cdot\hat{\mathbf s}_2)-\hat{\mathbf s}_1\times\hat{\mathbf s}_2\Big]\text{.}\label{eq:torq1}
\end{equation}
The expression of $\bm\Gamma_{12}$ is found by interchanging the subscripts ``1'' and ``2'' and changing the sign of $\hat{\mathbf n}$ in the expression of $\bm\Gamma_{21}$. We see that $\bm\Gamma_{21}\neq-\bm\Gamma_{12}$, in general.

We can infer from the use of \eqref{eq:e1de1} and \eqref{eq:torq1} that the magnitude of the spin is conserved during the motion (in accordance with the fact that we do not consider dissipation at the level of the rotational motion),
\begin{equation}
  \frac{\dd S_1}{\dd t}=\bm\Gamma_{21}\cdot\hat{\mathbf s}_1=0\text{.}
\end{equation}
The same is true for $S_2$. Therefore, the equations for $\hat{\mathbf s}_1$ and $\hat{\mathbf s}_2$ are given by the following expressions
\begin{equation}
  \frac{\dd\hat{\mathbf s}_1}{\dd t}=\frac{\bm\Gamma_{21}}{S_1}\text{,} \qquad \frac{\dd\hat{\mathbf s}_2}{\dd t}=\frac{\bm\Gamma_{12}}{S_2}\text{.}
  \label{eq:rotation}
\end{equation}

Eqs. \eqref{eq:motion} and \eqref{eq:rotation} represent the equations of motion that must be solved simultaneously for describing the dynamics of the binary system considering both GR effects (up to the 2.5PN approximation) and the magnetic dipole-dipole interaction.

\section{Secular equations of motion}
\label{sec:seceqs}

As discussed in Sect. \ref{sec:int}, in the context of LISA, GBs must be modeled beyond Keplerian motion. This is because the sensitivity of the instrument will potentially allow the measurement of several effects. The method of osculating elements offers a convenient framework to go beyond Keplerian motion \cite{1961mcm..book.....B,2014grav.book.....P}. We use it here in order to describe the secular evolution of the system.

\subsection{Homogeneous solutions}

The homogeneous solutions to the equation of motion~\eqref{eq:motion} correspond to the Keplerian motion \cite{1961mcm..book.....B,2014grav.book.....P,2000ssd..book.....M}. In the corotating frame it reads
\begin{equation}
  \mathbf{x}=r\hat{\mathbf n}\text{,} \qquad \mathbf{v}=v_n\hat{\mathbf n}+v_u\hat{\mathbf u}\text{,}
  \label{eq:solKep}
\end{equation}
with $r=p(1+e\cos f)^{-1}$ and 
\begin{equation}
  v_n=\sqrt{\frac{Gm}{p}}\,e\sin f\text{,} \qquad v_u=\sqrt{\frac{Gm}{p}}\,(1+e\cos f)\text{.}
\end{equation}
We recall that $p=a(1-e^2)$ is the semi-latus rectum, $a$ is the semi-major axis, and $e$ is the eccentricity. The solutions are computed at a given instant of time using the Kepler equation which relies on $\tau$, the time of pericenter passage, and the eccentric anomaly.

The solutions \eqref{eq:solKep} can be further specified within the source frame (cf. Fig. \ref{fig:orbit}) thanks to the inclination $\iota$, the longitude of the ascending node $\Omega$, and the argument of the pericenter $\omega$. Let us recall that $(a,e,\iota,\Omega,\omega,\tau)$ are constant for Kepler motion and are the so-called elliptic elements.

\subsection{Variation of arbitrary constants}

The right-hand side of the equation of motion \eqref{eq:motion} regroups the perturbing accelerations, namely the contribution from GR together with the magnetic dipole-dipole interaction. To handle them, we use the method of variation of arbitrary constants which allows us to reshape Eq. \eqref{eq:motion} as a set of six first-order differential equations for the elliptic elements. These equations are called the Lagrange planetary equations \cite{2014grav.book.....P} or the Gauss perturbation equations \cite{1961mcm..book.....B,2000ssd..book.....M} depending whether the perturbation is expressed as a gradient of a potential or not. The perturbation equations involve the components $\mathcal N$, $\mathcal U$, and $\mathcal S$ which are the projections of the perturbing accelerations on the unit-vectors $\hat{\mathbf n}$, $\hat{\mathbf u}$, and $\hat{\mathbf e}_z$, respectively. Thus, $\mathcal N$ is the radial component, $\mathcal U$ is the cross-track component, and $\mathcal S$ is the out-of-plane component.

The basic idea behind the method of variation of arbitrary constants is to consider that the Keplerian solutions \eqref{eq:solKep} are still correct, even beyond Keplerian motion. The apparent contradiction is evaded by allowing the elliptic elements to change with time. This description of motion will be particularly useful in Sect.~\ref{sec:GWsignal} when deriving the form of the GW mode polarizations while considering the perturbing effects of both GR and the magnetic dipole-dipole interaction.

The components of the perturbing acceleration due to GR corrections are given by (cf. Eq. \eqref{eq:compaccpertGR})
\begin{subequations}\label{eq:compertGR}
  \begin{align}
    \mathcal N_{\mathrm{GR}}&\equiv\mathbf{a}_{\mathrm{GR}}\cdot\hat{\mathbf n}=-a_{\mathrm{N}}\mathcal A\text{,}\\
    \mathcal U_{\mathrm{GR}}&\equiv\mathbf{a}_{\mathrm{GR}}\cdot\hat{\mathbf u}=-a_{\mathrm{N}}\mathcal B\text{,}
  \end{align}
\end{subequations}
with $\mathcal S_{\mathrm{GR}}\equiv\mathbf{a}_{\mathrm{GR}}\cdot\hat{\mathbf e}_z=0$. For the magnetic dipole-dipole interaction, the components of the perturbing acceleration read as follows (cf. Eq.~\eqref{eq:compaccpertM})
\begin{subequations}\label{eq:compertM}
  \begin{align}
    \mathcal N_{\mathrm{M}}&\equiv\mathbf{a}_{\mathrm M}\cdot\hat{\mathbf n}=-\frac{3\mu_0}{4\pi r^4}\frac{\mu_1\mu_2}{\eta m}\nonumber\\
    &\times\Big[\hat{\mathbf s}_1\cdot\hat{\mathbf s}_2-3(\hat{\mathbf n}\cdot\hat{\mathbf s}_1)(\hat{\mathbf n}\cdot\hat{\mathbf s}_2)\Big]\text{,}\\
    \mathcal U_{\mathrm{M}}&\equiv\mathbf{a}_{\mathrm M}\cdot\hat{\mathbf u}=-\frac{3\mu_0}{4\pi r^4}\frac{\mu_1\mu_2}{\eta m}\nonumber\\
    &\times\Big[(\hat{\mathbf n}\cdot\hat{\mathbf s}_1)(\hat{\mathbf u}\cdot\hat{\mathbf s}_2)+(\hat{\mathbf n}\cdot\hat{\mathbf s}_2)(\hat{\mathbf u}\cdot\hat{\mathbf s}_1)\Big]\text{,}\\
    \mathcal S_{\mathrm{M}}&\equiv\mathbf{a}_{\mathrm M}\cdot\hat{\mathbf e}_z=-\frac{3\mu_0}{4\pi r^4}\frac{\mu_1\mu_2}{\eta m}\nonumber\\
    &\times\Big[(\hat{\mathbf n}\cdot\hat{\mathbf s}_1)(\hat{\mathbf e}_z\cdot\hat{\mathbf s}_2)+(\hat{\mathbf n}\cdot\hat{\mathbf s}_2)(\hat{\mathbf e}_z\cdot\hat{\mathbf s}_1)\Big]\text{.}
  \end{align}
\end{subequations}

The 1PN perturbations to the Keplerian motion and the magnetic acceleration are of the order of
\begin{subequations}
\begin{align}
  \frac{|\mathbf a_{\mathrm{GR}}|}{a_{\mathrm{N}}}&\propto\frac{v^2}{c^2}\sim\frac{Gm}{c^2r}\text{,}\\
  \frac{|\bm a_{\mathrm{M}}|}{a_{\mathrm N}}&\propto \frac{3\mu_0}{4\pi r^2}\,\frac{\mu_1\mu_2}{Gm_1 m_2}\text{,}
\end{align}
\end{subequations}
respectively. Rough numerical estimates return
\begin{subequations}
\begin{align}
  \frac{|\mathbf a_{\mathrm{GR}}|}{a_{\mathrm{N}}}&\simeq 2.2\times10^{-4}\left(\frac{10^4\,\mathrm{km}}{r}\right)\left(\frac{m}{1.5\,\mathrm{M}_\odot}\right)\text{,}\\
  \frac{|\bm a_{\mathrm{M}}|}{a_{\mathrm N}}&\simeq 3.1\times 10^{-11}\left(\frac{\mu_1}{10^{30}\,\mathrm{A}\cdot\mathrm{m}^2}\right)\left(\frac{\mu_2}{10^{30}\,\mathrm{A}\cdot\mathrm{m}^2}\right)\nonumber\\
  &\times\left(\frac{1.2\,\mathrm{M}_\odot}{m_1}\right)\left(\frac{0.3\,\mathrm{M}_\odot}{m_2}\right)\left(\frac{10^4\,\mathrm{km}}{r}\right)^2\text{.}
\end{align}
\end{subequations}
These ratios show that, even for the most compact system of GBs that LISA can observe (i.e., $r\sim 10^4\,\mathrm{km}$ which corresponds to orbital frequency of the order of $10^{-1}\,\mathrm{Hz}$), the right-hand side of Eq. \eqref{eq:motion} can be treated as a perturbation to the Newtonian acceleration. In other words, we expect the changes in the orbital elements to be small.

Therefore, a simplified description of the motion can be achieved by inserting the constant zeroth-order values of the Keplerian elements in the right-hand side of the perturbation equations and by keeping first-order terms in the components of the perturbing accelerations. In this picture, it is convenient, for averaging purposes, to change the independent variable from time to angles that vary on short orbital timescales, such as the true anomaly. The system of first-order osculating equations eventually reads as
\begin{widetext}
\begin{subequations}\label{eq:pertsimp}
\begin{align}
  \frac{\dd a}{\dd f}&\simeq\frac{2a^3(1-e^2)}{Gm}\bigg[\frac{\mathcal Ne\sin f}{(1+e\cos f)^2}+\frac{\mathcal U}{1+e\cos f}\bigg]\text{,}\\
  \frac{\dd z}{\dd f}&\simeq-\frac{\mathrm iz}{e}\frac{p^2}{Gm}\Bigg[\frac{\mathcal N\mathrm e^{\mathrm i f}}{(1+e\cos f)^2}+\frac{2+e\big(\mathrm e^{-\mathrm i f}+\cos f\big)}{(1+e\cos f)^3}\,\mathrm i\mathcal U\mathrm e^{\mathrm if}+\frac{\mathrm i\mathcal S\big(\bar\zeta\, z\mathrm e^{\mathrm i f}-\zeta\,\bar z\mathrm e^{-\mathrm i f}\big)}{2\sqrt{1-\zeta\bar\zeta}(1+e\cos f)^3}\Bigg]\text{,}\label{eq:pertsimpz}\\
  \frac{\dd\zeta}{\dd f}&\simeq\frac{p^2}{Gm}\frac{\left[\left(2-\zeta\bar\zeta\right)z\mathrm e^{\mathrm if}-\zeta^2\,\bar z\mathrm e^{-\mathrm i f}\right]\mathcal S}{4e\sqrt{1-\zeta\bar\zeta}(1+e\cos f)^3}\text{,}\label{eq:pertsimpzeta}\\
  \frac{\dd L}{\dd f}&\simeq\frac{(1-e^2)^{3/2}}{(1+e\cos f)^2}-\frac{1}{e}\frac{p^2}{Gm}\Bigg\{\frac{2e\sqrt{1-e^2}}{(1+e\cos f)^3}\,\mathcal N\nonumber\\
  &+\bigg[1-\sqrt{1-e^2}+\frac{(1-e^2)^{3/2}}{(1+e\cos f)^2}\bigg]\bigg[\frac{\mathcal N\cos f}{(1+e\cos f)^2}-\frac{2+e\cos f}{(1+e\cos f)^3}\,\mathcal U\sin f\bigg]+\frac{\mathrm i\mathcal S\big(\bar\zeta\, z\mathrm e^{\mathrm i f}-\zeta\,\bar z\mathrm e^{-\mathrm i f}\big)}{2\sqrt{1-\zeta\bar\zeta}(1+e\cos f)^3}\Bigg\}\text{,}\label{eq:pertsimpL}
\end{align}
\end{subequations}
with the additional expression:
\begin{equation}
  \frac{\dd\varpi}{\dd f}\simeq-\frac{1}{e}\frac{p^2}{Gm}\bigg[\frac{\mathcal N\cos f}{(1+e\cos f)^2}-\frac{2+e\cos f}{(1+e\cos f)^3}\,\mathcal U\sin f+\frac{\mathrm i\mathcal S(\bar\zeta\, z\mathrm e^{\mathrm i f}-\zeta\,\bar z\mathrm e^{-\mathrm i f})}{2\sqrt{1-\zeta\bar\zeta}(1+e\cos f)^3}\bigg]\text{.}
  \label{eq:pertsimpvpi}
\end{equation}
\end{widetext}

In these expressions, we introduce the following regular parameters: $(a,z,\zeta,L)$; see \citet{2014grav.book.....P} for similar expressions in terms of the singular elements $(a,e,\iota,\Omega,\omega,\tau)$. The regular parameters are defined by
\begin{subequations}\label{eq:defregparam}
\begin{align}
z&=e\,\mathrm{e}^{\mathrm i\varpi}\text{,}\\
\zeta&=\sin\left(\frac{\iota}{2}\right)\mathrm{e}^{\mathrm i\Omega}\text{,}\\
L&=\varpi+M\text{,}
\end{align}
\end{subequations}
with $\mathrm i\equiv\sqrt{-1}$. The expression for $M$, the mean anomaly, is $M=n(t-\tau)$ with $n$ the mean motion, which is given by Kepler's third law: $n=(Gm/a^3)^{-1/2}$. The expression for $\varpi$, the longitude of the pericenter, is given by
\begin{equation}
  \varpi=\Omega+\omega\text{.}
\end{equation}
The complex variables $z$ and $\zeta$ represent the components of the eccentricity vector and the components of the longitude of the ascending node vector, respectively.

In the first-order perturbation Eqs. \eqref{eq:pertsimp}, we do not use $\Omega$, $\omega$, and $\tau$, which are singular when either the inclination or the eccentricity go to zero (see e.g. Eqs. (3.69) of \citet{2014grav.book.....P} and Eq. (2.167) of \citet{2000ssd..book.....M}). As a matter of fact, because the gravitational radiation efficiently circularizes the orbit (see e.g. \citet{2021PhRvD.104j4023T}, and see the discussion in the next section), most of the binary systems that LISA will observe are expected to be found in quasi-circular orbit within the frequency band from $10^{-4}\,\mathrm{Hz}$ to $10^{-1}\,\mathrm{Hz}$. This is the reason why we consider the set of non-singular elements: $(a,z,\zeta,L)$.

Let us emphasize that two additional equations, one for $\bar z$ and the other one for $\bar\zeta$, are derived straightforwardly from Eqs. \eqref{eq:pertsimpz} and \eqref{eq:pertsimpzeta}, where $\bar z$ and $\bar\zeta$ are the complex conjugate of $z$ and $\zeta$, respectively. Accordingly, the expression for $\varpi$ in Eq.~\eqref{eq:pertsimpvpi} is actually redundant since it can be inferred from Eq.~\eqref{eq:pertsimpz} and its complex conjugate. However, we provide it anyway for simplification purposes, as discussed in the next section. For the same reason, the eccentricity in Eqs. \eqref{eq:pertsimp} must actually be seen as a function of the complex variables $z$ and $\bar z$, namely $e=\sqrt{z\bar z}$.

Let us note that the first term on the right-hand side of Eq. \eqref{eq:pertsimpL} is a zeroth-order term, meaning that $L$ is not constant even for Kepler motion. This term corresponds to the product $n\,(\dd t/\dd f)$ expressed at zeroth-order in the components of the perturbation. A new convenient parameter, $\lambda$, is thus introduced such that
\begin{equation}
  \lambda(t)=L(t)-\int_0^tn(t')\dd t'\text{.}
  \label{eq:lambdadef}
\end{equation}
From this definition, it is clear that $\lambda$ coincides with the mean longitude at the instant $t=0$.

Hereafter, we employ the following non-singular orbital elements: $\mathbf{X}=(a,z,\zeta,\lambda)$.\\

\subsection{Secular motion}

We can expect that the solutions of the first-order perturbation equations will vary periodically, with a short orbital timescale on one hand and a long secular timescale on the other hand; this is a consequence of the smallness of the perturbing accelerations with respect to the Newtonian one. Accordingly, the two timescales can be treated as two independent variables, and conveniently for us, only the secular contribution can be kept from the perturbation equations.

In order to derive the secular components of the motion, the equations are averaged over the angle that varies on short orbital timescale, namely the true anomaly. Thus, to each non-singular element $\mathbf X$, we associate a secular time derivative defined such as
\begin{equation}
  \left\langle\frac{\dd\mathbf X}{\dd t}\right\rangle_{\mathrm{sec}}=\frac{n}{2\pi}\int_0^{2\pi}\frac{\dd\mathbf X}{\dd f}\,\dd f\text{.}
  \label{eq:secder}
\end{equation}

After substituting for $\mathcal N$, $\mathcal U$, and $\mathcal S$ from Eqs. \eqref{eq:compertGR} into Eqs. \eqref{eq:pertsimp} and using \eqref{eq:secder}, we derive the first-order secular equations describing the non-null contributions from GR (see also \citet{1990PhRvD..42.1123L})
\begin{widetext}
\begin{subequations}\label{eq:pertGR}
\begin{align}
  \left\langle\frac{\dd a}{\dd t}\right\rangle_{\mathrm{GR}}&=-\frac{64\eta}{5}\left(\frac{na}{1-e^2}\right)\bigg(\frac{G m}{c^2p}\bigg)^{5/2}\left(1+\frac{73}{24}e^2+\frac{37}{96}e^4\right)\text{,}\label{eq:pertGRdadt}\\
  \left\langle\frac{\dd\lambda}{\dd t}\right\rangle_{\mathrm{GR}}&=5n\bigg(\frac{Gm}{c^2p}\bigg)\left\{2\left(1-\frac{\eta}{2}\right)\frac{\left(1-\sqrt{1-e^2}\right)}{e^2}-\sqrt{1-e^2}-\frac{\eta}{10}\left(1-8\sqrt{1-e^2}\right)-\frac{7}{5}e^2\left(1-\frac{11\eta}{14}\right)\right\}\text{,}\label{eq:pertGRdLdt}\\
  \left\langle\frac{\dd z}{\dd t}\right\rangle_{\mathrm{GR}}&=\frac{z}{e}\left\langle\frac{\dd e}{\dd t}\right\rangle_{\mathrm{GR}}+\mathrm i z\left\langle\frac{\dd\varpi}{\dd t}\right\rangle_{\mathrm{GR}}\text{,}\label{eq:pertGRdzdt}
\end{align}
\end{subequations}
where we use the two following relationships:
\begin{equation}
  \left\langle\frac{\dd e}{\dd t}\right\rangle_{\mathrm{GR}}=-\frac{304\eta}{15}\,n e\,\bigg(\frac{G m}{c^2p}\bigg)^{5/2}\left(1+\frac{121}{304}e^2\right)\text{,}\qquad\qquad \left\langle\frac{\dd\varpi}{\dd t}\right\rangle_{\mathrm{GR}}=3n\left(\frac{Gm}{c^2p}\right)\text{.}\label{eq:pertGRdeodt}
\end{equation}
\end{widetext}

The orbital element $\zeta$ is the only element that is not secularly impacted by GR. Equations \eqref{eq:pertGRdadt} and $\left\langle\dd e/\dd t\right\rangle$ in \eqref{eq:pertGRdeodt} describe the secular changes in $a$ and $e$ due to the loss of orbital energy and angular momentum, respectively. Indeed, as stated previously, the gravitational radiation carries energy and angular momentum away from the source, causing a decrease in the orbit's semi-major axis and eccentricity. These effects are described by the 2.5PN order, namely the terms proportional to $\propto c^{-5}$. The equation for $\left\langle\dd\varpi/\dd t\right\rangle$ in \eqref{eq:pertGRdeodt} describes the secular change in the longitude of the pericenter. It contains the well-known pericenter advance which is described by the 1PN order, namely the term proportional to $\propto c^{-2}$. Equation \eqref{eq:pertGRdLdt} describes the secular change in the mean longitude minus the mean motion. It reduces to $n\eta(Gm/c^2a)$ at zeroth-order in the eccentricity, showing that the effect of GR does not cancel out for circular orbits. The parameter $\lambda$ is used to compute the mean longitude $L$ which is the parameter of interest in the discussion of Sect. \ref{sec:GWsignal}.

Let us emphasize that terms of order $c^{-4}$ are neglected in Eqs.~\eqref{eq:pertGRdLdt} and in $\left\langle\dd\varpi/\dd t\right\rangle$ in \eqref{eq:pertGRdeodt}. They are of two types. There are 2PN corrections arising from terms proportional to $c^{-4}$ in the expressions of Eq.~\eqref{eq:compertGR}.  There is also a second-order perturbation due to the 1PN corrections, since the terms neglected in the perturbation equations \eqref{eq:pertsimp} are quadratic in the components of the perturbing acceleration. Both these terms are negligible relative to the 1PN contribution, which represents the non-null dominant order.

After substituting for $\mathcal N$, $\mathcal U$, and $\mathcal S$ from Eqs. \eqref{eq:compertM} into Eqs. \eqref{eq:pertsimp} while considering \eqref{eq:secder}, we derive the first-order secular equations describing the non-null contributions from the dipole-dipole interaction
\begin{widetext}
\begin{subequations}\label{eq:pertM}
  \begin{align}
    \left\langle\frac{\dd\zeta}{\dd t}\right\rangle_{\mathrm{M}}&=-\frac{\nu}{4e\sqrt{1-\zeta\bar\zeta}}\Big\{\left[\left(2-\zeta\bar\zeta\right)z\mathrm{e}^{\mathrm i\beta_1}-\zeta^2\,\bar z\mathrm e^{-\mathrm i\beta_1}\right]\sin\epsilon_1\cos\epsilon_2+\left[\left(2-\zeta\bar\zeta\right)z\mathrm{e}^{\mathrm i\beta_2}-\zeta^2\,\bar z\mathrm e^{-\mathrm i\beta_2}\right]\cos\epsilon_1\sin\epsilon_2\Big\}\text{,}\\
    \left\langle\frac{\dd\lambda}{\dd t}\right\rangle_{\mathrm{M}}&=\left\langle\frac{\dd\varpi}{\dd t}\right\rangle_{\mathrm{M}}+\nu\sqrt{1-e^2}\big[2\cos\epsilon_1\cos\epsilon_2-\sin\epsilon_1\sin\epsilon_2\cos(\beta_1-\beta_2)\big]\nonumber\\*
    &+4\nu(1-e^2)\frac{\big[1-\sqrt{1-e^2}-e^2\big(1-\frac{1}{2}\sqrt{1-e^2}\big)\big]}{e^4}\sin\epsilon_1\sin\epsilon_2\cos(\beta_1+\beta_2)\text{,}\\
        \left\langle\frac{\dd z}{\dd t}\right\rangle_{\mathrm{M}}&=\mathrm i z\left\langle\frac{\dd\varpi}{\dd t}\right\rangle_{\mathrm{M}}\text{,}
  \end{align}
\end{subequations}
where the secular equation for the change in the longitude of the pericenter is given by
\begin{align}
  \left\langle\frac{\dd\varpi}{\dd t}\right\rangle_{\mathrm{M}}&=\nu\big[2\cos\epsilon_1\cos\epsilon_2-\sin\epsilon_1\sin\epsilon_2\cos(\beta_1-\beta_2)\big]\nonumber\\
  &+\frac{\mathrm i\nu}{2e\sqrt{1-\zeta\bar\zeta}}\Big[\big(\bar\zeta\,z\mathrm e^{\mathrm i\beta_1}-\zeta\,\bar z\mathrm e^{-\mathrm i\beta_1}\big)\sin\epsilon_1\cos\epsilon_2+\big(\bar\zeta\,z\mathrm e^{\mathrm i\beta_2}-\zeta\,\bar z\mathrm e^{-\mathrm i\beta_2}\big)\cos\epsilon_1\sin\epsilon_2\Big]\text{.}
\end{align}
\end{widetext}
In these expressions, we introduce $\nu$, the magnetic orbital frequency, defined by
\begin{equation}
  \nu=\frac{3\mu_0}{8\pi G}\,\frac{\mu_1\mu_2}{m_1m_2}\,\frac{n}{p^{2}}\text{.}
  \label{eq:nuM}
\end{equation}

The magnetic dipole-dipole interaction does not secularly affect the shape of the orbit (namely $a$ and $e$) but only its spatial orientation (namely $\varpi$ and $L$, and also $\iota$ and $\Omega$, through $\zeta$ and its complex conjugate). We saw in Eqs.~\eqref{eq:pertGR}, that the radiation-reaction terms do affect the shape of the orbit through a secular variation of the semi-major axis and eccentricity. The longitude of the pericenter and the mean longitude are simultaneously affected by both GR and magnetic perturbations.

In order to solve the secular Eqs. \eqref{eq:pertM}, we need solutions for the orientation of the magnetic moments, namely $\epsilon_1(t)$, $\epsilon_2(t)$, and $\beta_1(t)$, $\beta_2(t)$. After averaging Eqs.~\eqref{eq:rotation} over one orbital period and making use of the kinematic relationships in Eq.~\eqref{eq:epsbetkin}, we find
\begin{subequations}\label{eq:spin1}
\begin{align}
  \left\langle\frac{\dd\epsilon_1}{\dd t}\right\rangle_{\mathrm{M}}&=\nu_1\sin\epsilon_2\sin(\beta_1-\beta_2)\text{,}\\
  \sin\epsilon_1\left\langle\frac{\dd\beta_1}{\dd t}\right\rangle_{\mathrm{M}}&=2\nu_1\sin\epsilon_1\cos\epsilon_2\nonumber\\
  &+\nu_1\cos\epsilon_1\sin\epsilon_2\cos(\beta_1-\beta_2)\text{,}
\end{align}
\end{subequations}
where we have introduced $\nu_1$, the magnetic rotational frequency of the primary, defined by
\begin{equation}
  \nu_1=\frac{\mu_0}{8\pi}\,\frac{\mu_1\mu_2}{S_1}\,\frac{1}{a^3(1-e^2)^{3/2}}\text{.}
  \label{eq:nu1}
\end{equation}
There exist similar equations for the orientation of the secondary,
\begin{subequations}\label{eq:spin2}
\begin{align}
  \left\langle\frac{\dd\epsilon_2}{\dd t}\right\rangle_{\mathrm{M}}&=-\nu_2\sin\epsilon_1\sin(\beta_1-\beta_2)\text{,}\\
  \sin\epsilon_2\left\langle\frac{\dd\beta_2}{\dd t}\right\rangle_{\mathrm{M}}&=2\nu_2\cos\epsilon_1\sin\epsilon_2\nonumber\\
  &+\nu_2\sin\epsilon_1\cos\epsilon_2\cos(\beta_1-\beta_2)\text{,}
\end{align}
\end{subequations}
with
\begin{equation}
  \nu_2=\frac{\mu_0}{8\pi}\,\frac{\mu_1\mu_2}{S_2}\,\frac{1}{a^3(1-e^2)^{3/2}}\text{.}
  \label{eq:nu2}
\end{equation}

Equations \eqref{eq:pertGR}, \eqref{eq:pertM}, \eqref{eq:spin1}, and \eqref{eq:spin2} form a system of coupled first-order differential equations. This system describes the secular evolution of the orbital and the rotational motion of a binary system under gravitational and magnetic dipole-dipole interactions out of equilibrium. These equations are solved in the next section.

\section{Solutions}
\label{sec:sol}

In order to gain some insight into the motion, we would like to solve the secular Eqs. \eqref{eq:pertGR}, \eqref{eq:pertM}, \eqref{eq:spin1} and \eqref{eq:spin2} analytically.  In this way, we can derive scaling laws that can then be used while searching for magnetic signatures within the GW signal. The first-order analytic estimates that are derived hereafter cannot always be employed. For this reason, a numerical resolution of the secular equations of motion is also needed. In addition, the numerical solution can be used to verify the validity of the first-order analytic solutions.

\subsection{Numerical setup}

As one can see from the expressions for $\nu$, $\nu_1$, and $\nu_2$, the effects of the dipole-dipole interaction on the motion are proportional to the product of the magnetic moments of the stars (see Eq. \eqref{eq:magmom} for dimensional expression of the amplitude of the magnetic moment). Considering that magnetic fields can reach up to $10^9\,\mathrm{G}$ for the most magnetized WDs and up to $10^{15}\,\mathrm{G}$ for the most magnetized NSs, numerical rough estimates are as follows
\begin{equation}
  \mu_{\mathrm{WD}}\sim 10^{33}\,\mathrm{A}\cdot\mathrm{m}^2\left(\frac{R_{\mathrm{WD}}}{10^4\,\mathrm{km}}\right)^3\left(\frac{B_{\mathrm{WD}}}{10^9\,\mathrm{G}}\right)
  \label{eq:magmomWD}
\end{equation}
for WDs, and
\begin{equation}
\mu_{\mathrm{NS}}\sim 10^{30}\,\mathrm{A}\cdot\mathrm{m}^2\left(\frac{R_{\mathrm{NS}}}{10\,\mathrm{km}}\right)^3\left(\frac{B_{\mathrm{NS}}}{10^{15}\,\mathrm{G}}\right)
\end{equation}
for NSs. Therefore, even though the magnetic fields of highly magnetic NSs are several orders of magnitude higher than for highly magnetic WDs, their magnetic moments are smaller. Indeed, as seen from Eq. \eqref{eq:magmom}, the magnetic moment evolves as the cubic power of the radius whereas it is only linear in the magnitude of the magnetic field (see also \citet{2018ApJ...868...19W} and \citet{2021arXiv210910722M}). Therefore, we expect the dipole-dipole magnetic interaction to be the strongest for a binary of highly magnetic WDs. This is the case we focus on in the upcoming numerical applications.

We thus consider a double WD system where the mass of the primary is $m_1=1.2\,\mathrm{M}_{\odot}$ and the mass of the secondary is $m_2=0.3\,\mathrm{M}_{\odot}$ such that the total mass is $m=1.5\,\mathrm{M}_{\odot}$. Assuming that WDs are made of a cold Fermi gas in hydrostatic equilibrium \cite{1931ApJ....74...81C,1931MNRAS..91..456C,*1935MNRAS..95..207C,1967aits.book.....C}, we choose the radii according to the mass-radius relationship, so that we take $R_1=6\times 10^3\,\mathrm{km}$ and $R_2=15\times 10^3\,\mathrm{km}$. We consider a system with high magnetic fields, at the level of $B_1=B_2=10^{9}\,\mathrm{G}$.

We assume that the initial value of the semi-major axis is given by $a_0=(4Gm)^{1/3}\Phi_0{}^{-2/3}$, where $\Phi_0$ is the LISA main frequency for GBs (i.e., $\Phi_0=2n_0$ for a circular orbit). In order to probe the LISA frequency window, we consider three different cases where $\Phi_0=10^{-1}\,\mathrm{Hz}$, $10^{-2}\,\mathrm{Hz}$, and $10^{-3}\,\mathrm{Hz}$, which correspond to a semi-major axis at the level of $a_0=4.3\times 10^4\,\mathrm{km}$, $2\times 10^5\,\mathrm{km}$, and $9.2\times 10^5\,\mathrm{km}$, respectively. The initial conditions for the other orbital elements and angles for the orientation of the magnetic moments are reported in Tab.~\ref{tab:initcond}, where $D$ is the distance between the source of the gravitational radiations and the observer (see Sect. \ref{sec:GWsignal}).

\begin{table}
  \centering
  \caption[]{Numerical values and initial conditions.}
  \label{tab:initcond}
  \begin{tabularx}{\columnwidth}{l c| >{\centering\arraybackslash}X >{\centering\arraybackslash}X >{\centering\arraybackslash}X}
    \hline\hline
    Parameter & Unit & \multicolumn{3}{c}{Value}\\
    \hline
    \multicolumn{5}{c}{Physical parameters}\\
    \hline
    $m_1$ & $\mathrm{M}_\odot$ & \multicolumn{3}{c}{1.2}\\
    $m_2$ & $\mathrm{M}_\odot$ & \multicolumn{3}{c}{0.3}\\
    $R_1$ & km & \multicolumn{3}{c}{$6\times10^3$}\\
    $R_2$ & km & \multicolumn{3}{c}{$15\times 10^3$}\\
    $P_1$ & h & \multicolumn{3}{c}{1}\\
    $P_2$ & h & \multicolumn{3}{c}{10}\\
    $B_1$ & G & \multicolumn{3}{c}{$10^9$}\\
    $B_2$ & G & \multicolumn{3}{c}{$10^9$}\\
    $D$ & kpc & \multicolumn{3}{c}{1}\\
    \hline
    \multicolumn{5}{c}{LISA frequency}\\
    \hline
    $\Phi_0$ & Hz & $10^{-1}$ & $10^{-2}$ & $10^{-3}$\\
    $a_0$ & km & $4.3\times 10^4$ & $2.0\times 10^5$ & $9.2\times 10^5$\\
    \hline
    \multicolumn{5}{c}{Orbital parameters}\\
    \hline
    $e_0$ & - & \multicolumn{3}{c}{0.1}\\
    $\iota_0$ & deg & \multicolumn{3}{c}{45}\\
    $\Omega_0$ & deg & \multicolumn{3}{c}{0}\\
    $\omega_0$ & deg & \multicolumn{3}{c}{45}\\
    $\tau_0$ & s & \multicolumn{3}{c}{0}\\
    \hline
    \multicolumn{5}{c}{Rotational parameters}\\
    \hline
    $\epsilon_1$ & deg & \multicolumn{3}{c}{10}\\
    $\beta_1$ & deg & \multicolumn{3}{c}{10}\\
    $\epsilon_2$ & deg & \multicolumn{3}{c}{160}\\
    $\beta_2$ & deg & \multicolumn{3}{c}{20}\\
    \hline
  \end{tabularx}
\end{table}

We assume that both stars are spherically symmetric so that the magnitude of the angular momentum of the primary is given by
\begin{align}
  S_1&=\frac{4\pi}{5}\frac{m_1R_1^2}{P_1}\simeq6.0\times 10^{40}\,\mathrm{kg}\cdot\mathrm{m}^2\cdot\mathrm{s}^{-1}\nonumber\\
  &\times\left(\frac{m_1}{1.2\,\mathrm{M}_{\odot}}\right)\left(\frac{R_1}{6\times 10^3\,\mathrm{km}}\right)^2\left(\frac{1\,\mathrm{h}}{P_1}\right)\text{,}\label{eq:S1}
\end{align}
where $P_1$ is the period of the proper rotation. Similarly, for the secondary we have
\begin{align}
  S_2&=\frac{4\pi}{5}\frac{m_2R_2^2}{P_2}\simeq2.3\times 10^{39}\,\mathrm{kg}\cdot\mathrm{m}^2\cdot\mathrm{s}^{-1}\nonumber\\
  &\times\left(\frac{m_2}{0.3\,\mathrm{M}_{\odot}}\right)\left(\frac{R_2}{15\times 10^3\,\mathrm{km}}\right)^2\left(\frac{10\,\mathrm{h}}{P_2}\right)\text{,}\label{eq:S2}
\end{align}
where $P_2$ is the period of the proper rotation.

\subsection{Analytic estimates}

\begin{figure*}
  \begin{center}
    \includegraphics[scale=1]{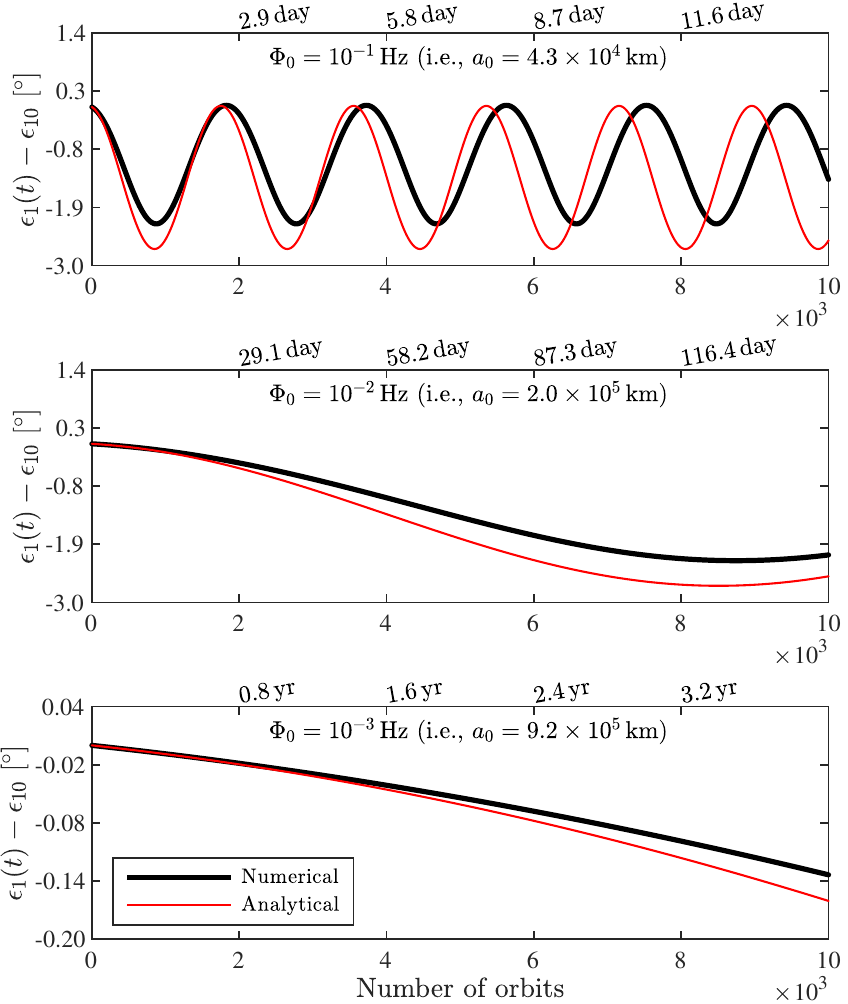}
    \hspace{0.2cm}\includegraphics[scale=1]{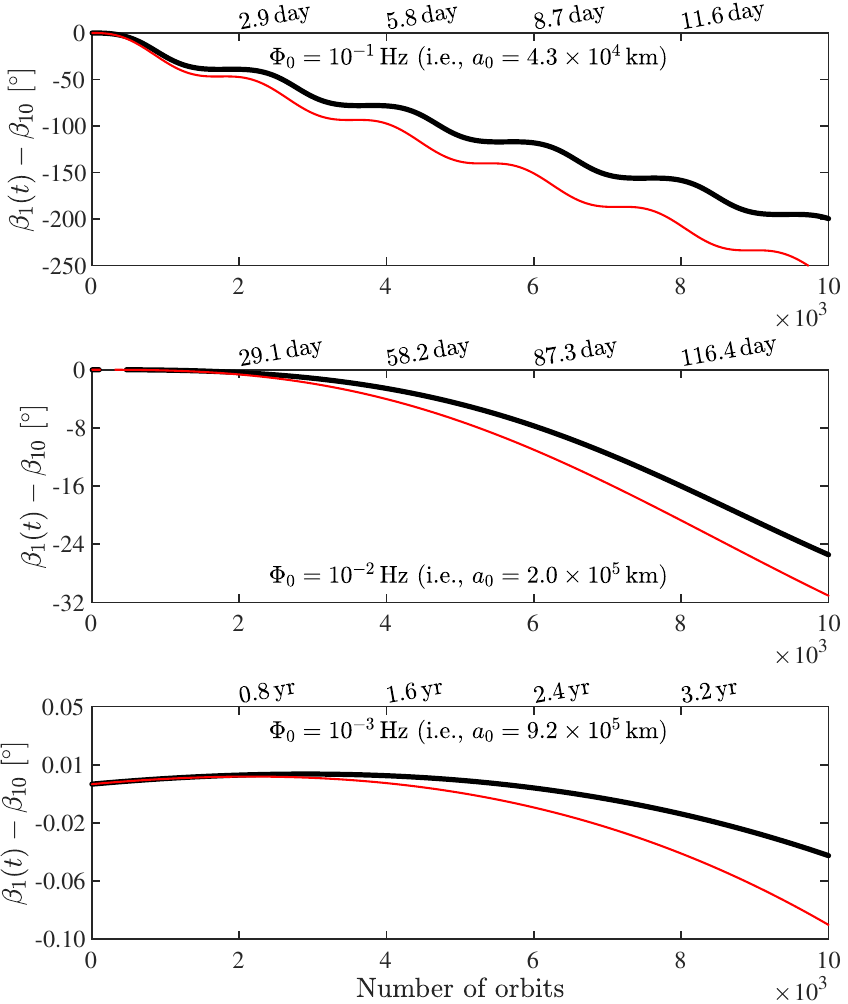}
  \end{center}
  \caption{Difference between analytical (thin red curves) and numerical solutions (thick black curves) for the evolution of $\epsilon_1$ (left-hand side) and $\beta_1$ (right-hand side) considering the magnetic dipole-dipole interaction. The initial values $\epsilon_{10}$ and $\beta_{10}$ have been removed for more readability. The evolutions are represented for different values of the LISA main frequency for GBs, namely $\Phi_0=10^{-1}\,\mathrm{Hz}$ (top panels), $\Phi_0=10^{-2}\,\mathrm{Hz}$ (middle panels), and $\Phi_0=10^{-3}\,\mathrm{Hz}$ (bottom panels). The bottom $x$-axis is the number of orbits and the top $x$-axis represents the elapsed time.}
  \label{fig:Diffepsbet1}
\end{figure*}

\begin{figure*}
  \begin{center}
    \includegraphics[scale=1]{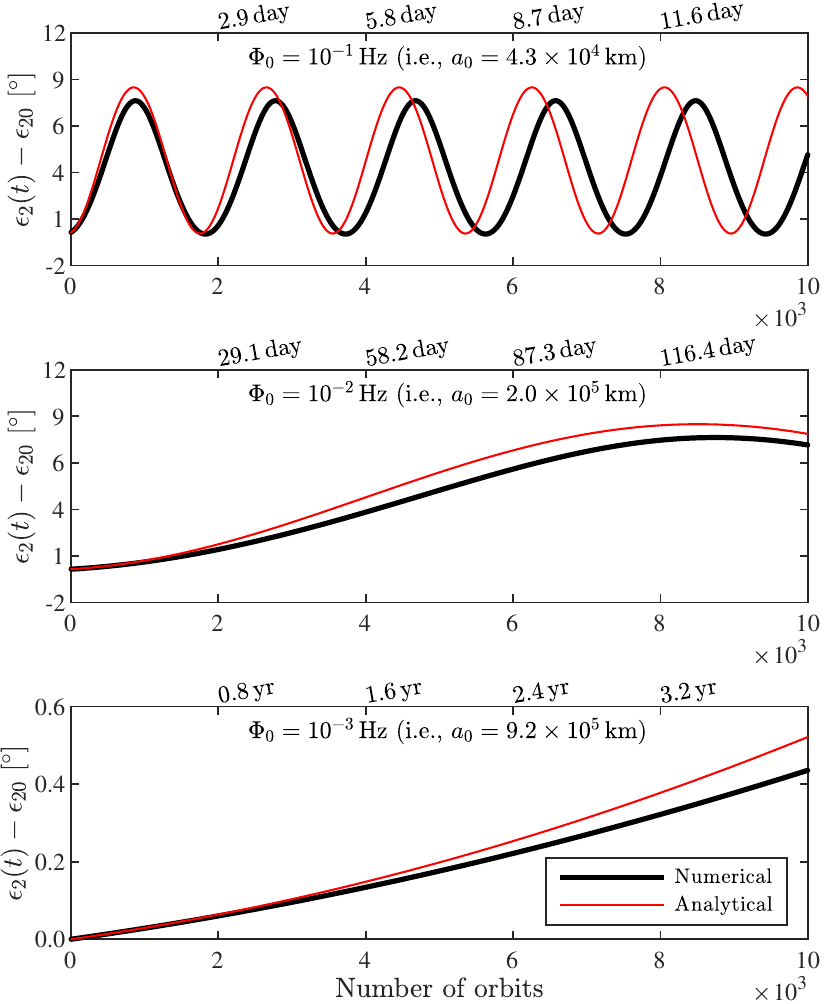}
    \hspace{0.2cm}\includegraphics[scale=1]{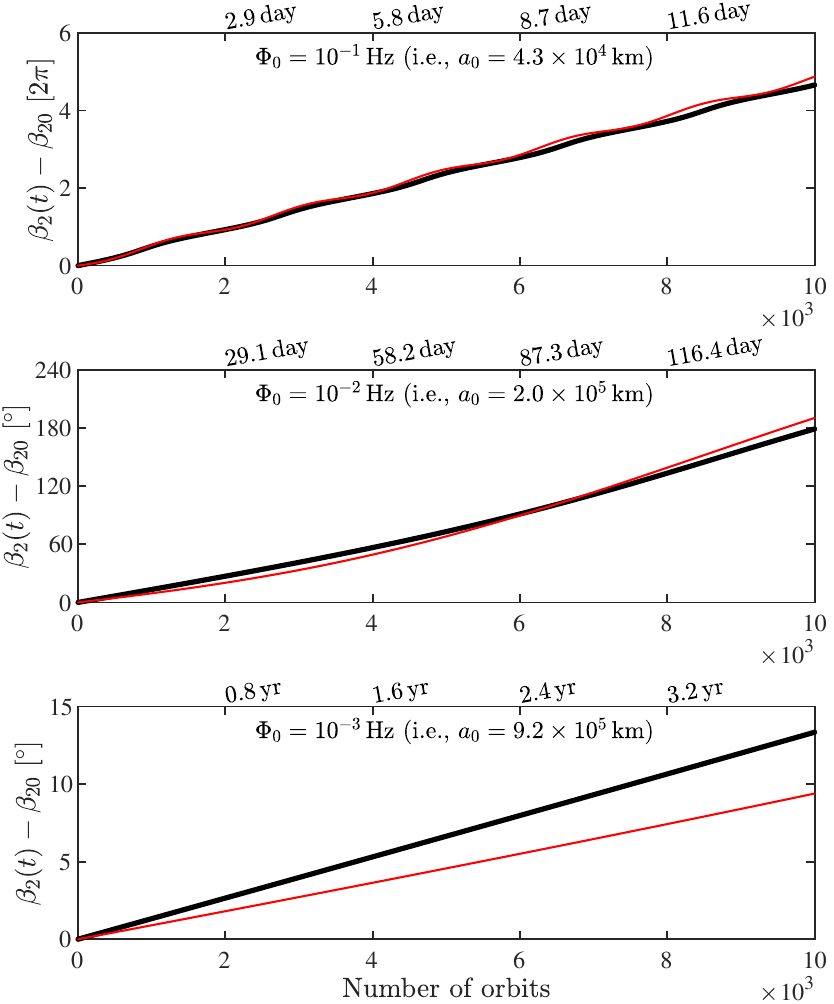}
  \end{center}
  \caption{Difference between analytical (thin red curves) and numerical solutions (thick black curves) for the evolution of $\epsilon_2$ (left-hand side) and $\beta_2$ (right-hand side) considering the magnetic dipole-dipole interaction. The initial values $\epsilon_{20}$ and $\beta_{20}$ have been removed for more readability. The axis are the same than in Fig.~\ref{fig:Diffepsbet1}.}
  \label{fig:Diffepsbet2}
\end{figure*}

\begin{figure*}
  \begin{center}
    \includegraphics[scale=1]{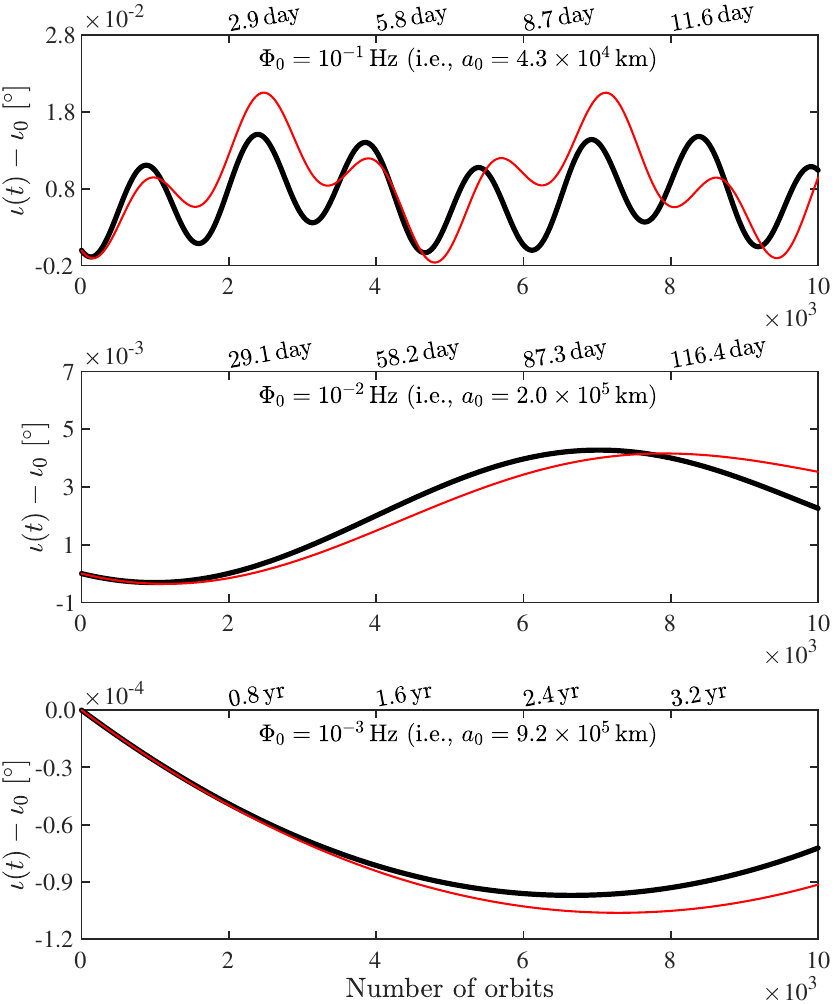}
    \hspace{0.1cm}\includegraphics[scale=1]{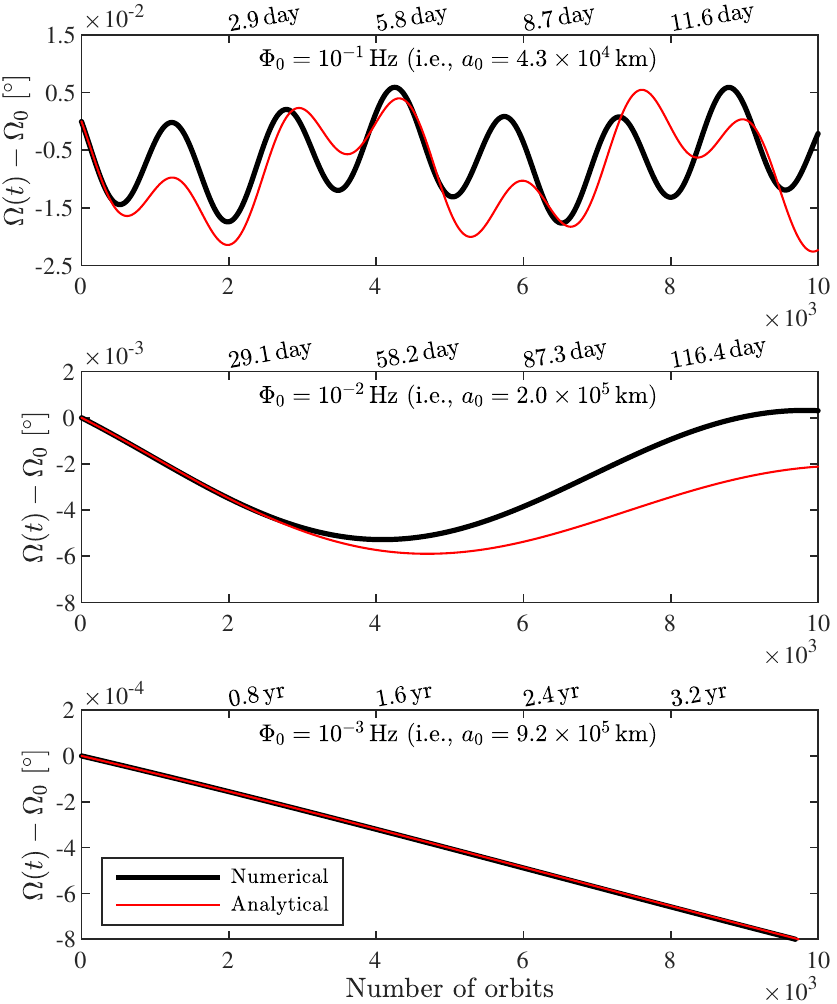}
  \end{center}
  \caption{Difference between analytical (thin red curves) and numerical solutions (thick black curves) for the evolution of $\iota$ (left-hand side) and $\Omega$ (right-hand side) considering the magnetic dipole-dipole interaction. The initial values $\iota_0$ and $\Omega_0$ have been removed for more readability. The axis are the same than in Fig.~\ref{fig:Diffepsbet1}.}
  \label{fig:DiffIncOmeMov}
\end{figure*}

\begin{figure*}
  \begin{center}
    \includegraphics[scale=1]{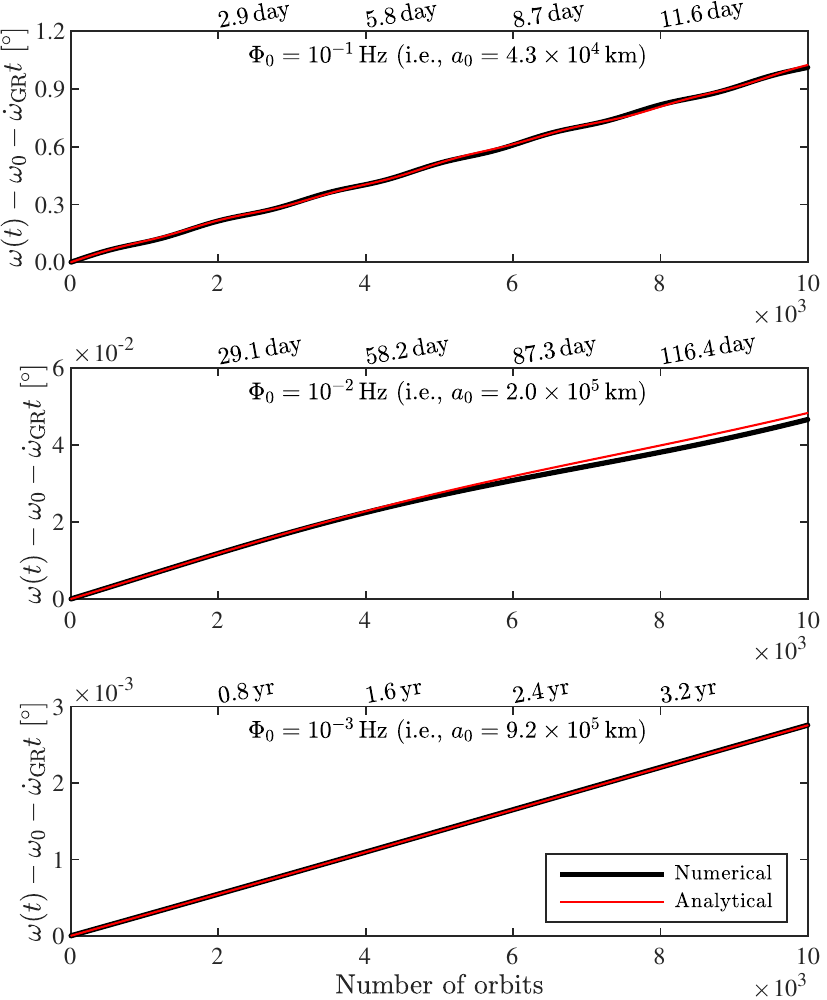}
    \hspace{0.1cm}\includegraphics[scale=1]{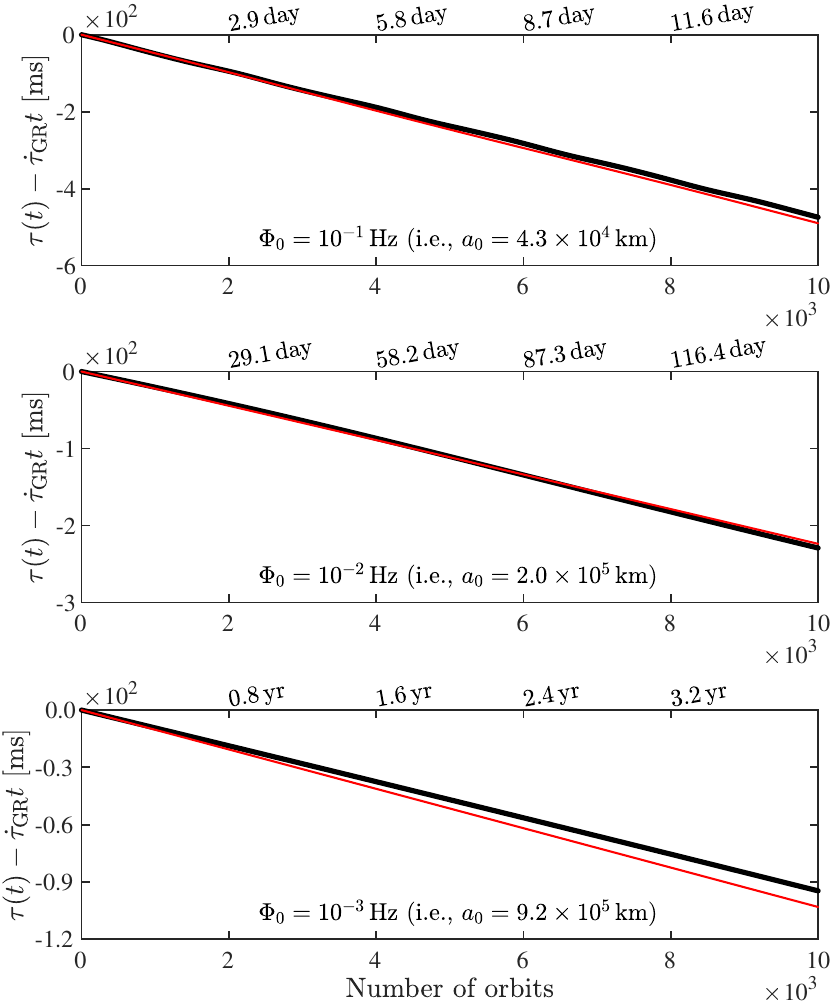}
  \end{center}
  \caption{Difference between analytical (thin red curves) and numerical solutions (thick black curves) for the evolution of $\omega$ (left-hand side) and $\tau$ (right-hand side) considering the magnetic dipole-dipole interaction. The initial value $\omega_0$ has been removed for more readability. The secular contribution from GR is removed too, so that the remaining secular effect is only due to magnetism. The axis are the same than in Fig.~\ref{fig:Diffepsbet1}.}
  \label{fig:DiffomeTauMov}
\end{figure*}

The first-order solutions for the GR contribution to the changes in the orbital elements, can be obtained by substituting the constant zeroth-order values of the non-singular elements $\mathbf X$ into the right-hand side of \eqref{eq:pertGR}. The solutions read as follows
\begin{subequations}\label{eq:GRsol}
\begin{align}
  a(t)&=a_0+\dot a_{\mathrm{GR}} t\text{,}\label{eq:sola}\\
  \lambda(t)&=\lambda_0+\dot\lambda_{\mathrm{GR}}t\text{,}\\
  z(t)&=z_0\mathrm{e}^{\dot e_{\mathrm{GR}}t/e_0}\mathrm{e}^{\mathrm i\dot\varpi_{\mathrm{GR}}t}\text{,}\label{eq:solz}
\end{align}
\end{subequations}
where $\dot a_{\mathrm{GR}}$, $\dot e_{\mathrm{GR}}$, $\dot\varpi_{\mathrm{GR}}$, and $\dot\lambda_{\mathrm{GR}}$ are shorthand notations referring to the secular time derivatives in Eqs. \eqref{eq:pertGR} and \eqref{eq:pertGRdeodt}, where the orbital elements in the right-hand sides, are replaced by their constant zeroth-order values.

The solutions \eqref{eq:GRsol} show that the rate of change of the shape of the orbit (namely $a$ and $e$), for typical inspiral of GBs between $\Phi_0=10^{-1}\,\mathrm{Hz}$ and $10^{-4}\,\mathrm{Hz}$, occurs on much longer timescale than the previsted time duration of the LISA mission (namely $\sim 4\,\mathrm{yr}$). Indeed, $t_{\mathrm{GW}}$, the characteristic time for the secular changes resulting from the gravitational radiation, is of the order of
\begin{equation}
  t_{\mathrm{GW}}\sim c^5(Gm)^{-5/3}\Phi_0{}^{-8/3}\text{.}
\end{equation}
For a binary with a total mass of $m=1.5\,\mathrm{M}_{\odot}$, this corresponds to $t_{\mathrm{GW}}\sim500\times 10^9\,\mathrm{yr}$ when $\Phi_0=10^{-4}\,\mathrm{Hz}$ and to $t_{\mathrm{GW}}\sim 5\,000\,\mathrm{yr}$ when $\Phi_0=10^{-1}\,\mathrm{Hz}$. For both cases, the characteristic time is much longer than the time duration of the LISA mission. Therefore, the change in the mean motion can safely be approximated by its first-order Taylor expansion, namely
\begin{equation}
  n(t)\simeq n_0-\frac{3n_0}{2a_0}\,\dot a_{\mathrm{GR}} t\text{.}
  \label{eq:nGR}
\end{equation}
From this last relationship, and from Eq. \eqref{eq:solz}, we can infer an approximate expression for the change in the eccentrity during the time needed to go from $\Phi_0$ to $\Phi$. The expression read as follows
\begin{equation}
e(\Phi)=e_0\,\mathrm{e}^{-\frac{19}{6}\frac{\left(1-\frac{173}{304}{e_0}^2-\frac{131}{304}{e_0}^4\right)}{\left(1+\frac{73}{24}{e_0}^2+\frac{37}{96}{e_0}^4\right)}\frac{(\Phi-\Phi_0)}{\Phi_0}}\text{,}
\end{equation}
with $e_0=e(\Phi_0)$. As an application, let us consider the following example. Let us assume an initial eccentricity at $e_0=0.7$, and let us compute the final eccentricity when $\Phi=10\Phi_0$; we find: $e(\Phi)=7\times 10^{-3}$. This shows that the gravitational radiation is an efficient mechanism for  orbit circularization. Therefore, for old inspiral binary systems, whose frequency has increased by one or several orders of magnitude since formation, we expect to observe mostly quasi-circular orbits. This justifies the use of the non-singular orbital elements that were introduced in the previous section.

Let us emphasize that the secular change in the longitude of the pericenter occurs on a much shorter timescale than the gravitational radiation. Indeed, the 1PN order perturbation's characteristic timescale reads
\begin{equation}
  t_{1\mathrm{PN}}\sim c^2(Gm)^{-2/3}\Phi_0{}^{-5/3}\text{.}
\end{equation}
For a binary with a total mass of $m=1.5\,\mathrm{M}_{\odot}$, this corresponds to $t_{1\mathrm{PN}}\sim 10^3\,\mathrm{yr}$ when $\Phi_0=10^{-4}\,\mathrm{Hz}$ and to $t_{1\mathrm{PN}}\sim 4\,\mathrm{day}$ when $\Phi_0=10^{-1}\,\mathrm{Hz}$.

We conclude that, in the context of the LISA mission, the rate of change of the shape of the orbit can be neglected while solving for the rotational motion. Accordingly, $\nu_1$ and $\nu_2$ can be considered constant. On the contrary, the 1PN effect must be accounted for, especially for the high frequency band, since it occurs on a timescale that is comparable to the time duration of the LISA mission.

The first-order solutions for the rotational motion can be derived by substituting the following first-order ansatz in the right-hand side of \eqref{eq:spin1} and \eqref{eq:spin2}:
\begin{equation}
  \epsilon_1=\epsilon_{10}\text{,}\qquad\beta_1=\beta_{10}+\dot\beta_1 t\text{,}
  \label{eq:ansrot}
\end{equation}
where $\epsilon_{10}$ and $\beta_{10}$ are two constants corresponding to the initial conditions for the orientations of the primary. Similar relationships are used for the secondary. The coefficient $\dot\beta_1$ corresponds to the rate of change of the precession angle and is determined by identification after integration. Note that substituting \eqref{eq:ansrot} in the right-hand side of \eqref{eq:spin1} and \eqref{eq:spin2} amounts to assuming small periodic variations.

Using \eqref{eq:ansrot} and integrating Eqs. \eqref{eq:spin1} and \eqref{eq:spin2}, the first-order solutions read
\begin{subequations}\label{eq:solspin1}
\begin{align}
  \epsilon_1(t)&=\epsilon_{10}+\widetilde\epsilon_1(t)-\widetilde\epsilon_1(0)\text{,}\\
  \beta_1(t)&=\beta_{10}+\widetilde\beta_1(t)-\widetilde\beta_1(0)+\dot\beta_1 t\text{,}  
\end{align}
\end{subequations}
where a ``tilde'' denotes a periodic contribution and a ``dot'' refers to a secular rate of change. We have similar relationships for the secondary. The secular precessing components are given by
\begin{subequations}
\begin{align}
  \dot\beta_1&=2\nu_{10}\cos\epsilon_{20}\text{,}\\
  \dot\beta_2&=2\nu_{20}\cos\epsilon_{10}\text{,}
\end{align}
\end{subequations}
and the periodic variations read as
\begin{subequations}\label{eq:oblprec1}
\begin{align}
  \widetilde\epsilon_1(t)&=-\frac{\nu_{10}}{\dot\beta_1-\dot\beta_2}\sin\epsilon_{20}\nonumber\\
  &\times\cos\Big[(\dot\beta_1-\dot\beta_2)t+\beta_{10}-\beta_{20}\Big]\text{,}\\
  \widetilde\beta_1(t)&=\frac{\nu_{10}}{\dot\beta_1-\dot\beta_2}\cot\epsilon_{10}\sin\epsilon_{20}\nonumber\\
  &\times\sin\Big[(\dot\beta_1-\dot\beta_2)t+\beta_{10}-\beta_{20}\Big]\text{,}
\end{align}
\end{subequations}
and
\begin{subequations}\label{eq:oblprec2}
\begin{align}
  \widetilde\epsilon_2(t)&=\frac{\nu_{20}}{\dot\beta_1-\dot\beta_2}\sin\epsilon_{10}\nonumber\\
  &\times\cos\Big[(\dot\beta_1-\dot\beta_2)t+\beta_{10}-\beta_{20}\Big]\text{,}\\
  \widetilde\beta_2(t)&=\frac{\nu_{20}}{\dot\beta_1-\dot\beta_2}\cot\epsilon_{20}\sin\epsilon_{10}\nonumber\\
  &\times\sin\Big[(\dot\beta_1-\dot\beta_2)t+\beta_{10}-\beta_{20}\Big]\text{.}
\end{align}
\end{subequations}
The frequencies $\nu_{10}$ and $\nu_{20}$ are obtained after substituting the constant zeroth-order orbital elements in the right-hand side of Eqs. \eqref{eq:nu1} and \eqref{eq:nu2}, respectively.

Let us emphasize that the first-order solutions cannot be employed near resonance, namely when $|\dot\beta_1|\sim|\dot\beta_2|$, that is to say when $S_1|\cos\epsilon_{10}|\sim S_2|\cos\epsilon_{20}|$. When the two rates of precession are similar, a more sophisticated method of resolution is needed, numerical integration for instance. This is the reason why the secular Eqs. \eqref{eq:pertGR}, \eqref{eq:pertM}, \eqref{eq:spin1}, and \eqref{eq:spin2} are also solved numerically with the \texttt{MATLAB} double precision variable order method \texttt{ode113} with variable step size and for a relative error tolerance equal to $10^{-12}$.

The first-order solutions for $\epsilon_1(t)$, $\beta_1(t)$, $\epsilon_2(t)$, and $\beta_2(t)$ are compared with the numerical ones in Figs. \ref{fig:Diffepsbet1} and \ref{fig:Diffepsbet2} for the three different initial values of the semi-major axis (cf. Tab. \ref{tab:initcond}). For the cases shown here, we have $|\cos\epsilon_{10}|/|\cos\epsilon_{20}|=1.0$ and $S_2/S_1=0.16$. This ensures that the rates of precession $\dot\beta_1$ and $\dot\beta_2$ are different and justifies the use of the first-order solutions.

In Figs. \ref{fig:Diffepsbet1} and \ref{fig:Diffepsbet2} it is shown that the precession angles $\beta_1$ and $\beta_2$ vary linearly with time while the obliquity angles $\epsilon_1$ and $\epsilon_2$ oscillate. In addition, we note that the amplitudes of the oscillations are independent of the star separation. This is confirmed by the analytic solutions in Eqs.~\eqref{eq:oblprec1} and \eqref{eq:oblprec2}. Indeed, after recalling that $S_1\gg S_2$, the amplitudes in Eqs.~\eqref{eq:oblprec1} and \eqref{eq:oblprec2} reduce to
\begin{equation}
  \frac{\nu_{10}}{\dot\beta_1-\dot\beta_2}\propto-\frac{S_2}{S_1}\text{,}\qquad\frac{\nu_{20}}{\dot\beta_1-\dot\beta_2}\propto-1\text{.}
\end{equation}
This shows that the amplitudes vary with the ratio between the magnitude of the spins. The frequency of the oscillations changes with star separation. 

It is now possible to estimate the effect of the dipole-dipole interaction on the orbital motion. In order to further simplify the integration of Eqs. \eqref{eq:pertM}, we only consider the secular variations in the precession angles $\beta_1$ and $\beta_2$ and neglect the oscillations. In addition, we account for GR by substituting $z$ with the 1PN solution (i.e., by taking the limit $\dot e_{\mathrm{GR}}\rightarrow 0$ in Eq. \eqref{eq:solz}) into the right-hand side of Eqs.~\eqref{eq:pertM}. The 2.5PN contribution is re-inserted after integration for completness. Since GR has no effect on the inclination nor the longitude of the node, we replace $\zeta$ by $\zeta_0$ in Eqs. \eqref{eq:pertM}.

After integrating the secular equations with respect to time, the total first-order solutions for the orbital motion of the binary reads as follows: 
\begin{subequations}\label{eq:solorb}
\begin{align}
  \zeta(t)&=\zeta_0+\widetilde\zeta_{\mathrm{M}}(t)-\widetilde\zeta_{\mathrm{M}}(0)\text{,}\\
  \lambda(t)&=\lambda_0+\widetilde\lambda_{\mathrm{M}}(t)-\widetilde\lambda_{\mathrm{M}}(0)+(\dot\lambda_{\mathrm{GR}}+\dot\lambda_{\mathrm{M}})t\text{,}\label{eq:solorbtau}\\
  z(t)&=z_0\mathrm{e}^{\dot e_{\mathrm{GR}}t/e_0}\mathrm{e}^{\mathrm{i}(\varpi(t)-\varpi_0)}\text{.}\label{eq:solorbz}
\end{align}
\end{subequations}
The solution for $a$ is the same than in Eq. \eqref{eq:sola} since the dipole-dipole interaction has no secular effect on the semi-major axis evolution. The expression for the longitude of the pericenter (in Eq. \eqref{eq:solorbz}) is given by
\begin{equation}
  \varpi(t)=\varpi_0+\widetilde\varpi_{\mathrm{M}}(t)-\widetilde\varpi_{\mathrm{M}}(0)+(\dot\varpi_{\mathrm{GR}}+\dot\varpi_{\mathrm{M}})t\text{.}\label{eq:solorbome}
\end{equation}
From Eqs. \eqref{eq:solorbtau}, we can compute the secular evolution of the mean longitude. Indeed, after substituting for $n(t)$ from Eq.~\eqref{eq:nGR} into \eqref{eq:lambdadef}, we find
\begin{align}
  L(t)&=L_0+\widetilde\lambda_{\mathrm{M}}(t)-\widetilde\lambda_{\mathrm{M}}(0)\nonumber\\
  &+(n_0+\dot\lambda_{\mathrm{GR}}+\dot\lambda_{\mathrm{M}})t-\frac{3n_0}{4a_0}\,\dot{a}_{\mathrm{GR}}t^2\label{eq:solL}
\end{align}
with $L_0=\lambda_0$. The secular contributions $\dot\varpi_{\mathrm{M}}$ and $\dot\lambda_{\mathrm{M}}$ are, respectively, given by
\begin{subequations}\label{eq:dotomeTauM}
\begin{align}
  \dot\varpi_{\mathrm{M}}&=2\nu_{0}\cos\epsilon_{10}\cos\epsilon_{20}\text{,}\label{eq:dotTauM}\\
  \dot\lambda_{\mathrm{M}}&=\dot\varpi_{\mathrm{M}}\Big(1+\sqrt{1-{e_0}^2}\Big)\text{,}\label{eq:dotomeM}
\end{align}
\end{subequations}
where the frequency $\nu_0$ is determined by substituting the constant zeroth-order orbital elements into the right-hand side of Eq. \eqref{eq:nuM}.

The expressions of the periodic contributions $\widetilde\zeta_{\mathrm{M}}(t)$, $\widetilde\lambda_{\mathrm{M}}(t)$, and $\widetilde\varpi_{\mathrm{M}}(t)$ are given explicitly in Eqs. \eqref{eq:solorbM}. With these, we can now compute the evolution of the regular elements, using Eqs. \eqref{eq:sola}, \eqref{eq:solorb}, \eqref{eq:solorbome}, and \eqref{eq:solL}, together with the secular pieces in Eqs. \eqref{eq:dotomeTauM}.

We recall that, when the eccentricity and the inclination are different from zero, it is straightforward to re-express the solutions \eqref{eq:solorb} in terms of the more familiar but singular elements $(a,e,\iota,\Omega,\omega,\tau)$. In Figs. \ref{fig:DiffIncOmeMov} and \ref{fig:DiffomeTauMov}, we present the comparison between the analytic estimates (given in terms of the singular elements) and the results of a numerical integration for the three different initial values of the LISA main frequency (cf. Tab. \ref{tab:initcond}).

As it can be seen from the analytic solutions, the evolution of the inclination and the longitude of the node (which are determined from $\zeta$ and its complex conjugate) is a sum of two periodic oscillations. The longest periodic oscillation possesses an amplitude $\Theta_1$, while the other one, with the shortest period, has the amplitude $\Theta_2$ (cf. Eqs.~\eqref{eq:ampTheta1} and \eqref{eq:ampTheta2} for the expressions of $\Theta_1$ and $\Theta_2$, respectively). In addition, one can see from Fig. \ref{fig:DiffIncOmeMov} that the amplitudes $\Theta_1$ and $\Theta_2$ increase when the semi-major axis decreases. This behavior is highlighted in Fig.~\ref{fig:Amplitude}, which shows the evolution of $\Theta_1$ and $\Theta_2$ with respect to the semi-major axis. It is shown that the amplitudes of the oscillations are actually negligible relative to the secular variations and for the LISA frequency band. This point is further discussed in appendix~\ref{sec:Amposc}.

In conclusion, the overall magnetic effect that must be eventually considered are the secular change of the mean longitude (i.e., terms $\propto t$ and $\propto t^2$ in Eq.~\eqref{eq:solL}) and the secular change in the longitude of the pericenter (i.e., terms $\propto t$ in Eq.~\eqref{eq:solorbome}). The solution for the mean longitude in Eq. \eqref{eq:solL} is used in the following section to model the combined effects of GR and magnetism on the GW mode polarizations at zeroth-order in eccentricity. The solution for the longitude of the pericenter in Eq.~\eqref{eq:solorbome}, is used to model the GW mode polarizations at the first-order in eccentricity.

\section{Impact of the dipole-dipole interaction on the GW strain}
\label{sec:GWsignal}

In this section, we derive the expressions of the mode polarizations up to the first-order in eccentricity in agreement with the assumption that GBs are in quasi-circular orbit. Mode decomposition is performed in the source frame in order to be coherent with the conventions of the LISA Data Challenge (LDC) \cite{LDCGroup}. We then use the method of variation of arbitrary constants to account for the combined effects of GR and magnetic perturbations on the orbital dynamics. We make use of the secular solutions derived in Sect. \ref{sec:sol}. Finally, we discuss the effects of the perturbations, in both the time and frequency domains, and in the context of the future LISA mission.

\subsection{Gravitational radiation from quasi-circular binary system}

We now suppose that the observer is in the far-away wave zone. Hence, the field point is considered far from the source point in the sense that the separation between the two is much larger than the characteristic wavelength of the GW emitted by the binary system. Accordingly, the relative motion of the observer can be neglected. We assume for convenience that the $\hat{\mathbf{e}}_Z$-axis is aligned with the direction of the observer. In contrast with LDC conventions \cite{LDCGroup}, we do not suppose that the $\hat{\mathbf{e}}_X$-axis is aligned with the direction of the ascending node. This choice is motivated by the fact that, in general, $\Omega$ is time-dependent when perturbations are considered. In order to recover LDC conventions, one has to take the limit $\Omega\rightarrow 0$.

For an observer in the far-away wave zone, it is well-known (see e.g., \citet{1963PhRv..131..435P}, \citet{2014grav.book.....P}) that the GW mode polarizations $h_+$ and $h_\times$ are given by the following expressions
\begin{subequations}\label{eq:GWmodes}
\begin{align}
  h_{+}&=\frac{1}{2}\left[(\hat{\mathbf e}_Y)_j(\hat{\mathbf e}_Y)_k-(\hat{\mathbf e}_X)_j(\hat{\mathbf e}_X)_k\right]h^{jk}\text{,}\\
  h_{\times}&=\frac{1}{2}\left[(\hat{\mathbf e}_X)_j(\hat{\mathbf e}_Y)_k+(\hat{\mathbf e}_Y)_j(\hat{\mathbf e}_X)_k\right]h^{jk}\text{,}  
\end{align}
\end{subequations}
in the source frame. The components $h^{jk}$ are defined by the well-known quadrupole formula
\begin{equation}
  h^{jk}(t,\mathbf x)=\frac{2G}{c^4D}\,\ddot{I}^{jk}(t_*)\text{.}
  \label{eq:potquad}
\end{equation}
In the latter expression, $D$ is the distance between the source and the field points, namely $D=|\mathbf x|$, and $t_*$ is the retarded time, namely $t_*=t-D/c$. The last term in Eq. \eqref{eq:potquad} is the second time derivative of the quadrupole moment of inertia evaluated at the retarded time. The components of the quadrupole moment of inertia are defined by
\begin{equation}
  I^{jk}(t)=\int\rho(t,\mathbf x){x}^j{x}^k\,\dd^3 x\text{.}
  \label{eq:quadInert}
\end{equation}
For a binary formed by two point-masses, the expression of the second time derivative of \eqref{eq:quadInert} can be derived straightforwardly from the tensor virial theorem. It involves the Kepler solution (cf. Eqs. \eqref{eq:solKep}). Therefore, the mode polarizations \eqref{eq:GWmodes} are conveniently expressed in term of the orbital elements $(a,e,\iota,\Omega,\omega,f)$, where the true anomaly is the angle varying on short timescale \cite{1963PhRv..131..435P,2012ApJ...752...67K}. Here, we give the expression of the mode polarizations up to first-order in eccentricity and in terms of non-singular elements $(a,z,\zeta,L)$. Hence, we use the mean longitude instead of the true anomaly, given that the former is still defined for quasi-circular orbits. The mode polarizations $h_+$ and $h_\times$ are conveniently given by the following Fourier series:
\begin{equation}
  h_+-\mathrm{i}h_\times=h(a)\sum_{k=-3}^{+3}c_k(z,\zeta)\,\mathrm{e}^{\mathrm i kL}\text{,}
  \label{eq:h+hx}
\end{equation}
where $c_k$, the Fourier coefficients, are given by
\begin{subequations}\label{eq:FourierCoef}
\begin{align}
  c_{+3}&=9\bar z\left(1-\zeta\bar\zeta\right){}^2\text{,}\\
  c_{+2}&=4\left(1-\zeta\bar\zeta\right){}^2\text{,}\\
  c_{+1}&=-3z\left(1-\zeta\bar\zeta\right){}^2-2\bar z\,\zeta^2\left(1-\zeta\bar\zeta\right)\text{,}\\
  c_{-1}&=-3\bar z\,\zeta^4-2z\,\zeta^2\left(1-\zeta\bar\zeta\right)\text{,}\\
  c_{-2}&=4\zeta^4\text{,}\\
  c_{-3}&=9z\,\zeta^4\text{.}
\end{align}
\end{subequations}
with $c_0=0$. Let us remind that the complex variables $z$ and $\zeta$ are defined in Eqs. \eqref{eq:defregparam}. The GW strain amplitude, $h$, is function of the semi-major axis and is given by
\begin{equation}
  h=\eta\,\left(\frac{a}{D}\right)\left(\frac{Gm}{c^2a}\right)^{2}\text{.}
\end{equation}

According to the method of variation of arbitrary constants, Eq. \eqref{eq:h+hx} is also valid beyond Kepler motion. Hence, in order to compute the combined effects of GR (up to the 2.5PN order) and magnetism on the mode polarizations, we just have to insert the first-order solutions \eqref{eq:sola}, \eqref{eq:solorb}, and \eqref{eq:solL} into the right-hand side of Eq.~\eqref{eq:h+hx}. This shows that, as the semi-major axis decreases because of the energy loss due to the gravitational radiation, the GW strain amplitude increases as $1/a(t)$, giving rise to the so-called ``chirp''. On the other hand, the magnetic dipole-dipole interaction secularly affects the mean longitude and the longitude of the pericenter. Because the latter only appears in Eq. \eqref{eq:h+hx} at first-order in eccentricity, it can be neglected for quasi-circular orbits. Hence, only the secular drift on the mean longitude need to be kept for circular orbit. Therefore, we anticipate that magnetism slightly changes the frequency of the mode polarizations with respect to the frequency that would be expected for two point-masses in circular orbit (without GR corrections).

\subsection{Time evolution}

The effect of the magnetic interaction on the mode polarizations can be shown in the time domain. For this, we compute the relative error made when evaluating Eqs.~\eqref{eq:h+hx} without the dipole-dipole interaction. The relative error on the mode polarization $h_+$ reads as
\begin{equation}
  \mathrm{err}(h_+)=\frac{|(h_+)_{\mathrm{GR}+\mathrm{M}}-(h_+)_{\mathrm{GR}}|}{(h_+)_{\mathrm{GR}+\mathrm{M}}}\text{,}
\end{equation}
where $(h_+)_{\mathrm{GR}+\mathrm{M}}$ is the ``$+$'' polarization computed with both GR and magnetism, and $(h_+)_{\mathrm{GR}}$ contains the gravitational contribution only. There exists a similar relationship for the ``$\times$'' polarization.

\begin{figure}
  \begin{center}
    \includegraphics[scale=0.95]{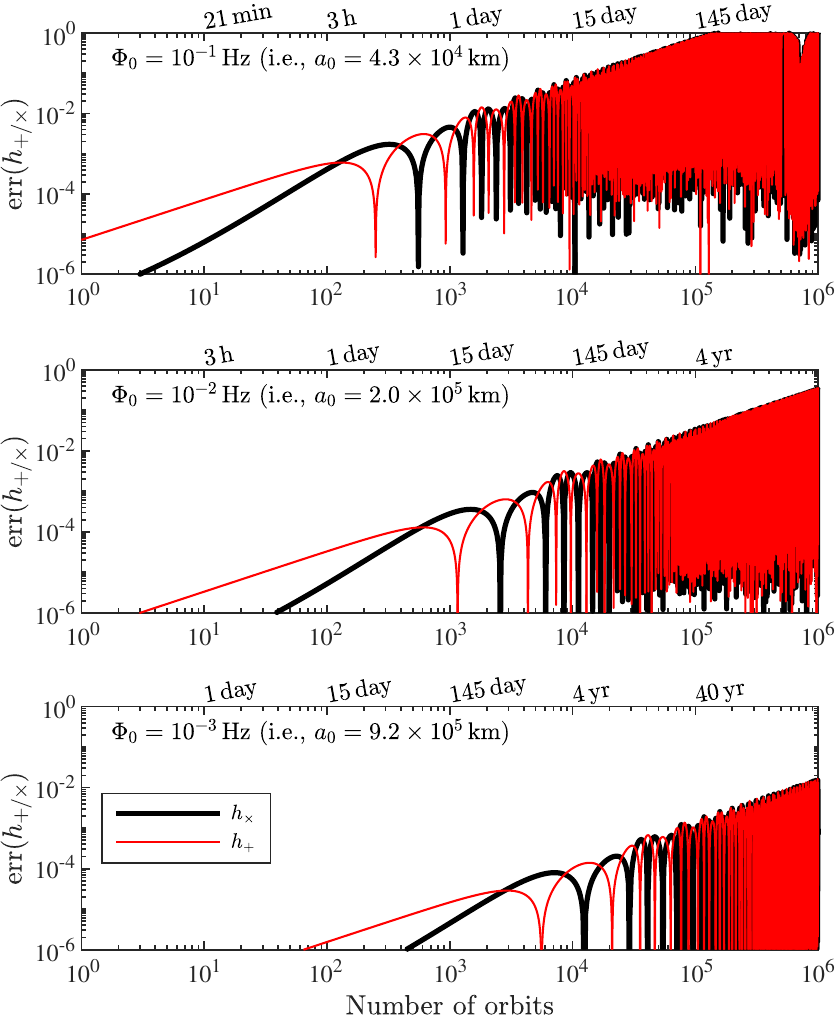}
  \end{center}
  \caption{Relative errors caused by the fact of neglecting the dipole-dipole interaction in the computation of the mode polarizations $h_+$ (thin red curve) and $h_\times$ (thick black curve). The evolutions are represented for different values of the LISA main frequency for GBs, namely $\Phi_0=10^{-1}\,\mathrm{Hz}$ (top panel), $10^{-2}\,\mathrm{Hz}$ (middle panel), and $10^{-3}\,\mathrm{Hz}$ (bottom panel). The bottom $x$-axis is the number of orbits and the top $x$-axis represents the elapsed time.}
  \label{fig:modeCPdipdip}
\end{figure}

The evolutions of $\mathrm{err}(h_+)$ and $\mathrm{err}(h_\times)$ are depicted in Fig. \ref{fig:modeCPdipdip}. It can be seen that, in the low frequency band of LISA (i.e., from $10^{-3}\,\mathrm{Hz}$ and below), which corresponds to a semi-major axis of the order of $9.2\times 10^5\,\mathrm{km}$, neglecting the magnetic dipole-dipole interaction for a binary system made of highly magnetic WDs generates a relative error of the order of $1\%$ after $400\,\mathrm{yr}$. This effect is probably not detectable in the time-span of the LISA mission. For the intermediate frequency band (i.e., around $10^{-2}\,\mathrm{Hz}$), which corresponds to a semi-major axis of the order of $2.0\times 10^5\,\mathrm{km}$, the relative error reaches $1\%$ after $4\,\mathrm{yr}$ and could potentially be observed for highly magnetic binaries. From $10^{-2}\,\mathrm{Hz}$ to $10^{-1}\,\mathrm{Hz}$, which corresponds to a semi-major axis of the order of $4.3\times 10^4\,\mathrm{km}$, the effect of the magnetic interaction on the GW mode polarizations becomes significant in a time much shorter than mission duration. Indeed, the relative errors reaches $100\%$ in only 145 days for a binary in close orbit and composed of two highly magnetic WDs.

This behavior is retrieved from the scaling laws derived in Sect. \ref{sec:sol}. Indeed, we have shown in Eqs. \eqref{eq:dotomeTauM} that the dipole-dipole interaction induces a precession motion of the mean longitude and the longitude of the pericenter. The rate of precession is proportional to $\nu_0$, namely $\propto\mu_1\mu_2a_0{}^{-7/2}$.

In order to determine whether or not LISA could detect magnetic effects, it is more appropriate to decompose the GW strain signal in terms of its fundamental harmonics, to be able to identify the mesurable parameters. This is the topic of the next section.

\subsection{Frequency shift}

In order to track the effect of magnetism on the GW mode polarizations, let us substitute the first order solutions \eqref{eq:sola}, \eqref{eq:solorb}, and \eqref{eq:solL} into the right-hand side of Eq.~\eqref{eq:h+hx}. At the zeroth-order in eccentricity, we find
\begin{subequations}\label{eq:h+hxsimp}
\begin{align}
  h_+&=2h_0(1+\cos^2\iota_0)\cos(\phi+\Phi t_*+\dot\Phi t_*^2)\text{,}\\
  h_\times&=-4h_0\cos\iota_0\sin(\phi+\Phi t_*+\dot\Phi t_*^2)\text{,}
\end{align}
\end{subequations}
where $h_0$ is the amplitude of the GW signal at $t_*=0$, $\phi$ is the initial phase of the signal, $\Phi$ is the main frequency, and $\dot\Phi$ is the frequency shift.

According to the results presented in Sect. \ref{sec:sol}, these quantities are directly linked to the dynamics of the binary system, that is to say
\begin{subequations}\label{eq:PhiPhi0}
\begin{align}
  \Phi&=2n_0\left(1+\frac{\dot\lambda_{\mathrm{GR}}}{n_0}+\frac{\dot\lambda_{\mathrm{M}}}{n_0}\right)\text{,}\label{eq:Phi}\\
  \dot\Phi&=-\frac{3n_0}{4a_0}\,\dot{a}_{\mathrm{GR}}\text{,}
\end{align}
\end{subequations}
with $\phi=2L_0$. We have used Eq. \eqref{eq:solL} for $L(t_*)$, omitting the periodic oscillations and considering the secular variations only. For clarity and without loss of generality, we have used the LDC conventions by imposing $\Omega_0=0$.

It can be seen from Eqs. \eqref{eq:Phi} that when GR and magnetic effects are negligible, the main frequency $\Phi$ reduces to $\Phi_0$, namely $2n_0$, as it might be expected for a circular orbit \cite{LDCGroup}. In the frequency domain, the circular case thus corresponds to a main peak at frequency $2n_0$. However, if the orbit is elliptic, we expect the GW signal to have a discrete frequency decomposition. Indeed, as one might infer from Eqs.~\eqref{eq:h+hx} and \eqref{eq:FourierCoef}, the secondary peaks are expected with harmonic frequencies at $n_0$ and $3n_0$, at linear order in eccentricity.

Now, if the contributions from GR and magnetism are too important to be neglected, we see from Eq. \eqref{eq:Phi} that the main frequency is shifted by the amount $2\dot\lambda_{\mathrm{GR}}$ and $2\dot\lambda_{\mathrm{M}}$ due to GR and magnetism, respectively. Therefore, we expect that magnetic effects should be accounted for while interpreting the main frequency that is measured by LISA, only if $\sigma_\Phi$, the uncertainty in the main frequency, satisfies a relation as follows
\begin{equation}
  \frac{\sigma_\Phi}{\Phi}<\frac{\dot\lambda_{\mathrm{M}}}{n_0}\text{,}
\end{equation}
where $\dot\lambda_{\mathrm{M}}$ is given by Eqs. \eqref{eq:dotomeTauM}. A numerical estimate yields
\begin{align}
  \frac{\sigma_\Phi}{\Phi}&<6.8\times 10^{-7}\left(\frac{\Phi_0}{10^{-1}\,\mathrm{Hz}}\right)^{4/3}\nonumber\\
  &\times\left(\frac{1.2\,\mathrm{M}_\odot}{m_1}\right)
\left(\frac{0.3\,\mathrm{M}_\odot}{m_2}\right)\left(\frac{B_1}{10^9\,\mathrm{G}}\right)\left(\frac{B_2}{10^9\,\mathrm{G}}\right)\nonumber\\
  &\times\left(\frac{R_1}{6\times10^3\,\mathrm{km}}\right)^3\left(\frac{R_2}{15\times10^3\,\mathrm{km}}\right)^3\text{,}\label{eq:sigmamag}
\end{align}
where the values of $e_0$, $\epsilon_{10}$, and $\epsilon_{20}$ are taken from Tab.~\ref{tab:initcond}. This relation can be used as a threshold to determine which sources of gravitational waves might necessitate to carefully account for the magnetic dipole-dipole effect while attempting to interpret the physical content behind the measured frequency. As a matter of fact, most of the current verification binaries\footnote{See a list of the verification binaries with measurement of their frequency evolution (i.e., the SNR and the uncertainty in the main frequency) here: \url{https://apc.u-paris.fr/~lejeune/lisa/fom-8/report_SO1a_detectability_vgb_6_yr/}.} are actually known with relative uncertainties ranging from $10^{-6}$ to $10^{-9}$ and we expect LISA to be able to determine the main frequencies of GBs with a better accuracy \cite{LDCGroup}.

\section{Conclusion}
\label{sec:ccl}

Observations have shown that WDs or NSs can develop large scale magnetic fields at the level of $10^9\,\mathrm{G}$ and $10^{15}\,\mathrm{G}$, respectively. In addition, there should exist, in the galaxy, hundreds of millions of WD-WD binary systems and millions of NS-WD binaries. In this context, we have aimed at quantifying the effect of the magnetic interaction on the generation of GWs by compact GBs. In this work, we modeled a well-separated binary system composed of WDs or NSs considering both the orbital and rotational motion of the degenerate stars. We used the magnetostatic approximation, in accordance with the fossil-field hypothesis. This enabled us to assume that the magnetic fields of both stars in the binary system are dominated by their dipole moments. In addition, we supposed, for simplicity, that the direction of the magnetic moments are aligned with the spin axis direction of the stars. We employed a post-Newtonian description of the point-mass interaction up to terms proportional to $c^{-5}$. Within this framework, we derived the secular equations governing the orbital and rotational motion of the binary system. We showed that the rotational motion can be mainly decoupled from the orbital motion. We provided first analytical estimates that we validated by comparison to results of a numerical integration of the equations of motion for the orbit and rotation. Then, we solved for the orbital motion and showed that the longitude of the pericenter and the mean longitude are the only orbital elements being secularly impacted by the dipole-dipole interaction. The rate of precessions are given by (cf. Eqs.~\eqref{eq:dotomeTauM})
\begin{subequations}
\begin{align}
  \dot\varpi_{\mathrm{M}}&=\frac{3\mu_0}{4\pi\sqrt{G}}\frac{\sqrt{m_1+m_2}}{m_1m_2}\frac{\mu_1\mu_2}{a_0^{7/2}}\frac{\cos\epsilon_{10}\cos\epsilon_{20}}{(1-{e_0}^2)^2}\text{,}\label{eq:secvarpi}\\
  \dot\lambda_{\mathrm{M}}&=\dot\varpi_{\mathrm{M}}\left(1+\sqrt{1-{e_0}^2}\right)\text{.}
\end{align}
\end{subequations}
This shows that a system of double WD in a closed orbit is more likely to feel the effect of the magnetic interaction since it is proportional to $\mu_1\mu_2$ and evolves as the inverse of the semi-major axis raised to a power $7/2$. The inclination and the longitude of the node are varying periodically with an amplitude which remains negligible. We evaluated the relative error that is generated when computing the GW mode polarizations without taking into account the secular drift of the longitude of the pericenter due to the dipole-dipole interaction. We showed that neglecting magnetism can generate a relative error of the order of $1\%$ after $4\,\mathrm{yr}$, and $100\%$ after $145\,\mathrm{days}$ for typical frequencies at $10^{-2}\,\mathrm{Hz}$ and $10^{-1}\,\mathrm{Hz}$, respectively. Finally, we demonstrated that, at leading order in eccentricity, the magnetic effect shifts the frequency $\Phi_0$ (with $\Phi_0=2n_0$) by the amount $2\dot\varpi_{\mathrm{M}}(1+\sqrt{1-{e_0}^2})$. Hence, if one wants to interpret the circular frequency measured by LISA in terms of its physical contents, one has to worry about magnetism if the main frequency is determined with sufficient accuracy (cf. Eq.~\eqref{eq:sigmamag}).

Because LISA will directly determines $\Phi$ and not $\Phi_0$, we can conclude that magnetism is totally degenerated with the determination of the main frequency at zeroth-order in eccentricity (cf. Eqs.~\eqref{eq:h+hxsimp}). In other words, LISA observations alone cannot disentangle between the contribution of magnetism and the total mass within the determination of the frequency for circular orbit. Combining LISA with EM observations (e.g., using spectropolorimetric observations) could help to determine the masses and the amplitude of the magnetic moments unambiguously. The degeneracy can be broken in the case where the binary system is in non-circular orbit. Indeed, the eccentricity gives rise to an additional sinusoidal signal with the phase $3L-\varpi$ (cf. Eqs.~\eqref{eq:h+hx} and \eqref{eq:FourierCoef}). We thus expect magnetism to shift the expected frequency of the new signal (i.e., $\Phi_0'=3\Phi_0/2=3n_0$) by the amount $\dot\varpi_{\mathrm{M}}(2+3\sqrt{1-{e_0}^2})$. Therefore, by combining linearly $\Phi$, the measured  main frequency, and $\Phi'$, the measured frequency of the first harmonic, the product $\mu_1\mu_2$ can, in principle, be directly inferred. As a matter of fact, the following linear combination: $3\Phi/2-\Phi'$, allows to directly determine $\dot\varpi_{\mathrm{M}}$ (cf. Eq.~\eqref{eq:secvarpi}) from the measured values of $\Phi$ and $\Phi'$. This point will be further investigated in a future work by making use of the LDC tools \cite{LDCGroup} for circular GBs, that we will adapt to the case of magnetic GBs in eccentric orbits.

Other planned future work is two-fold. Firstly, we will further improve the magnetic field's ``static'' picture that is presented in this paper. Indeed, by focusing on the inspiral phase of GBs, we implicitly assumed that the internal physics is decoupled from the orbital dynamics, and so we have considered that the direction, the structure, and the magnitude of the magnetic fields were frozen and independent of time. Within this ``static'' picture, several improvements can be made. Given that some stars present non-axisymmetric magnetic field configurations, even when they bear strong magnetic fields \cite{2006MNRAS.370..629D,2007A&A...463..647B,2015SSRv..191..111F}, one possibility is to investigate the effect of higher multipole structures on the GW strain. Another interesting perspective is to include the effect of a misalignment between the magnetic moments and the direction of the spins for future applications to the dynamics of pulsar stars.

The second step will be to explore the ``dynamical'' picture, where internal physics is treated simultaneously with the orbital dynamics. The idea is to build a coherent model for the dynamics and GW strain of GBs, accounting for dissipation through magnetohydrodynamic processes. Within the ``dynamical'' picture, we first plan to investigate the effect of energy dissipation through the unipolar induction mechanism \cite{1969ApJ...156...59G,2012ApJ...755...80P,2012ApJ...757L...3L,2018ApJ...868...19W}, whose EM energy dissipation may potentially compete with the loss of energy caused by gravitational radiation. In addition, given that WDs and NSs can develop strong magnetic fields \cite{2005MNRAS.356..615F}, and can efficiently dissipate energy through internal gravity waves excited by tides \cite{2012MNRAS.421..426F}, we plan to investigate the impact of magnetism in the modeling of internal magneto-gravito-inertial waves \cite{2011A&A...526A..65M,2012A&A...540A..37M}. Then, the backreaction on the orbital dynamics and on the GW signal will be investigated.


\section*{Acknowledgments}

A.B. is grateful to the Centre National d'\'Etudes Spatiales (CNES) for financial support. Authors are thankful to Q. Baghi, S. Brun, T. Foglizzo, A. Petiteau, E. Savalle, and A. Strugarek for interesting discussions. A.B. is grateful to  A. Hees and M. Lilley for useful comments about a preliminary version of this work.

\appendix

\section{Post-Newtonian motion}
\label{sec:PNcorr}

\citet{2014LRR....17....2B} gives the 3PN equations of motion for a binary system (see also \cite{2021PhRvD.104j4023T}). We consider the expansion up to the 2.5PN approximation. It involves two coefficients $\mathcal A'$ and $\mathcal B'$ (without primes in Blanchet's paper) multiplying $\hat{\mathbf{n}}$ and $\mathbf v$, respectively. By making use of the method of variation of arbitrary constants, we can always enforce the solution \eqref{eq:solKep} for the velocity $\mathbf v$, so that the GR contribution can be written as in Eq. \eqref{eq:compaccpertGR} where the dimensionless coefficients $\mathcal A$ and $\mathcal B$ are given by
\begin{subequations}
\begin{align}
    \mathcal A&=\mathcal A_{\mathrm{1PN}}+\mathcal A_{\mathrm{2PN}}+\mathcal A_{\mathrm{2.5PN}}\text{,}\\
    \mathcal B&=\mathcal B_{\mathrm{1PN}}+\mathcal B_{\mathrm{2PN}}+\mathcal B_{\mathrm{2.5PN}}\text{,}   
\end{align}
\end{subequations}
with
\begin{subequations}
  \begin{align}
    \mathcal A_{\mathrm{1PN}}&=\frac{1}{c^2}\bigg\{\bigg(\frac{\eta}{2}-4\bigg)v_n^2\nonumber\\
    &+(1+3\eta)v^2-\frac{Gm}{r}(4+2\eta)\bigg\}\text{,}\\
    \mathcal A_{\mathrm{2PN}}&=\frac{1}{c^4}\bigg\{\bigg(\frac{51}{8}-\frac{21\eta}{8}\bigg)\eta v_n^4-(12-4\eta)\eta v_n^2v^2\nonumber\\
    &+(3-4\eta)\eta v^4-\frac{Gm}{r}\bigg[\bigg(\frac{9}{2}-2\eta\bigg)\eta v_n^2\nonumber\\
    &+\bigg(\frac{13}{2}-2\eta\bigg)\eta v^2\bigg]+\frac{G^2m^2}{r^2}\bigg(9+\frac{87\eta}{4}\bigg)\bigg\}\text{,}\\
    \mathcal A_{\mathrm{2.5PN}}&=-\frac{1}{c^5}\bigg\{\frac{16\eta v_nv^2}{5}\frac{Gm}{r}+\frac{64\eta v_n}{15}\frac{G^2m^2}{r^2}\bigg\}\text{,}
  \end{align}
  and
  \begin{align}  
    \mathcal B_{\mathrm{1PN}}&=-\frac{1}{c^2}\bigg\{(4-2\eta)v_nv_u\bigg\}\text{,}\\
    \mathcal B_{\mathrm{2PN}}&=\frac{1}{c^4}\bigg\{\bigg(\frac{9}{2}+3\eta\bigg)\eta v_n^3v_u-\bigg(\frac{15}{2}+2\eta\bigg)\eta v_nv_uv^2\nonumber\\
    &+\frac{Gm}{r}\bigg(2+\frac{41\eta}{2}+4\eta^2\bigg)v_nv_u\bigg\}\text{,}\\
    \mathcal B_{\mathrm{2.5PN}}&=\frac{1}{c^5}\bigg\{\frac{8\eta v_uv^2}{5}\frac{Gm}{r}+\frac{24\eta v_u}{5}\frac{G^2m^2}{r^2}\bigg\}\text{.}
  \end{align}
\end{subequations}

Similarly, the coefficients for finding the individual positions from the relative position can be determined by substituting for $\mathbf x$ and $\mathbf v$ from Eq. \eqref{eq:solKep} into Eq. (216) of \citet{2014LRR....17....2B}. After some algebra, we find Eqs. \eqref{eq:posabs}, where the dimensionless coefficients $\mathcal P$ and $\mathcal Q$ are given by
\begin{subequations}
  \begin{align}
    \mathcal P&=\mathcal P_{\mathrm{1PN}}+\mathcal P_{\mathrm{2PN}}\text{,}\\
    \mathcal Q&=\mathcal Q_{\mathrm{2PN}}+\mathcal Q_{\mathrm{2.5PN}}\text{,}   
  \end{align}
\end{subequations}
with
\begin{subequations}
  \begin{align}
    \mathcal P_{\mathrm{1PN}}&=\frac{1}{c^2}\bigg\{\frac{v^2}{2}-\frac{Gm}{2r}\bigg\}\text{,}\\
    \mathcal P_{\mathrm{2PN}}&=\frac{1}{c^4}\bigg\{\bigg(\frac{3}{8}-\frac{3\eta}{2}\bigg)v^4-\frac{Gm}{r}\bigg[\bigg(\frac{15}{8}-\frac{3\eta}{4}\bigg)v_n^2\nonumber\\
                             &-\bigg(\frac{19}{8}+\frac{3\eta}{2}\bigg)v^2\bigg]+\frac{G^2m^2}{r^2}\bigg(\frac{7}{4}-\frac{\eta}{2}\bigg)\bigg\}\text{,}
  \end{align}
  and
  \begin{align}                      
    \mathcal Q_{\mathrm{2PN}}&=-\frac{1}{c^4}\bigg\{\frac{7v_nv_u}{4}\frac{Gm}{r}\bigg\}\text{,}\\
    \mathcal Q_{\mathrm{2.5PN}}&=\frac{1}{c^5}\bigg\{\frac{4v_uv^2}{5}\frac{Gm}{r}-\frac{8v_u}{5}\frac{G^2m^2}{r^2}\bigg\}\text{.}
  \end{align}
\end{subequations}\\

\section{Amplitudes of oscillations}
\label{sec:Amposc}

Three of the non-singular orbital parameters present periodic variations following the magnetic dipole-dipole perturbation, namely $z$, the complex eccentricity vector, $\zeta$, the complex longitude of the node vector, and $\lambda$, the mean longitude.

Oscillating signatures occurring on $z$ are actually caused by oscillations of $\varpi$, the longitude of the pericenter. After integrating the secular equations of motion with respect to time, we find that the magnetic periodic variations, $\widetilde\zeta_{\mathrm{M}}(t)$, $\widetilde\lambda_{\mathrm{M}}(t)$, and $\widetilde\varpi_{\mathrm{M}}(t)$, are given by
\begin{widetext}
\begin{subequations}\label{eq:solorbM}
\begin{align}
  \widetilde\zeta_{\mathrm{M}}(t)&=\frac{1}{2e_0\sqrt{1-\zeta_0\bar\zeta_0}}\sum_{k=1}^2\mathrm{i}\Theta_k\left\{\left(1-\frac{\zeta_0\bar\zeta_0}{2}\right)z_0\mathrm{e}^{\mathrm i[(\dot\varpi_{\mathrm{GR}}+\dot\varpi_{\mathrm{M}}+\dot\beta_k)t+\beta_{k0}]}+\frac{\zeta_0^2}{2}\,\bar z_0\mathrm{e}^{-\mathrm i[(\dot\varpi_{\mathrm{GR}}+\dot\varpi_{\mathrm{M}}+\dot\beta_k)t+\beta_{k0}]}\right\}\\[8pt]
  \widetilde\lambda_{\mathrm{M}}(t)&=\widetilde\varpi_{\mathrm{M}}(t)-\Psi_{(-)}\sqrt{1-e^2}\sin\big[(\dot\beta_1-\dot\beta_2)t+\beta_{10}-\beta_{20}\big]\nonumber\\
  &+4\Psi_{(+)}(1-e^2)\frac{\big[1-\sqrt{1-e^2}-e^2\big(1-\frac{1}{2}\sqrt{1-e^2}\big)\big]}{e^4}\sin\big[(\dot\beta_1+\dot\beta_2)t+\beta_{10}+\beta_{20}\big]\text{,}\\
  \widetilde\varpi_{\mathrm{M}}(t)&=-\Psi_{(-)}\sin\big[(\dot\beta_1-\dot\beta_2)t+\beta_{10}-\beta_{20}\big]+\sum_{k=1}^2\Theta_k\tan\left(\frac{\iota_0}{2}\right)\cos\big[(\dot{\varpi}_{\mathrm{GR}}+\dot{\varpi}_{\mathrm{M}}+\dot\beta_k)t+\beta_{k0}+\omega_0\big]\text{.}
\end{align}
\end{subequations}
\end{widetext}
In these expressions, we introduce the amplitudes $\Theta_1$, $\Theta_2$, and $\Psi_{(\pm)}$ which are defined by
\begin{subequations}
\begin{align}
  \Theta_1&=\left(\frac{\nu_{0}}{\dot{\varpi}_{\mathrm{GR}}+\dot{\varpi}_{\mathrm{M}}+\dot\beta_1}\right)\sin\epsilon_{10}\cos\epsilon_{20}\text{,}\label{eq:ampTheta1}\\
  \Theta_2&=\left(\frac{\nu_{0}}{\dot{\varpi}_{\mathrm{GR}}+\dot{\varpi}_{\mathrm{M}}+\dot\beta_2}\right)\cos\epsilon_{10}\sin\epsilon_{20}\text{,}\label{eq:ampTheta2}\\
  \Psi_{(\pm)}&=\left(\frac{\nu_{0}}{\dot\beta_1\pm\dot\beta_2}\right)\sin\epsilon_{10}\sin\epsilon_{20}\text{.}
\end{align}
\end{subequations}

\begin{figure}
  \begin{center}
    \includegraphics[scale=0.98]{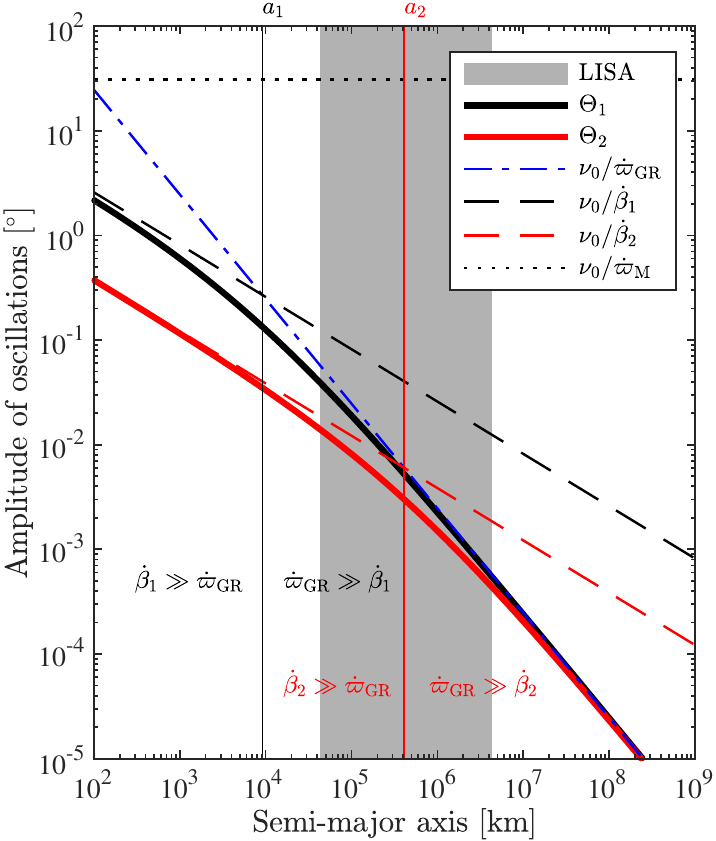}
  \end{center}
  \caption{Evolution of the amplitude $\Theta_1$ (thick black line) and $\Theta_2$ (thick red line) with the semi-major axis.}
  \label{fig:Amplitude}
\end{figure}

The amplitudes $\Theta_1$ and $\Theta_2$ increase when the semi-major axis decreases (cf. Fig.~\ref{fig:Amplitude}). For the set of numerical values that we selected, and for the LISA frequency band (i.e., between $\Phi_0=10^{-4}\,\mathrm{Hz}$ and $10^{-1}\,\mathrm{Hz}$), we can approximate $\Theta_1$ by $\nu_{0}/\dot\omega_{\mathrm{GR}}$, namely
\begin{equation}
  \Theta_1\simeq\frac{\mu_0c^2}{8\pi G^2}\frac{\mu_1\mu_2}{\eta m^3}\frac{\sin\epsilon_{10}\cos\epsilon_{20}}{a_0(1-{e_0}^2)}\text{.}\label{eq:Theta1simp}
\end{equation}

For the high frequency part of the LISA band (i.e., $\Phi_0=10^{-1}\,\mathrm{Hz}$), we can approximate $\Theta_2$ by $\nu_{0}/\dot\beta_2$, that is to say
\begin{equation}
  \Theta_2\simeq\frac{3}{2\sqrt{G}}\frac{\sqrt{m}}{m_1m_2}\frac{S_2\sin\epsilon_{20}}{\sqrt{a_0(1-{e_0}^2)}}\text{.}
\end{equation}
However, as seen from Fig. \ref{fig:Amplitude}, the transition between the regimes $\dot\omega_{\mathrm{GR}}\gg\dot\beta_2$ and $\dot\omega_{\mathrm{GR}}\ll\dot\beta_2$ occurs precisely within the LISA frequency band meaning that the general expression \eqref{eq:ampTheta2} must be favored between $\Phi_0=10^{-2}\,\mathrm{Hz}$ to $10^{-3}\,\mathrm{Hz}$. From $\Phi_0=10^{-3}\,\mathrm{Hz}$ to $10^{-4}\,\mathrm{Hz}$, $\Theta_2$ can be approximated by $\nu_{0}/\dot\omega_{\mathrm{GR}}$ (cf. Eq. \eqref{eq:Theta1simp}).

For the amplitude $\Theta_1$, the transition between the regimes where $\dot\omega_{\mathrm{GR}}\gg\dot\beta_1$ and $\dot\omega_{\mathrm{GR}}\ll\dot\beta_1$ occurs for $a=a_1$ with $a_1$ being the abscissa of the intersection between the curves $\nu_{0}/\dot\beta_1$ and $\nu_{0}/\dot\omega_{\mathrm{GR}}$, as depicted in Fig. \ref{fig:Amplitude}. The expression of $a_1$ is given by 
\begin{equation}
  a_1=\frac{\mu_0^2c^4}{144\pi^2G^3}\frac{\mu_1^2\mu_2^2}{S_1^2m^3}\frac{\cos^2\epsilon_{20}}{(1-{e_0}^2)}\text{.}
\end{equation}
Similarly, for the amplitude $\Theta_2$, the transition between the regimes where $\dot\omega_{\mathrm{GR}}\gg\dot\beta_2$ and $\dot\omega_{\mathrm{GR}}\ll\dot\beta_2$ occurs for $a=a_2$ where the expression for $a_2$ is found by interchanging subscripts ``1'' and ``2'' in the expression of $a_1$. Both $a_1$ and $a_2$ are depicted as vertical lines in Fig. \ref{fig:Amplitude} (black line for $a_1$ and red line for $a_2$). The relative separation between $a_1$ and $a_2$ is thus mainly dependent of the ratio $(S_2\cos\epsilon_{20})/(S_1\cos\epsilon_{10})$. 

Higher amplitudes of oscillation can be reached in regions $\dot\omega_{\mathrm{GR}}\gg\dot\beta_1$ or $\dot\omega_{\mathrm{GR}}\gg\dot\beta_2$ if the ratio $\nu_{0}/\dot\omega_{\mathrm{GR}}$ increases. This can be done, for instance, with a higher value of the product $\mu_1\mu_2$. The amplitudes $\Theta_1$ and $\Theta_2$ can also be higher in regions $\dot\beta_1\gg\dot\omega_{\mathrm{GR}}$ or $\dot\beta_2\gg\dot\omega_{\mathrm{GR}}$, respectively, if the ratios $\nu_{0}/\dot\beta_1$ and $\nu_{0}/\dot\beta_2$ increase. This can happen, for instance, when the magnitudes of the angular momentum $S_1$ and $S_2$ are higher than in Eqs.~\eqref{eq:S1} and \eqref{eq:S2}. However, the values that we selected for the magnetic moments and the magnitudes of the angular momentum, are already in the upper part of WDs population studies \cite{2015SSRv..191..111F}.

\begin{table*}[h]
  \caption[]{List of notations used in this paper.}
  \label{tab:notations}
  \begin{tabularx}{\linewidth}{@{} l X @{}}
    \hline\hline
    \multicolumn{2}{c}{Physical constants}\\
    \hline
    $\mu_0$ & Permeability of vacuum\\
    $c$ & Speed of light in vacuum\\
    $G$ & Gravitational constant\\
    \hline
    \multicolumn{2}{c}{Physical parameters}\\
    \hline
    $m_{1/2}$ & Mass of the primary/secondary\\
    $m$ & Total mass of the binary\\
    $\Delta$ & Relative mass difference\\
    $\eta$ & Symmetric mass ratio\\
    $\mu_{1/2}$ & Magnitude of the magnetic moment of the primary/secondary\\
    $B_{1/2}$ & Magnitude of the magnetic field of the primary/secondary\\
    $S_{1/2}$ & Magnitude of the spin of rotation of the primary/secondary\\ 
    $R_{1/2}$ & Equatorial radius of the primary/secondary\\
    $P_{1/2}$ & Period of proper rotation of the primary/secondary\\
    \hline
    \multicolumn{2}{c}{Unit-vectors}\\
    \hline
    $(\hat{\mathbf{e}}_X,\hat{\mathbf{e}}_Y,\hat{\mathbf{e}}_Z)$ & Vectorial basis for the source frame\\
    $(\hat{\mathbf{e}}_x,\hat{\mathbf{e}}_y,\hat{\mathbf{e}}_z)$ & Vectorial basis for the orbit frame\\
    $(\hat{\mathbf{n}},\hat{\mathbf{u}},\hat{\mathbf{e}}_z)$ & Vectorial basis for the corotating frame\\
    $\hat{\mathbf{N}}$ & Unit-vector for the direction of the observer\\
    $\hat{\mathbf{s}}_{1/2}$ & Unit-vector for the direction of the magnetic moment of the primary/secondary\\
    \hline
    \multicolumn{2}{c}{Keplerian solution and orientation of the magnetic moments}\\
    \hline
    $p$ & Semi-latus rectum\\
    $a$ & Semi-major axis\\
    $e$ & Eccentricity\\
    $\iota$ & Inclination\\
    $\Omega$ & Longitude of the ascending node\\
    $\omega$ & Argument of the pericenter\\
    $\tau$ & Time of pericenter passage\\
    $z$ & Imaginary eccentricity vector\\
    $\zeta$ & Imaginary longitude of the ascending node vector\\
    $f$ & True anomaly\\
    $M$ & Mean anomaly\\
    $\varpi$ & Longitude of the pericenter\\
    $L$ & Mean longitude\\
    $n$ & Mean motion\\
    $P$ & Orbital period\\
    $r$ & Relative separation between the stars\\
    $v$ & Magnitude of the relative velocity\\
    $v_n$ & Component of the relative velocity along $\hat{\mathbf{n}}$\\
    $v_u$ & Component of the relative velocity along $\hat{\mathbf{u}}$\\
    $\epsilon_{1/2}$ & Obliquity of the magnetic moment of the primary/secondary\\
    $\beta_{1/2}$ & Precession angle of the magnetic moment of the primary/secondary\\ 
    \hline
    \multicolumn{2}{c}{Perturbations}\\
    \hline
    $\mathcal{N}$ & Component of the perturbation along $\hat{\mathbf{n}}$\\
    $\mathcal{U}$ & Component of the perturbation along $\hat{\mathbf{u}}$\\
    $\mathcal{S}$ & Component of the perturbation along $\hat{\mathbf{e}}_z$\\
    \hline
    \multicolumn{2}{c}{Frequencies and rate of changes}\\
    \hline
    $\Phi$ & Mean frequency of the GW strain\\
    $\nu$ & Orbital frequency due to the magnetic dipole-dipole interaction\\
    $\nu_{1/2}$ & Rotational frequency of the primary/secondary due to the magnetic dipole-dipole interaction\\
    $\dot\beta_{1/2}$ & Rate of precession of the magnetic moments of the primary/secondary due to the magnetic dipole-dipole interaction\\
    $\dot\lambda_{\mathrm{M}}$ & Rate of precession of the mean longitude due to the magnetic dipole-dipole interaction\\
    $\dot\lambda_{\mathrm{GR}}$ & Rate of precession of the mean longitude due to general relativity at 1PN approximation\\
    $\dot a_{\mathrm{GR}}$ & Rate of change of the semi-major axis due to general relativity at 2.5PN approximation\\
    $\dot e_{\mathrm{GR}}$ & Rate of change of the eccentricity due to general relativity at 2.5PN approximation\\
    \hline
  \end{tabularx}
\end{table*}

\bibliographystyle{apsrev4-1}
\bibliography{magnetism_WD}

\begin{thebibliography}{70}%
\makeatletter
\providecommand \@ifxundefined [1]{%
 \@ifx{#1\undefined}
}%
\providecommand \@ifnum [1]{%
 \ifnum #1\expandafter \@firstoftwo
 \else \expandafter \@secondoftwo
 \fi
}%
\providecommand \@ifx [1]{%
 \ifx #1\expandafter \@firstoftwo
 \else \expandafter \@secondoftwo
 \fi
}%
\providecommand \natexlab [1]{#1}%
\providecommand \enquote  [1]{``#1''}%
\providecommand \bibnamefont  [1]{#1}%
\providecommand \bibfnamefont [1]{#1}%
\providecommand \citenamefont [1]{#1}%
\providecommand \href@noop [0]{\@secondoftwo}%
\providecommand \href [0]{\begingroup \@sanitize@url \@href}%
\providecommand \@href[1]{\@@startlink{#1}\@@href}%
\providecommand \@@href[1]{\endgroup#1\@@endlink}%
\providecommand \@sanitize@url [0]{\catcode `\\12\catcode `\$12\catcode
  `\&12\catcode `\#12\catcode `\^12\catcode `\_12\catcode `\%12\relax}%
\providecommand \@@startlink[1]{}%
\providecommand \@@endlink[0]{}%
\providecommand \url  [0]{\begingroup\@sanitize@url \@url }%
\providecommand \@url [1]{\endgroup\@href {#1}{\urlprefix }}%
\providecommand \urlprefix  [0]{URL }%
\providecommand \Eprint [0]{\href }%
\providecommand \doibase [0]{http://dx.doi.org/}%
\providecommand \selectlanguage [0]{\@gobble}%
\providecommand \bibinfo  [0]{\@secondoftwo}%
\providecommand \bibfield  [0]{\@secondoftwo}%
\providecommand \translation [1]{[#1]}%
\providecommand \BibitemOpen [0]{}%
\providecommand \bibitemStop [0]{}%
\providecommand \bibitemNoStop [0]{.\EOS\space}%
\providecommand \EOS [0]{\spacefactor3000\relax}%
\providecommand \BibitemShut  [1]{\csname bibitem#1\endcsname}%
\let\auto@bib@innerbib\@empty
\bibitem [{LIS(2020)}]{LISAcoll}%
  \BibitemOpen
  \href@noop {} {\enquote {\bibinfo {title} {{The LISA collaboration}},}\
  }\bibinfo {howpublished} {\url{https://www.lisamission.org/}} (\bibinfo
  {year} {2020})\BibitemShut {NoStop}%
\bibitem [{\citenamefont {{Amaro-Seoane}}\ \emph {et~al.}(2017)\citenamefont
  {{Amaro-Seoane}}, \citenamefont {{Audley}}, \citenamefont {{Babak}},
  \citenamefont {{Baker}}, \citenamefont {{Barausse}}, \citenamefont
  {{Bender}}, \citenamefont {{Berti}}, \citenamefont {{Binetruy}},
  \citenamefont {{Born}}, \citenamefont {{Bortoluzzi}},\ and\ \citenamefont
  {{et. al}}}]{2017arXiv170200786A}%
  \BibitemOpen
  \bibfield  {author} {\bibinfo {author} {\bibfnamefont {P.}~\bibnamefont
  {{Amaro-Seoane}}}, \bibinfo {author} {\bibfnamefont {H.}~\bibnamefont
  {{Audley}}}, \bibinfo {author} {\bibfnamefont {S.}~\bibnamefont {{Babak}}},
  \bibinfo {author} {\bibfnamefont {J.}~\bibnamefont {{Baker}}}, \bibinfo
  {author} {\bibfnamefont {E.}~\bibnamefont {{Barausse}}}, \bibinfo {author}
  {\bibfnamefont {P.}~\bibnamefont {{Bender}}}, \bibinfo {author}
  {\bibfnamefont {E.}~\bibnamefont {{Berti}}}, \bibinfo {author} {\bibfnamefont
  {P.}~\bibnamefont {{Binetruy}}}, \bibinfo {author} {\bibfnamefont
  {M.}~\bibnamefont {{Born}}}, \bibinfo {author} {\bibfnamefont
  {D.}~\bibnamefont {{Bortoluzzi}}}, \ and\ \bibinfo {author} {\bibnamefont
  {{et. al}}},\ }\href@noop {} {\bibfield  {journal} {\bibinfo  {journal}
  {arXiv e-prints}\ ,\ \bibinfo {eid} {arXiv:1702.00786}} (\bibinfo {year}
  {2017})},\ \Eprint {http://arxiv.org/abs/1702.00786} {arXiv:1702.00786
  [astro-ph.IM]} \BibitemShut {NoStop}%
\bibitem [{\citenamefont {{Hils}}\ \emph {et~al.}(1990)\citenamefont {{Hils}},
  \citenamefont {{Bender}},\ and\ \citenamefont
  {{Webbink}}}]{1990ApJ...360...75H}%
  \BibitemOpen
  \bibfield  {author} {\bibinfo {author} {\bibfnamefont {D.}~\bibnamefont
  {{Hils}}}, \bibinfo {author} {\bibfnamefont {P.~L.}\ \bibnamefont
  {{Bender}}}, \ and\ \bibinfo {author} {\bibfnamefont {R.~F.}\ \bibnamefont
  {{Webbink}}},\ }\href {\doibase 10.1086/169098} {\bibfield  {journal}
  {\bibinfo  {journal} {\apj}\ }\textbf {\bibinfo {volume} {360}},\ \bibinfo
  {pages} {75} (\bibinfo {year} {1990})}\BibitemShut {NoStop}%
\bibitem [{\citenamefont {Timpano}\ \emph {et~al.}(2006)\citenamefont
  {Timpano}, \citenamefont {Rubbo},\ and\ \citenamefont
  {Cornish}}]{PhysRevD.73.122001}%
  \BibitemOpen
  \bibfield  {author} {\bibinfo {author} {\bibfnamefont {S.~E.}\ \bibnamefont
  {Timpano}}, \bibinfo {author} {\bibfnamefont {L.~J.}\ \bibnamefont {Rubbo}},
  \ and\ \bibinfo {author} {\bibfnamefont {N.~J.}\ \bibnamefont {Cornish}},\
  }\href {\doibase 10.1103/PhysRevD.73.122001} {\bibfield  {journal} {\bibinfo
  {journal} {Phys. Rev. D}\ }\textbf {\bibinfo {volume} {73}},\ \bibinfo
  {pages} {122001} (\bibinfo {year} {2006})}\BibitemShut {NoStop}%
\bibitem [{LIG(2020)}]{LIGOcoll}%
  \BibitemOpen
  \href@noop {} {\enquote {\bibinfo {title} {{The LIGO Collaboration}},}\
  }\bibinfo {howpublished} {\url{https://www.ligo.caltech.edu/}} (\bibinfo
  {year} {2020})\BibitemShut {NoStop}%
\bibitem [{Vir(2020)}]{Virgocoll}%
  \BibitemOpen
  \href@noop {} {\enquote {\bibinfo {title} {{The Virgo Collaboration}},}\
  }\bibinfo {howpublished} {\url{https://www.virgo-gw.eu/}} (\bibinfo {year}
  {2020})\BibitemShut {NoStop}%
\bibitem [{KRA(2020)}]{KRAGAcoll}%
  \BibitemOpen
  \href@noop {} {\enquote {\bibinfo {title} {{The KAGRA Collaboration}},}\
  }\bibinfo {howpublished} {\url{https://gwcenter.icrr.u-tokyo.ac.jp/en/}}
  (\bibinfo {year} {2020})\BibitemShut {NoStop}%
\bibitem [{ETc(2020)}]{ETcoll}%
  \BibitemOpen
  \href@noop {} {\enquote {\bibinfo {title} {{The ET collaboration}},}\
  }\bibinfo {howpublished} {\url{http://www.et-gw.eu/}} (\bibinfo {year}
  {2020})\BibitemShut {NoStop}%
\bibitem [{\citenamefont {Abbott}\ \emph
  {et~al.}(2017{\natexlab{a}})\citenamefont {Abbott}, \citenamefont {Abbott},
  \citenamefont {Abbott}, \citenamefont {Acernese}, \citenamefont {Ackley},
  \citenamefont {Adams}, \citenamefont {Adams}, \citenamefont {Addesso},
  \citenamefont {Adhikari}, \citenamefont {Adya},\ and\ \citenamefont
  {et~al.}}]{PhysRevLett.119.161101}%
  \BibitemOpen
  \bibfield  {author} {\bibinfo {author} {\bibfnamefont {B.~P.}\ \bibnamefont
  {Abbott}}, \bibinfo {author} {\bibfnamefont {R.}~\bibnamefont {Abbott}},
  \bibinfo {author} {\bibfnamefont {T.~D.}\ \bibnamefont {Abbott}}, \bibinfo
  {author} {\bibfnamefont {F.}~\bibnamefont {Acernese}}, \bibinfo {author}
  {\bibfnamefont {K.}~\bibnamefont {Ackley}}, \bibinfo {author} {\bibfnamefont
  {C.}~\bibnamefont {Adams}}, \bibinfo {author} {\bibfnamefont
  {T.}~\bibnamefont {Adams}}, \bibinfo {author} {\bibfnamefont
  {P.}~\bibnamefont {Addesso}}, \bibinfo {author} {\bibfnamefont {R.~X.}\
  \bibnamefont {Adhikari}}, \bibinfo {author} {\bibfnamefont {V.~B.}\
  \bibnamefont {Adya}}, \ and\ \bibinfo {author} {\bibnamefont {et~al.}}
  (\bibinfo {collaboration} {LIGO Scientific Collaboration and Virgo
  Collaboration}),\ }\href {\doibase 10.1103/PhysRevLett.119.161101} {\bibfield
   {journal} {\bibinfo  {journal} {Phys. Rev. Lett.}\ }\textbf {\bibinfo
  {volume} {119}},\ \bibinfo {pages} {161101} (\bibinfo {year}
  {2017}{\natexlab{a}})}\BibitemShut {NoStop}%
\bibitem [{\citenamefont {Abbott}\ \emph
  {et~al.}(2017{\natexlab{b}})\citenamefont {Abbott}, \citenamefont {Abbott},
  \citenamefont {Abbott}, \citenamefont {Acernese}, \citenamefont {Ackley},
  \citenamefont {Adams}, \citenamefont {Adams}, \citenamefont {Addesso},
  \citenamefont {Adhikari}, \citenamefont {Adya},\ and\ \citenamefont
  {et~al.}}]{2017ApJ...848L..12A}%
  \BibitemOpen
  \bibfield  {author} {\bibinfo {author} {\bibfnamefont {B.~P.}\ \bibnamefont
  {Abbott}}, \bibinfo {author} {\bibfnamefont {R.}~\bibnamefont {Abbott}},
  \bibinfo {author} {\bibfnamefont {T.~D.}\ \bibnamefont {Abbott}}, \bibinfo
  {author} {\bibfnamefont {F.}~\bibnamefont {Acernese}}, \bibinfo {author}
  {\bibfnamefont {K.}~\bibnamefont {Ackley}}, \bibinfo {author} {\bibfnamefont
  {C.}~\bibnamefont {Adams}}, \bibinfo {author} {\bibfnamefont
  {T.}~\bibnamefont {Adams}}, \bibinfo {author} {\bibfnamefont
  {P.}~\bibnamefont {Addesso}}, \bibinfo {author} {\bibfnamefont {R.~X.}\
  \bibnamefont {Adhikari}}, \bibinfo {author} {\bibfnamefont {V.~B.}\
  \bibnamefont {Adya}}, \ and\ \bibinfo {author} {\bibnamefont {et~al.}},\
  }\href {\doibase 10.3847/2041-8213/aa91c9} {\bibfield  {journal} {\bibinfo
  {journal} {\apjl}\ }\textbf {\bibinfo {volume} {848}},\ \bibinfo {eid} {L12}
  (\bibinfo {year} {2017}{\natexlab{b}})}\BibitemShut {NoStop}%
\bibitem [{\citenamefont {{Abbott}}\ \emph {et~al.}(2018)\citenamefont
  {{Abbott}}, \citenamefont {{Abbott}}, \citenamefont {{Abbott}}, \citenamefont
  {{Acernese}}, \citenamefont {{Ackley}}, \citenamefont {{Adams}},
  \citenamefont {{Adams}}, \citenamefont {{Addesso}}, \citenamefont
  {{Adhikari}},\ and\ \citenamefont {et~al.}}]{2018PhRvL.121p1101A}%
  \BibitemOpen
  \bibfield  {author} {\bibinfo {author} {\bibfnamefont {B.~P.}\ \bibnamefont
  {{Abbott}}}, \bibinfo {author} {\bibfnamefont {R.}~\bibnamefont {{Abbott}}},
  \bibinfo {author} {\bibfnamefont {T.~D.}\ \bibnamefont {{Abbott}}}, \bibinfo
  {author} {\bibfnamefont {F.}~\bibnamefont {{Acernese}}}, \bibinfo {author}
  {\bibfnamefont {K.}~\bibnamefont {{Ackley}}}, \bibinfo {author}
  {\bibfnamefont {C.}~\bibnamefont {{Adams}}}, \bibinfo {author} {\bibfnamefont
  {T.}~\bibnamefont {{Adams}}}, \bibinfo {author} {\bibfnamefont
  {P.}~\bibnamefont {{Addesso}}}, \bibinfo {author} {\bibfnamefont {R.~X.}\
  \bibnamefont {{Adhikari}}}, \ and\ \bibinfo {author} {\bibnamefont
  {et~al.}},\ }\href {\doibase 10.1103/PhysRevLett.121.161101} {\bibfield
  {journal} {\bibinfo  {journal} {\prl}\ }\textbf {\bibinfo {volume} {121}},\
  \bibinfo {eid} {161101} (\bibinfo {year} {2018})}\BibitemShut {NoStop}%
\bibitem [{\citenamefont {{Abbott}}\ \emph {et~al.}(2019)\citenamefont
  {{Abbott}}, \citenamefont {{Abbott}}, \citenamefont {{Abbott}}, \citenamefont
  {{Acernese}}, \citenamefont {{Ackley}}, \citenamefont {{Adams}},
  \citenamefont {{Adams}}, \citenamefont {{Addesso}}, \citenamefont
  {{Adhikari}},\ and\ \citenamefont {et~al.}}]{2019PhRvX...9a1001A}%
  \BibitemOpen
  \bibfield  {author} {\bibinfo {author} {\bibfnamefont {B.~P.}\ \bibnamefont
  {{Abbott}}}, \bibinfo {author} {\bibfnamefont {R.}~\bibnamefont {{Abbott}}},
  \bibinfo {author} {\bibfnamefont {T.~D.}\ \bibnamefont {{Abbott}}}, \bibinfo
  {author} {\bibfnamefont {F.}~\bibnamefont {{Acernese}}}, \bibinfo {author}
  {\bibfnamefont {K.}~\bibnamefont {{Ackley}}}, \bibinfo {author}
  {\bibfnamefont {C.}~\bibnamefont {{Adams}}}, \bibinfo {author} {\bibfnamefont
  {T.}~\bibnamefont {{Adams}}}, \bibinfo {author} {\bibfnamefont
  {P.}~\bibnamefont {{Addesso}}}, \bibinfo {author} {\bibfnamefont {R.~X.}\
  \bibnamefont {{Adhikari}}}, \ and\ \bibinfo {author} {\bibnamefont
  {et~al.}},\ }\href {\doibase 10.1103/PhysRevX.9.011001} {\bibfield  {journal}
  {\bibinfo  {journal} {Physical Review X}\ }\textbf {\bibinfo {volume} {9}},\
  \bibinfo {eid} {011001} (\bibinfo {year} {2019})}\BibitemShut {NoStop}%
\bibitem [{\citenamefont {Abbott}\ \emph
  {et~al.}(2017{\natexlab{c}})\citenamefont {Abbott}, \citenamefont {Abbott},
  \citenamefont {Abbott}, \citenamefont {Acernese}, \citenamefont {Ackley},
  \citenamefont {Adams}, \citenamefont {Adams}, \citenamefont {Addesso},
  \citenamefont {Adhikari}, \citenamefont {Adya},\ and\ \citenamefont
  {et~al.}}]{Abbott_2017}%
  \BibitemOpen
  \bibfield  {author} {\bibinfo {author} {\bibfnamefont {B.~P.}\ \bibnamefont
  {Abbott}}, \bibinfo {author} {\bibfnamefont {R.}~\bibnamefont {Abbott}},
  \bibinfo {author} {\bibfnamefont {T.~D.}\ \bibnamefont {Abbott}}, \bibinfo
  {author} {\bibfnamefont {F.}~\bibnamefont {Acernese}}, \bibinfo {author}
  {\bibfnamefont {K.}~\bibnamefont {Ackley}}, \bibinfo {author} {\bibfnamefont
  {C.}~\bibnamefont {Adams}}, \bibinfo {author} {\bibfnamefont
  {T.}~\bibnamefont {Adams}}, \bibinfo {author} {\bibfnamefont
  {P.}~\bibnamefont {Addesso}}, \bibinfo {author} {\bibfnamefont {R.~X.}\
  \bibnamefont {Adhikari}}, \bibinfo {author} {\bibfnamefont {V.~B.}\
  \bibnamefont {Adya}}, \ and\ \bibinfo {author} {\bibnamefont {et~al.}},\
  }\href {\doibase 10.3847/2041-8213/aa920c} {\bibfield  {journal} {\bibinfo
  {journal} {The Astrophysical Journal}\ }\textbf {\bibinfo {volume} {848}},\
  \bibinfo {pages} {L13} (\bibinfo {year} {2017}{\natexlab{c}})}\BibitemShut
  {NoStop}%
\bibitem [{\citenamefont {{Nelemans}}(2009)}]{2009CQGra..26i4030N}%
  \BibitemOpen
  \bibfield  {author} {\bibinfo {author} {\bibfnamefont {G.}~\bibnamefont
  {{Nelemans}}},\ }\href {\doibase 10.1088/0264-9381/26/9/094030} {\bibfield
  {journal} {\bibinfo  {journal} {Classical and Quantum Gravity}\ }\textbf
  {\bibinfo {volume} {26}},\ \bibinfo {eid} {094030} (\bibinfo {year}
  {2009})}\BibitemShut {NoStop}%
\bibitem [{\citenamefont {{Ferrario}}\ and\ \citenamefont
  {{Wickramasinghe}}(2005)}]{2005MNRAS.356..615F}%
  \BibitemOpen
  \bibfield  {author} {\bibinfo {author} {\bibfnamefont {L.}~\bibnamefont
  {{Ferrario}}}\ and\ \bibinfo {author} {\bibfnamefont {D.~T.}\ \bibnamefont
  {{Wickramasinghe}}},\ }\href {\doibase 10.1111/j.1365-2966.2004.08474.x}
  {\bibfield  {journal} {\bibinfo  {journal} {\mnras}\ }\textbf {\bibinfo
  {volume} {356}},\ \bibinfo {pages} {615} (\bibinfo {year}
  {2005})}\BibitemShut {NoStop}%
\bibitem [{\citenamefont {{Kawka}}\ \emph {et~al.}(2007)\citenamefont
  {{Kawka}}, \citenamefont {{Vennes}}, \citenamefont {{Schmidt}}, \citenamefont
  {{Wickramasinghe}},\ and\ \citenamefont {{Koch}}}]{2007ApJ...654..499K}%
  \BibitemOpen
  \bibfield  {author} {\bibinfo {author} {\bibfnamefont {A.}~\bibnamefont
  {{Kawka}}}, \bibinfo {author} {\bibfnamefont {S.}~\bibnamefont {{Vennes}}},
  \bibinfo {author} {\bibfnamefont {G.~D.}\ \bibnamefont {{Schmidt}}}, \bibinfo
  {author} {\bibfnamefont {D.~T.}\ \bibnamefont {{Wickramasinghe}}}, \ and\
  \bibinfo {author} {\bibfnamefont {R.}~\bibnamefont {{Koch}}},\ }\href
  {\doibase 10.1086/509072} {\bibfield  {journal} {\bibinfo  {journal} {\apj}\
  }\textbf {\bibinfo {volume} {654}},\ \bibinfo {pages} {499} (\bibinfo {year}
  {2007})}\BibitemShut {NoStop}%
\bibitem [{\citenamefont {{Ferrario}}\ \emph {et~al.}(2015)\citenamefont
  {{Ferrario}}, \citenamefont {{de Martino}},\ and\ \citenamefont
  {{G{\"a}nsicke}}}]{2015SSRv..191..111F}%
  \BibitemOpen
  \bibfield  {author} {\bibinfo {author} {\bibfnamefont {L.}~\bibnamefont
  {{Ferrario}}}, \bibinfo {author} {\bibfnamefont {D.}~\bibnamefont {{de
  Martino}}}, \ and\ \bibinfo {author} {\bibfnamefont {B.~T.}\ \bibnamefont
  {{G{\"a}nsicke}}},\ }\href {\doibase 10.1007/s11214-015-0152-0} {\bibfield
  {journal} {\bibinfo  {journal} {\ssr}\ }\textbf {\bibinfo {volume} {191}},\
  \bibinfo {pages} {111} (\bibinfo {year} {2015})}\BibitemShut {NoStop}%
\bibitem [{\citenamefont {{Ferrario}}\ \emph {et~al.}(2020)\citenamefont
  {{Ferrario}}, \citenamefont {{Wickramasinghe}},\ and\ \citenamefont
  {{Kawka}}}]{2020AdSpR..66.1025F}%
  \BibitemOpen
  \bibfield  {author} {\bibinfo {author} {\bibfnamefont {L.}~\bibnamefont
  {{Ferrario}}}, \bibinfo {author} {\bibfnamefont {D.}~\bibnamefont
  {{Wickramasinghe}}}, \ and\ \bibinfo {author} {\bibfnamefont
  {A.}~\bibnamefont {{Kawka}}},\ }\href {\doibase 10.1016/j.asr.2019.11.012}
  {\bibfield  {journal} {\bibinfo  {journal} {Advances in Space Research}\
  }\textbf {\bibinfo {volume} {66}},\ \bibinfo {pages} {1025} (\bibinfo {year}
  {2020})}\BibitemShut {NoStop}%
\bibitem [{\citenamefont {{Tout}}\ \emph {et~al.}(2008)\citenamefont {{Tout}},
  \citenamefont {{Wickramasinghe}}, \citenamefont {{Liebert}}, \citenamefont
  {{Ferrario}},\ and\ \citenamefont {{Pringle}}}]{2008MNRAS.387..897T}%
  \BibitemOpen
  \bibfield  {author} {\bibinfo {author} {\bibfnamefont {C.~A.}\ \bibnamefont
  {{Tout}}}, \bibinfo {author} {\bibfnamefont {D.~T.}\ \bibnamefont
  {{Wickramasinghe}}}, \bibinfo {author} {\bibfnamefont {J.}~\bibnamefont
  {{Liebert}}}, \bibinfo {author} {\bibfnamefont {L.}~\bibnamefont
  {{Ferrario}}}, \ and\ \bibinfo {author} {\bibfnamefont {J.~E.}\ \bibnamefont
  {{Pringle}}},\ }\href {\doibase 10.1111/j.1365-2966.2008.13291.x} {\bibfield
  {journal} {\bibinfo  {journal} {\mnras}\ }\textbf {\bibinfo {volume} {387}},\
  \bibinfo {pages} {897} (\bibinfo {year} {2008})}\BibitemShut {NoStop}%
\bibitem [{\citenamefont {{Bagnulo}}\ and\ \citenamefont
  {{Landstreet}}(2021)}]{2021MNRAS.507.5902B}%
  \BibitemOpen
  \bibfield  {author} {\bibinfo {author} {\bibfnamefont {S.}~\bibnamefont
  {{Bagnulo}}}\ and\ \bibinfo {author} {\bibfnamefont {J.~D.}\ \bibnamefont
  {{Landstreet}}},\ }\href {\doibase 10.1093/mnras/stab2046} {\bibfield
  {journal} {\bibinfo  {journal} {\mnras}\ }\textbf {\bibinfo {volume} {507}},\
  \bibinfo {pages} {5902} (\bibinfo {year} {2021})}\BibitemShut {NoStop}%
\bibitem [{\citenamefont {{Ferrario}}\ \emph {et~al.}(1997)\citenamefont
  {{Ferrario}}, \citenamefont {{Vennes}}, \citenamefont {{Wickramasinghe}},
  \citenamefont {{Bailey}},\ and\ \citenamefont
  {{Christian}}}]{1997MNRAS.292..205F}%
  \BibitemOpen
  \bibfield  {author} {\bibinfo {author} {\bibfnamefont {L.}~\bibnamefont
  {{Ferrario}}}, \bibinfo {author} {\bibfnamefont {S.}~\bibnamefont
  {{Vennes}}}, \bibinfo {author} {\bibfnamefont {D.~T.}\ \bibnamefont
  {{Wickramasinghe}}}, \bibinfo {author} {\bibfnamefont {J.~A.}\ \bibnamefont
  {{Bailey}}}, \ and\ \bibinfo {author} {\bibfnamefont {D.~J.}\ \bibnamefont
  {{Christian}}},\ }\href {\doibase 10.1093/mnras/292.2.205} {\bibfield
  {journal} {\bibinfo  {journal} {\mnras}\ }\textbf {\bibinfo {volume} {292}},\
  \bibinfo {pages} {205} (\bibinfo {year} {1997})}\BibitemShut {NoStop}%
\bibitem [{\citenamefont {{Landstreet}}\ and\ \citenamefont
  {{Bagnulo}}(2020)}]{2020A&A...634L..10L}%
  \BibitemOpen
  \bibfield  {author} {\bibinfo {author} {\bibfnamefont {J.~D.}\ \bibnamefont
  {{Landstreet}}}\ and\ \bibinfo {author} {\bibfnamefont {S.}~\bibnamefont
  {{Bagnulo}}},\ }\href {\doibase 10.1051/0004-6361/201937301} {\bibfield
  {journal} {\bibinfo  {journal} {\aap}\ }\textbf {\bibinfo {volume} {634}},\
  \bibinfo {eid} {L10} (\bibinfo {year} {2020})}\BibitemShut {NoStop}%
\bibitem [{\citenamefont {{Duncan}}\ and\ \citenamefont
  {{Thompson}}(1992)}]{1992ApJ...392L...9D}%
  \BibitemOpen
  \bibfield  {author} {\bibinfo {author} {\bibfnamefont {R.~C.}\ \bibnamefont
  {{Duncan}}}\ and\ \bibinfo {author} {\bibfnamefont {C.}~\bibnamefont
  {{Thompson}}},\ }\href {\doibase 10.1086/186413} {\bibfield  {journal}
  {\bibinfo  {journal} {\apjl}\ }\textbf {\bibinfo {volume} {392}},\ \bibinfo
  {pages} {L9} (\bibinfo {year} {1992})}\BibitemShut {NoStop}%
\bibitem [{\citenamefont {{Charbonneau}}\ and\ \citenamefont
  {{MacGregor}}(2001)}]{2001ApJ...559.1094C}%
  \BibitemOpen
  \bibfield  {author} {\bibinfo {author} {\bibfnamefont {P.}~\bibnamefont
  {{Charbonneau}}}\ and\ \bibinfo {author} {\bibfnamefont {K.~B.}\ \bibnamefont
  {{MacGregor}}},\ }\href {\doibase 10.1086/322417} {\bibfield  {journal}
  {\bibinfo  {journal} {\apj}\ }\textbf {\bibinfo {volume} {559}},\ \bibinfo
  {pages} {1094} (\bibinfo {year} {2001})}\BibitemShut {NoStop}%
\bibitem [{\citenamefont {{Raynaud}}\ \emph {et~al.}(2020)\citenamefont
  {{Raynaud}}, \citenamefont {{Guilet}}, \citenamefont {{Janka}},\ and\
  \citenamefont {{Gastine}}}]{2020SciA....6.2732R}%
  \BibitemOpen
  \bibfield  {author} {\bibinfo {author} {\bibfnamefont {R.}~\bibnamefont
  {{Raynaud}}}, \bibinfo {author} {\bibfnamefont {J.}~\bibnamefont {{Guilet}}},
  \bibinfo {author} {\bibfnamefont {H.-T.}\ \bibnamefont {{Janka}}}, \ and\
  \bibinfo {author} {\bibfnamefont {T.}~\bibnamefont {{Gastine}}},\ }\href
  {\doibase 10.1126/sciadv.aay2732} {\bibfield  {journal} {\bibinfo  {journal}
  {Science Advances}\ }\textbf {\bibinfo {volume} {6}},\ \bibinfo {pages}
  {eaay2732} (\bibinfo {year} {2020})},\ \Eprint
  {http://arxiv.org/abs/2003.06662} {arXiv:2003.06662 [astro-ph.HE]}
  \BibitemShut {NoStop}%
\bibitem [{\citenamefont {{Reboul-Salze}}\ \emph {et~al.}(2021)\citenamefont
  {{Reboul-Salze}}, \citenamefont {{Guilet}}, \citenamefont {{Raynaud}},\ and\
  \citenamefont {{Bugli}}}]{2021arXiv211102148R}%
  \BibitemOpen
  \bibfield  {author} {\bibinfo {author} {\bibfnamefont {A.}~\bibnamefont
  {{Reboul-Salze}}}, \bibinfo {author} {\bibfnamefont {J.}~\bibnamefont
  {{Guilet}}}, \bibinfo {author} {\bibfnamefont {R.}~\bibnamefont {{Raynaud}}},
  \ and\ \bibinfo {author} {\bibfnamefont {M.}~\bibnamefont {{Bugli}}},\
  }\href@noop {} {\bibfield  {journal} {\bibinfo  {journal} {arXiv e-prints}\
  ,\ \bibinfo {eid} {arXiv:2111.02148}} (\bibinfo {year} {2021})},\ \Eprint
  {http://arxiv.org/abs/2111.02148} {arXiv:2111.02148 [astro-ph.HE]}
  \BibitemShut {NoStop}%
\bibitem [{\citenamefont {{Cowling}}(1945)}]{1945MNRAS.105..166C}%
  \BibitemOpen
  \bibfield  {author} {\bibinfo {author} {\bibfnamefont {T.~G.}\ \bibnamefont
  {{Cowling}}},\ }\href {\doibase 10.1093/mnras/105.3.166} {\bibfield
  {journal} {\bibinfo  {journal} {\mnras}\ }\textbf {\bibinfo {volume} {105}},\
  \bibinfo {pages} {166} (\bibinfo {year} {1945})}\BibitemShut {NoStop}%
\bibitem [{\citenamefont {{Moss}}(1987)}]{1987MNRAS.226..297M}%
  \BibitemOpen
  \bibfield  {author} {\bibinfo {author} {\bibfnamefont {D.}~\bibnamefont
  {{Moss}}},\ }\href {\doibase 10.1093/mnras/226.2.297} {\bibfield  {journal}
  {\bibinfo  {journal} {\mnras}\ }\textbf {\bibinfo {volume} {226}},\ \bibinfo
  {pages} {297} (\bibinfo {year} {1987})}\BibitemShut {NoStop}%
\bibitem [{\citenamefont {{Wickramasinghe}}\ and\ \citenamefont
  {{Ferrario}}(2005)}]{2005MNRAS.356.1576W}%
  \BibitemOpen
  \bibfield  {author} {\bibinfo {author} {\bibfnamefont {D.~T.}\ \bibnamefont
  {{Wickramasinghe}}}\ and\ \bibinfo {author} {\bibfnamefont {L.}~\bibnamefont
  {{Ferrario}}},\ }\href {\doibase 10.1111/j.1365-2966.2004.08603.x} {\bibfield
   {journal} {\bibinfo  {journal} {\mnras}\ }\textbf {\bibinfo {volume}
  {356}},\ \bibinfo {pages} {1576} (\bibinfo {year} {2005})}\BibitemShut
  {NoStop}%
\bibitem [{\citenamefont {{Hu}}\ and\ \citenamefont
  {{Lou}}(2009)}]{2009MNRAS.396..878H}%
  \BibitemOpen
  \bibfield  {author} {\bibinfo {author} {\bibfnamefont {R.-Y.}\ \bibnamefont
  {{Hu}}}\ and\ \bibinfo {author} {\bibfnamefont {Y.-Q.}\ \bibnamefont
  {{Lou}}},\ }\href {\doibase 10.1111/j.1365-2966.2009.14648.x} {\bibfield
  {journal} {\bibinfo  {journal} {\mnras}\ }\textbf {\bibinfo {volume} {396}},\
  \bibinfo {pages} {878} (\bibinfo {year} {2009})}\BibitemShut {NoStop}%
\bibitem [{\citenamefont {{Braithwaite}}\ and\ \citenamefont
  {{Spruit}}(2004)}]{2004Natur.431..819B}%
  \BibitemOpen
  \bibfield  {author} {\bibinfo {author} {\bibfnamefont {J.}~\bibnamefont
  {{Braithwaite}}}\ and\ \bibinfo {author} {\bibfnamefont {H.~C.}\ \bibnamefont
  {{Spruit}}},\ }\href {\doibase 10.1038/nature02934} {\bibfield  {journal}
  {\bibinfo  {journal} {\nat}\ }\textbf {\bibinfo {volume} {431}},\ \bibinfo
  {pages} {819} (\bibinfo {year} {2004})}\BibitemShut {NoStop}%
\bibitem [{\citenamefont {{Duez}}\ and\ \citenamefont
  {{Mathis}}(2010)}]{2010A&A...517A..58D}%
  \BibitemOpen
  \bibfield  {author} {\bibinfo {author} {\bibfnamefont {V.}~\bibnamefont
  {{Duez}}}\ and\ \bibinfo {author} {\bibfnamefont {S.}~\bibnamefont
  {{Mathis}}},\ }\href {\doibase 10.1051/0004-6361/200913496} {\bibfield
  {journal} {\bibinfo  {journal} {\aap}\ }\textbf {\bibinfo {volume} {517}},\
  \bibinfo {eid} {A58} (\bibinfo {year} {2010})}\BibitemShut {NoStop}%
\bibitem [{\citenamefont {{Braithwaite}}(2008)}]{2008MNRAS.386.1947B}%
  \BibitemOpen
  \bibfield  {author} {\bibinfo {author} {\bibfnamefont {J.}~\bibnamefont
  {{Braithwaite}}},\ }\href {\doibase 10.1111/j.1365-2966.2008.13218.x}
  {\bibfield  {journal} {\bibinfo  {journal} {\mnras}\ }\textbf {\bibinfo
  {volume} {386}},\ \bibinfo {pages} {1947} (\bibinfo {year}
  {2008})}\BibitemShut {NoStop}%
\bibitem [{\citenamefont {{Donati}}\ \emph {et~al.}(2006)\citenamefont
  {{Donati}}, \citenamefont {{Howarth}}, \citenamefont {{Jardine}},
  \citenamefont {{Petit}}, \citenamefont {{Catala}}, \citenamefont
  {{Landstreet}}, \citenamefont {{Bouret}}, \citenamefont {{Alecian}},
  \citenamefont {{Barnes}}, \citenamefont {{Forveille}}, \citenamefont
  {{Paletou}},\ and\ \citenamefont {{Manset}}}]{2006MNRAS.370..629D}%
  \BibitemOpen
  \bibfield  {author} {\bibinfo {author} {\bibfnamefont {J.~F.}\ \bibnamefont
  {{Donati}}}, \bibinfo {author} {\bibfnamefont {I.~D.}\ \bibnamefont
  {{Howarth}}}, \bibinfo {author} {\bibfnamefont {M.~M.}\ \bibnamefont
  {{Jardine}}}, \bibinfo {author} {\bibfnamefont {P.}~\bibnamefont {{Petit}}},
  \bibinfo {author} {\bibfnamefont {C.}~\bibnamefont {{Catala}}}, \bibinfo
  {author} {\bibfnamefont {J.~D.}\ \bibnamefont {{Landstreet}}}, \bibinfo
  {author} {\bibfnamefont {J.~C.}\ \bibnamefont {{Bouret}}}, \bibinfo {author}
  {\bibfnamefont {E.}~\bibnamefont {{Alecian}}}, \bibinfo {author}
  {\bibfnamefont {J.~R.}\ \bibnamefont {{Barnes}}}, \bibinfo {author}
  {\bibfnamefont {T.}~\bibnamefont {{Forveille}}}, \bibinfo {author}
  {\bibfnamefont {F.}~\bibnamefont {{Paletou}}}, \ and\ \bibinfo {author}
  {\bibfnamefont {N.}~\bibnamefont {{Manset}}},\ }\href {\doibase
  10.1111/j.1365-2966.2006.10558.x} {\bibfield  {journal} {\bibinfo  {journal}
  {\mnras}\ }\textbf {\bibinfo {volume} {370}},\ \bibinfo {pages} {629}
  (\bibinfo {year} {2006})}\BibitemShut {NoStop}%
\bibitem [{\citenamefont {{Beuermann}}\ \emph {et~al.}(2007)\citenamefont
  {{Beuermann}}, \citenamefont {{Euchner}}, \citenamefont {{Reinsch}},
  \citenamefont {{Jordan}},\ and\ \citenamefont
  {{G{\"a}nsicke}}}]{2007A&A...463..647B}%
  \BibitemOpen
  \bibfield  {author} {\bibinfo {author} {\bibfnamefont {K.}~\bibnamefont
  {{Beuermann}}}, \bibinfo {author} {\bibfnamefont {F.}~\bibnamefont
  {{Euchner}}}, \bibinfo {author} {\bibfnamefont {K.}~\bibnamefont
  {{Reinsch}}}, \bibinfo {author} {\bibfnamefont {S.}~\bibnamefont {{Jordan}}},
  \ and\ \bibinfo {author} {\bibfnamefont {B.~T.}\ \bibnamefont
  {{G{\"a}nsicke}}},\ }\href {\doibase 10.1051/0004-6361:20066332} {\bibfield
  {journal} {\bibinfo  {journal} {\aap}\ }\textbf {\bibinfo {volume} {463}},\
  \bibinfo {pages} {647} (\bibinfo {year} {2007})}\BibitemShut {NoStop}%
\bibitem [{\citenamefont {Cornish}\ and\ \citenamefont
  {Littenberg}(2007)}]{PhysRevD.76.083006}%
  \BibitemOpen
  \bibfield  {author} {\bibinfo {author} {\bibfnamefont {N.~J.}\ \bibnamefont
  {Cornish}}\ and\ \bibinfo {author} {\bibfnamefont {T.~B.}\ \bibnamefont
  {Littenberg}},\ }\href {\doibase 10.1103/PhysRevD.76.083006} {\bibfield
  {journal} {\bibinfo  {journal} {Phys. Rev. D}\ }\textbf {\bibinfo {volume}
  {76}},\ \bibinfo {pages} {083006} (\bibinfo {year} {2007})}\BibitemShut
  {NoStop}%
\bibitem [{\citenamefont {{Bbak}}\ \emph {et~al.}(2020)\citenamefont {{Bbak}},
  \citenamefont {{Le Jeune}}, \citenamefont {{Petiteau}},\ and\ \citenamefont
  {{Vallisneri}}}]{LDCGroup}%
  \BibitemOpen
  \bibfield  {author} {\bibinfo {author} {\bibfnamefont {S.}~\bibnamefont
  {{Bbak}}}, \bibinfo {author} {\bibfnamefont {M.}~\bibnamefont {{Le Jeune}}},
  \bibinfo {author} {\bibfnamefont {A.}~\bibnamefont {{Petiteau}}}, \ and\
  \bibinfo {author} {\bibfnamefont {M.}~\bibnamefont {{Vallisneri}}},\ }\href
  {https://lisa-ldc.lal.in2p3.fr/static/data/pdf/LDC-manual-Sangria.pdf}
  {\bibfield  {journal} {\bibinfo  {journal} {LISA-LCST-SGS-MAN-001, Rev. 1}\ }
  (\bibinfo {year} {2020})}\BibitemShut {NoStop}%
\bibitem [{\citenamefont {{Lai}}(1997)}]{1997ApJ...490..847L}%
  \BibitemOpen
  \bibfield  {author} {\bibinfo {author} {\bibfnamefont {D.}~\bibnamefont
  {{Lai}}},\ }\href {\doibase 10.1086/304899} {\bibfield  {journal} {\bibinfo
  {journal} {\apj}\ }\textbf {\bibinfo {volume} {490}},\ \bibinfo {pages} {847}
  (\bibinfo {year} {1997})}\BibitemShut {NoStop}%
\bibitem [{\citenamefont {{Willems}}\ \emph {et~al.}(2010)\citenamefont
  {{Willems}}, \citenamefont {{Deloye}},\ and\ \citenamefont
  {{Kalogera}}}]{2010ApJ...713..239W}%
  \BibitemOpen
  \bibfield  {author} {\bibinfo {author} {\bibfnamefont {B.}~\bibnamefont
  {{Willems}}}, \bibinfo {author} {\bibfnamefont {C.~J.}\ \bibnamefont
  {{Deloye}}}, \ and\ \bibinfo {author} {\bibfnamefont {V.}~\bibnamefont
  {{Kalogera}}},\ }\href {\doibase 10.1088/0004-637X/713/1/239} {\bibfield
  {journal} {\bibinfo  {journal} {\apj}\ }\textbf {\bibinfo {volume} {713}},\
  \bibinfo {pages} {239} (\bibinfo {year} {2010})}\BibitemShut {NoStop}%
\bibitem [{\citenamefont {{Fuller}}\ and\ \citenamefont
  {{Lai}}(2011)}]{2011MNRAS.412.1331F}%
  \BibitemOpen
  \bibfield  {author} {\bibinfo {author} {\bibfnamefont {J.}~\bibnamefont
  {{Fuller}}}\ and\ \bibinfo {author} {\bibfnamefont {D.}~\bibnamefont
  {{Lai}}},\ }\href {\doibase 10.1111/j.1365-2966.2010.18017.x} {\bibfield
  {journal} {\bibinfo  {journal} {\mnras}\ }\textbf {\bibinfo {volume} {412}},\
  \bibinfo {pages} {1331} (\bibinfo {year} {2011})}\BibitemShut {NoStop}%
\bibitem [{\citenamefont {{Fuller}}\ and\ \citenamefont
  {{Lai}}(2012)}]{2012MNRAS.421..426F}%
  \BibitemOpen
  \bibfield  {author} {\bibinfo {author} {\bibfnamefont {J.}~\bibnamefont
  {{Fuller}}}\ and\ \bibinfo {author} {\bibfnamefont {D.}~\bibnamefont
  {{Lai}}},\ }\href {\doibase 10.1111/j.1365-2966.2011.20320.x} {\bibfield
  {journal} {\bibinfo  {journal} {\mnras}\ }\textbf {\bibinfo {volume} {421}},\
  \bibinfo {pages} {426} (\bibinfo {year} {2012})}\BibitemShut {NoStop}%
\bibitem [{\citenamefont {{Fuller}}\ and\ \citenamefont
  {{Lai}}(2013)}]{2013MNRAS.430..274F}%
  \BibitemOpen
  \bibfield  {author} {\bibinfo {author} {\bibfnamefont {J.}~\bibnamefont
  {{Fuller}}}\ and\ \bibinfo {author} {\bibfnamefont {D.}~\bibnamefont
  {{Lai}}},\ }\href {\doibase 10.1093/mnras/sts606} {\bibfield  {journal}
  {\bibinfo  {journal} {\mnras}\ }\textbf {\bibinfo {volume} {430}},\ \bibinfo
  {pages} {274} (\bibinfo {year} {2013})}\BibitemShut {NoStop}%
\bibitem [{\citenamefont {{Fuller}}\ and\ \citenamefont
  {{Lai}}(2014)}]{2014MNRAS.444.3488F}%
  \BibitemOpen
  \bibfield  {author} {\bibinfo {author} {\bibfnamefont {J.}~\bibnamefont
  {{Fuller}}}\ and\ \bibinfo {author} {\bibfnamefont {D.}~\bibnamefont
  {{Lai}}},\ }\href {\doibase 10.1093/mnras/stu1698} {\bibfield  {journal}
  {\bibinfo  {journal} {\mnras}\ }\textbf {\bibinfo {volume} {444}},\ \bibinfo
  {pages} {3488} (\bibinfo {year} {2014})}\BibitemShut {NoStop}%
\bibitem [{\citenamefont {{McNeill}}\ \emph {et~al.}(2020)\citenamefont
  {{McNeill}}, \citenamefont {{Mardling}},\ and\ \citenamefont
  {{M{\"u}ller}}}]{2020MNRAS.491.3000M}%
  \BibitemOpen
  \bibfield  {author} {\bibinfo {author} {\bibfnamefont {L.~O.}\ \bibnamefont
  {{McNeill}}}, \bibinfo {author} {\bibfnamefont {R.~A.}\ \bibnamefont
  {{Mardling}}}, \ and\ \bibinfo {author} {\bibfnamefont {B.}~\bibnamefont
  {{M{\"u}ller}}},\ }\href {\doibase 10.1093/mnras/stz3215} {\bibfield
  {journal} {\bibinfo  {journal} {\mnras}\ }\textbf {\bibinfo {volume} {491}},\
  \bibinfo {pages} {3000} (\bibinfo {year} {2020})}\BibitemShut {NoStop}%
\bibitem [{\citenamefont {{Xu}}\ and\ \citenamefont
  {{Lai}}(2017)}]{2017PhRvD..96h3005X}%
  \BibitemOpen
  \bibfield  {author} {\bibinfo {author} {\bibfnamefont {W.}~\bibnamefont
  {{Xu}}}\ and\ \bibinfo {author} {\bibfnamefont {D.}~\bibnamefont {{Lai}}},\
  }\href {\doibase 10.1103/PhysRevD.96.083005} {\bibfield  {journal} {\bibinfo
  {journal} {\prd}\ }\textbf {\bibinfo {volume} {96}},\ \bibinfo {eid} {083005}
  (\bibinfo {year} {2017})}\BibitemShut {NoStop}%
\bibitem [{\citenamefont {{Vick}}\ and\ \citenamefont
  {{Lai}}(2019)}]{2019PhRvD.100f3001V}%
  \BibitemOpen
  \bibfield  {author} {\bibinfo {author} {\bibfnamefont {M.}~\bibnamefont
  {{Vick}}}\ and\ \bibinfo {author} {\bibfnamefont {D.}~\bibnamefont {{Lai}}},\
  }\href {\doibase 10.1103/PhysRevD.100.063001} {\bibfield  {journal} {\bibinfo
   {journal} {\prd}\ }\textbf {\bibinfo {volume} {100}},\ \bibinfo {eid}
  {063001} (\bibinfo {year} {2019})}\BibitemShut {NoStop}%
\bibitem [{\citenamefont {{Wang}}\ and\ \citenamefont
  {{Lai}}(2020)}]{2020PhRvD.102h3005W}%
  \BibitemOpen
  \bibfield  {author} {\bibinfo {author} {\bibfnamefont {J.-S.}\ \bibnamefont
  {{Wang}}}\ and\ \bibinfo {author} {\bibfnamefont {D.}~\bibnamefont {{Lai}}},\
  }\href {\doibase 10.1103/PhysRevD.102.083005} {\bibfield  {journal} {\bibinfo
   {journal} {\prd}\ }\textbf {\bibinfo {volume} {102}},\ \bibinfo {eid}
  {083005} (\bibinfo {year} {2020})}\BibitemShut {NoStop}%
\bibitem [{\citenamefont {{Tucker}}\ and\ \citenamefont
  {{Will}}(2021)}]{2021PhRvD.104j4023T}%
  \BibitemOpen
  \bibfield  {author} {\bibinfo {author} {\bibfnamefont {A.}~\bibnamefont
  {{Tucker}}}\ and\ \bibinfo {author} {\bibfnamefont {C.~M.}\ \bibnamefont
  {{Will}}},\ }\href {\doibase 10.1103/PhysRevD.104.104023} {\bibfield
  {journal} {\bibinfo  {journal} {\prd}\ }\textbf {\bibinfo {volume} {104}},\
  \bibinfo {eid} {104023} (\bibinfo {year} {2021})}\BibitemShut {NoStop}%
\bibitem [{\citenamefont {{Poisson}}\ and\ \citenamefont
  {{Will}}(2014)}]{2014grav.book.....P}%
  \BibitemOpen
  \bibfield  {author} {\bibinfo {author} {\bibfnamefont {E.}~\bibnamefont
  {{Poisson}}}\ and\ \bibinfo {author} {\bibfnamefont {C.~M.}\ \bibnamefont
  {{Will}}},\ }\href {http://adsabs.harvard.edu/abs/2014grav.book.....P} {\emph
  {\bibinfo {title} {{Gravity}}}}\ (\bibinfo  {publisher} {Cambridge University
  Press},\ \bibinfo {year} {2014})\BibitemShut {NoStop}%
\bibitem [{\citenamefont {{Lincoln}}\ and\ \citenamefont
  {{Will}}(1990)}]{1990PhRvD..42.1123L}%
  \BibitemOpen
  \bibfield  {author} {\bibinfo {author} {\bibfnamefont {C.~W.}\ \bibnamefont
  {{Lincoln}}}\ and\ \bibinfo {author} {\bibfnamefont {C.~M.}\ \bibnamefont
  {{Will}}},\ }\href {\doibase 10.1103/PhysRevD.42.1123} {\bibfield  {journal}
  {\bibinfo  {journal} {\prd}\ }\textbf {\bibinfo {volume} {42}},\ \bibinfo
  {pages} {1123} (\bibinfo {year} {1990})}\BibitemShut {NoStop}%
\bibitem [{\citenamefont {{Blanchet}}(2014)}]{2014LRR....17....2B}%
  \BibitemOpen
  \bibfield  {author} {\bibinfo {author} {\bibfnamefont {L.}~\bibnamefont
  {{Blanchet}}},\ }\href {http://adsabs.harvard.edu/abs/2014LRR....17....2B}
  {\bibfield  {journal} {\bibinfo  {journal} {Living Reviews in Relativity}\
  }\textbf {\bibinfo {volume} {17}} (\bibinfo {year} {2014})}\BibitemShut
  {NoStop}%
\bibitem [{\citenamefont {{Duez}}\ \emph {et~al.}(2010)\citenamefont {{Duez}},
  \citenamefont {{Braithwaite}},\ and\ \citenamefont
  {{Mathis}}}]{2010ApJ...724L..34D}%
  \BibitemOpen
  \bibfield  {author} {\bibinfo {author} {\bibfnamefont {V.}~\bibnamefont
  {{Duez}}}, \bibinfo {author} {\bibfnamefont {J.}~\bibnamefont
  {{Braithwaite}}}, \ and\ \bibinfo {author} {\bibfnamefont {S.}~\bibnamefont
  {{Mathis}}},\ }\href {\doibase 10.1088/2041-8205/724/1/L34} {\bibfield
  {journal} {\bibinfo  {journal} {\apjl}\ }\textbf {\bibinfo {volume} {724}},\
  \bibinfo {pages} {L34} (\bibinfo {year} {2010})}\BibitemShut {NoStop}%
\bibitem [{\citenamefont {{King}}\ \emph {et~al.}(1990)\citenamefont {{King}},
  \citenamefont {{Frank}},\ and\ \citenamefont
  {{Whitehurst}}}]{1990MNRAS.244..731K}%
  \BibitemOpen
  \bibfield  {author} {\bibinfo {author} {\bibfnamefont {A.~R.}\ \bibnamefont
  {{King}}}, \bibinfo {author} {\bibfnamefont {J.}~\bibnamefont {{Frank}}}, \
  and\ \bibinfo {author} {\bibfnamefont {R.}~\bibnamefont {{Whitehurst}}},\
  }\href {https://ui.adsabs.harvard.edu/abs/1990MNRAS.244..731K} {\bibfield
  {journal} {\bibinfo  {journal} {\mnras}\ }\textbf {\bibinfo {volume} {244}},\
  \bibinfo {pages} {731} (\bibinfo {year} {1990})}\BibitemShut {NoStop}%
\bibitem [{\citenamefont {{Pablo}}\ \emph {et~al.}(2019)\citenamefont
  {{Pablo}}, \citenamefont {{Shultz}}, \citenamefont {{Fuller}}, \citenamefont
  {{Wade}}, \citenamefont {{Paunzen}}, \citenamefont {{Mathis}}, \citenamefont
  {{Le Bouquin}}, \citenamefont {{Pigulski}}, \citenamefont {{Handler}},
  \citenamefont {{Alecian}}, \citenamefont {{Kuschnig}}, \citenamefont
  {{Moffat}}, \citenamefont {{Neiner}}, \citenamefont {{Popowicz}},
  \citenamefont {{Rucinski}}, \citenamefont {{Smolec}}, \citenamefont
  {{Weiss}}, \citenamefont {{Zwintz}},\ and\ \citenamefont {{BinaMIcS
  Collaboration}}}]{2019MNRAS.488...64P}%
  \BibitemOpen
  \bibfield  {author} {\bibinfo {author} {\bibfnamefont {H.}~\bibnamefont
  {{Pablo}}}, \bibinfo {author} {\bibfnamefont {M.}~\bibnamefont {{Shultz}}},
  \bibinfo {author} {\bibfnamefont {J.}~\bibnamefont {{Fuller}}}, \bibinfo
  {author} {\bibfnamefont {G.~A.}\ \bibnamefont {{Wade}}}, \bibinfo {author}
  {\bibfnamefont {E.}~\bibnamefont {{Paunzen}}}, \bibinfo {author}
  {\bibfnamefont {S.}~\bibnamefont {{Mathis}}}, \bibinfo {author}
  {\bibfnamefont {J.~B.}\ \bibnamefont {{Le Bouquin}}}, \bibinfo {author}
  {\bibfnamefont {A.}~\bibnamefont {{Pigulski}}}, \bibinfo {author}
  {\bibfnamefont {G.}~\bibnamefont {{Handler}}}, \bibinfo {author}
  {\bibfnamefont {E.}~\bibnamefont {{Alecian}}}, \bibinfo {author}
  {\bibfnamefont {R.}~\bibnamefont {{Kuschnig}}}, \bibinfo {author}
  {\bibfnamefont {A.~F.~J.}\ \bibnamefont {{Moffat}}}, \bibinfo {author}
  {\bibfnamefont {C.}~\bibnamefont {{Neiner}}}, \bibinfo {author}
  {\bibfnamefont {A.}~\bibnamefont {{Popowicz}}}, \bibinfo {author}
  {\bibfnamefont {S.}~\bibnamefont {{Rucinski}}}, \bibinfo {author}
  {\bibfnamefont {R.}~\bibnamefont {{Smolec}}}, \bibinfo {author}
  {\bibfnamefont {W.}~\bibnamefont {{Weiss}}}, \bibinfo {author} {\bibfnamefont
  {K.}~\bibnamefont {{Zwintz}}}, \ and\ \bibinfo {author} {\bibnamefont
  {{BinaMIcS Collaboration}}},\ }\href {\doibase 10.1093/mnras/stz1661}
  {\bibfield  {journal} {\bibinfo  {journal} {\mnras}\ }\textbf {\bibinfo
  {volume} {488}},\ \bibinfo {pages} {64} (\bibinfo {year} {2019})}\BibitemShut
  {NoStop}%
\bibitem [{\citenamefont {{Shultz}}\ \emph {et~al.}(2015)\citenamefont
  {{Shultz}}, \citenamefont {{Wade}}, \citenamefont {{Alecian}},\ and\
  \citenamefont {{BinaMIcS Collaboration}}}]{2015MNRAS.454L...1S}%
  \BibitemOpen
  \bibfield  {author} {\bibinfo {author} {\bibfnamefont {M.}~\bibnamefont
  {{Shultz}}}, \bibinfo {author} {\bibfnamefont {G.~A.}\ \bibnamefont
  {{Wade}}}, \bibinfo {author} {\bibfnamefont {E.}~\bibnamefont {{Alecian}}}, \
  and\ \bibinfo {author} {\bibnamefont {{BinaMIcS Collaboration}}},\ }\href
  {\doibase 10.1093/mnrasl/slv096} {\bibfield  {journal} {\bibinfo  {journal}
  {\mnras}\ }\textbf {\bibinfo {volume} {454}},\ \bibinfo {pages} {L1}
  (\bibinfo {year} {2015})},\ \Eprint {http://arxiv.org/abs/1507.05084}
  {arXiv:1507.05084 [astro-ph.SR]} \BibitemShut {NoStop}%
\bibitem [{\citenamefont {{Brouwer}}\ and\ \citenamefont
  {{Clemence}}(1961)}]{1961mcm..book.....B}%
  \BibitemOpen
  \bibfield  {author} {\bibinfo {author} {\bibfnamefont {D.}~\bibnamefont
  {{Brouwer}}}\ and\ \bibinfo {author} {\bibfnamefont {G.~M.}\ \bibnamefont
  {{Clemence}}},\ }\href {http://adsabs.harvard.edu/abs/1961mcm..book.....B}
  {\emph {\bibinfo {title} {{Methods of celestial mechanics}}}}\ (\bibinfo
  {publisher} {New York; London: Academic press},\ \bibinfo {year}
  {1961})\BibitemShut {NoStop}%
\bibitem [{\citenamefont {{Murray}}\ and\ \citenamefont
  {{Dermott}}(2000)}]{2000ssd..book.....M}%
  \BibitemOpen
  \bibfield  {author} {\bibinfo {author} {\bibfnamefont {C.~D.}\ \bibnamefont
  {{Murray}}}\ and\ \bibinfo {author} {\bibfnamefont {S.~F.}\ \bibnamefont
  {{Dermott}}},\ }\href {http://adsabs.harvard.edu/abs/2000ssd..book.....M}
  {\emph {\bibinfo {title} {{Solar System Dynamics}}}}\ (\bibinfo  {publisher}
  {Cambridge University Press},\ \bibinfo {year} {2000})\BibitemShut {NoStop}%
\bibitem [{\citenamefont {{Wang}}\ \emph {et~al.}(2018)\citenamefont {{Wang}},
  \citenamefont {{Peng}}, \citenamefont {{Wu}},\ and\ \citenamefont
  {{Dai}}}]{2018ApJ...868...19W}%
  \BibitemOpen
  \bibfield  {author} {\bibinfo {author} {\bibfnamefont {J.-S.}\ \bibnamefont
  {{Wang}}}, \bibinfo {author} {\bibfnamefont {F.-K.}\ \bibnamefont {{Peng}}},
  \bibinfo {author} {\bibfnamefont {K.}~\bibnamefont {{Wu}}}, \ and\ \bibinfo
  {author} {\bibfnamefont {Z.-G.}\ \bibnamefont {{Dai}}},\ }\href {\doibase
  10.3847/1538-4357/aae531} {\bibfield  {journal} {\bibinfo  {journal} {\apj}\
  }\textbf {\bibinfo {volume} {868}},\ \bibinfo {eid} {19} (\bibinfo {year}
  {2018})}\BibitemShut {NoStop}%
\bibitem [{\citenamefont {{Mik{\'o}czi}}(2021)}]{2021arXiv210910722M}%
  \BibitemOpen
  \bibfield  {author} {\bibinfo {author} {\bibfnamefont {B.}~\bibnamefont
  {{Mik{\'o}czi}}},\ }\href@noop {} {\bibfield  {journal} {\bibinfo  {journal}
  {arXiv e-prints}\ ,\ \bibinfo {eid} {arXiv:2109.10722}} (\bibinfo {year}
  {2021})},\ \Eprint {http://arxiv.org/abs/2109.10722} {arXiv:2109.10722
  [gr-qc]} \BibitemShut {NoStop}%
\bibitem [{\citenamefont
  {{Chandrasekhar}}(1931{\natexlab{a}})}]{1931ApJ....74...81C}%
  \BibitemOpen
  \bibfield  {author} {\bibinfo {author} {\bibfnamefont {S.}~\bibnamefont
  {{Chandrasekhar}}},\ }\href {\doibase 10.1086/143324} {\bibfield  {journal}
  {\bibinfo  {journal} {\apj}\ }\textbf {\bibinfo {volume} {74}},\ \bibinfo
  {pages} {81} (\bibinfo {year} {1931}{\natexlab{a}})}\BibitemShut {NoStop}%
\bibitem [{\citenamefont
  {{Chandrasekhar}}(1931{\natexlab{b}})}]{1931MNRAS..91..456C}%
  \BibitemOpen
  \bibfield  {author} {\bibinfo {author} {\bibfnamefont {S.}~\bibnamefont
  {{Chandrasekhar}}},\ }\href {\doibase 10.1093/mnras/91.5.456} {\bibfield
  {journal} {\bibinfo  {journal} {\mnras}\ }\textbf {\bibinfo {volume} {91}},\
  \bibinfo {pages} {456} (\bibinfo {year} {1931}{\natexlab{b}})}\BibitemShut
  {NoStop}%
\bibitem [{\citenamefont {{Chandrasekhar}}(1935)}]{1935MNRAS..95..207C}%
  \BibitemOpen
  \bibfield  {author} {\bibinfo {author} {\bibfnamefont {S.}~\bibnamefont
  {{Chandrasekhar}}},\ }\href {\doibase 10.1093/mnras/95.3.207} {\bibfield
  {journal} {\bibinfo  {journal} {\mnras}\ }\textbf {\bibinfo {volume} {95}},\
  \bibinfo {pages} {207} (\bibinfo {year} {1935})}\BibitemShut {NoStop}%
\bibitem [{\citenamefont {{Chandrasekhar}}(1967)}]{1967aits.book.....C}%
  \BibitemOpen
  \bibfield  {author} {\bibinfo {author} {\bibfnamefont {S.}~\bibnamefont
  {{Chandrasekhar}}},\ }\href@noop {} {\emph {\bibinfo {title} {{An
  introduction to the study of stellar structure}}}}\ (\bibinfo  {publisher}
  {University of Chicago Press, 1939},\ \bibinfo {year} {1967})\BibitemShut
  {NoStop}%
\bibitem [{\citenamefont {{Peters}}\ and\ \citenamefont
  {{Mathews}}(1963)}]{1963PhRv..131..435P}%
  \BibitemOpen
  \bibfield  {author} {\bibinfo {author} {\bibfnamefont {P.~C.}\ \bibnamefont
  {{Peters}}}\ and\ \bibinfo {author} {\bibfnamefont {J.}~\bibnamefont
  {{Mathews}}},\ }\href {\doibase 10.1103/PhysRev.131.435} {\bibfield
  {journal} {\bibinfo  {journal} {Physical Review}\ }\textbf {\bibinfo {volume}
  {131}},\ \bibinfo {pages} {435} (\bibinfo {year} {1963})}\BibitemShut
  {NoStop}%
\bibitem [{\citenamefont {{Kocsis}}\ \emph {et~al.}(2012)\citenamefont
  {{Kocsis}}, \citenamefont {{Ray}},\ and\ \citenamefont {{Portegies
  Zwart}}}]{2012ApJ...752...67K}%
  \BibitemOpen
  \bibfield  {author} {\bibinfo {author} {\bibfnamefont {B.}~\bibnamefont
  {{Kocsis}}}, \bibinfo {author} {\bibfnamefont {A.}~\bibnamefont {{Ray}}}, \
  and\ \bibinfo {author} {\bibfnamefont {S.}~\bibnamefont {{Portegies
  Zwart}}},\ }\href {\doibase 10.1088/0004-637X/752/1/67} {\bibfield  {journal}
  {\bibinfo  {journal} {\apj}\ }\textbf {\bibinfo {volume} {752}},\ \bibinfo
  {eid} {67} (\bibinfo {year} {2012})}\BibitemShut {NoStop}%
\bibitem [{\citenamefont {{Goldreich}}\ and\ \citenamefont
  {{Lynden-Bell}}(1969)}]{1969ApJ...156...59G}%
  \BibitemOpen
  \bibfield  {author} {\bibinfo {author} {\bibfnamefont {P.}~\bibnamefont
  {{Goldreich}}}\ and\ \bibinfo {author} {\bibfnamefont {D.}~\bibnamefont
  {{Lynden-Bell}}},\ }\href {\doibase 10.1086/149947} {\bibfield  {journal}
  {\bibinfo  {journal} {\apj}\ }\textbf {\bibinfo {volume} {156}},\ \bibinfo
  {pages} {59} (\bibinfo {year} {1969})}\BibitemShut {NoStop}%
\bibitem [{\citenamefont {{Piro}}(2012)}]{2012ApJ...755...80P}%
  \BibitemOpen
  \bibfield  {author} {\bibinfo {author} {\bibfnamefont {A.~L.}\ \bibnamefont
  {{Piro}}},\ }\href {\doibase 10.1088/0004-637X/755/1/80} {\bibfield
  {journal} {\bibinfo  {journal} {\apj}\ }\textbf {\bibinfo {volume} {755}},\
  \bibinfo {eid} {80} (\bibinfo {year} {2012})}\BibitemShut {NoStop}%
\bibitem [{\citenamefont {{Lai}}(2012)}]{2012ApJ...757L...3L}%
  \BibitemOpen
  \bibfield  {author} {\bibinfo {author} {\bibfnamefont {D.}~\bibnamefont
  {{Lai}}},\ }\href {\doibase 10.1088/2041-8205/757/1/L3} {\bibfield  {journal}
  {\bibinfo  {journal} {\apjl}\ }\textbf {\bibinfo {volume} {757}},\ \bibinfo
  {eid} {L3} (\bibinfo {year} {2012})}\BibitemShut {NoStop}%
\bibitem [{\citenamefont {{Mathis}}\ and\ \citenamefont {{de
  Brye}}(2011)}]{2011A&A...526A..65M}%
  \BibitemOpen
  \bibfield  {author} {\bibinfo {author} {\bibfnamefont {S.}~\bibnamefont
  {{Mathis}}}\ and\ \bibinfo {author} {\bibfnamefont {N.}~\bibnamefont {{de
  Brye}}},\ }\href {\doibase 10.1051/0004-6361/201015571} {\bibfield  {journal}
  {\bibinfo  {journal} {\aap}\ }\textbf {\bibinfo {volume} {526}},\ \bibinfo
  {eid} {A65} (\bibinfo {year} {2011})}\BibitemShut {NoStop}%
\bibitem [{\citenamefont {{Mathis}}\ and\ \citenamefont {{de
  Brye}}(2012)}]{2012A&A...540A..37M}%
  \BibitemOpen
  \bibfield  {author} {\bibinfo {author} {\bibfnamefont {S.}~\bibnamefont
  {{Mathis}}}\ and\ \bibinfo {author} {\bibfnamefont {N.}~\bibnamefont {{de
  Brye}}},\ }\href {\doibase 10.1051/0004-6361/201118322} {\bibfield  {journal}
  {\bibinfo  {journal} {\aap}\ }\textbf {\bibinfo {volume} {540}},\ \bibinfo
  {eid} {A37} (\bibinfo {year} {2012})}\BibitemShut {NoStop}%
\end{thebibliography}%

\end{document}